\pdfoutput=1
\documentclass[11pt, reqno,preprint]{article}
\usepackage[utf8]{inputenc}
\usepackage{jheppub}
\usepackage{epsfig}
\usepackage{amssymb}
\usepackage{amsmath}
\usepackage{mathrsfs}
\usepackage{hyperref}
\usepackage{multirow}
\usepackage{feynmp-auto}
\usepackage[makeroom]{cancel}

\def\be{\begin{equation}}
\def\ee{\end{equation}}
\def\ba{\begin{eqnarray}}
\def\ea{\end{eqnarray}}
\newcommand{\bea}{\begin{eqnarray}}
\newcommand{\eea}{\end{eqnarray}}

\def\Li{\textrm{Li}}

\def\Hhex{{\cal H}^{\rm hex}}
\def\Hhexn{{\cal H}^{\rm hex}_n}
\def\Hzeta{{\cal H}^{\zeta}}
\def\EE{\mathcal{E}}
\def\Et{\tilde{E}}
\def\Disc{{\rm Disc}}
\def\Gcusp{\Gamma_{\rm cusp}}

\newcommand{\fwboxL}[2]{\text{\makebox[#1][l]{$#2$}}}

\font \rus= wncyr10
\newcommand{\shuffle}{\, \hbox{\rus x} \,} % stolen from 1102.1310

\def\eqn#1{eq.~(\ref{#1})}
\def\Eqn#1{Equation~(\ref{#1})}
\def\eqns#1#2{eqs.~(\ref{#1}) and~(\ref{#2})}

\def\Tab#1{Table~{\ref{#1}}}

\def\Eqn#1{Equation~(\ref{#1})}
\def\eqn#1{eq.~(\ref{#1})}
\def\eqns#1#2{eqs.~(\ref{#1}) and~(\ref{#2})}

\DeclareMathOperator{\sign}{sign}

\newcommand{\cA}{\begin{cal}A\end{cal}}

\newcommand{\cE}{\begin{cal}E\end{cal}}

\newcommand{\cN}{\begin{cal}N\end{cal}}

\newcommand{\cP}{\begin{cal}P\end{cal}}

\newcommand{\cR}{\begin{cal}R\end{cal}}
\newcommand{\cS}{\begin{cal}S\end{cal}}

\def\green#1{{\color{green}#1}}

\usepackage{tikz}
\usetikzlibrary{tikzmark}

\usetikzlibrary{decorations.pathreplacing,calc,arrows,decorations.markings}

\title{The Cosmic Galois Group and Extended Steinmann Relations
for Planar ${\cal N} = 4$ SYM Amplitudes}

\author{Simon~Caron-Huot,$^1$}
\author{Lance~J.~Dixon,$^{2,3,4,5}$}
\author{Falko Dulat,$^{2}$}
\author{Matt~von~Hippel,$^{6,7}$}
\author{Andrew~J.~McLeod$^{2,3,7}$}
\author{and Georgios~Papathanasiou$^{3,8}$}

\affiliation{$^1$ Department of Physics, McGill University, 
3600 Rue University, Montr\'eal, QC Canada H3A 2T8}

\affiliation{$^2$ SLAC National Accelerator Laboratory,
Stanford University, Stanford, CA 94309, USA}

\affiliation{$^3$ Kavli Institute for Theoretical Physics, 
UC Santa Barbara, Santa Barbara, CA 93106, USA}

\affiliation{$^4$ Institut f\"ur Physik and IRIS Adlershof,
Humboldt-Universit\"at zu Berlin, \\
Zum Gro\ss en Windkanal 6, D-12489 Berlin, Germany}

\affiliation{$^5$ Pauli Center, ETH Z\"urich and University of Z\"urich,
Z\"urich, Switzerland}

\affiliation{$^6$ Perimeter Institute for Theoretical Physics, 
Waterloo, Ontario N2L 2Y5, Canada}

\affiliation{$^7$ Niels Bohr International Academy, Blegdamsvej 17, 2100 Copenhagen, Denmark}

\affiliation{$^8$ DESY Theory Group, DESY Hamburg, Notkestra{\ss}e 85, D-22607 Hamburg, Germany}

\abstract{We describe the minimal space of polylogarithmic functions that is required to express the six-particle amplitude in planar ${\cal N}=4$ super-Yang-Mills theory through six and seven loops, in the NMHV and MHV sectors respectively. This space respects a set of extended Steinmann relations that restrict the iterated discontinuity structure of the amplitude, as well as a cosmic Galois coaction principle that constrains the functions and the transcendental numbers that can appear in the amplitude at special kinematic points. To put the amplitude into this space, we must divide it by the BDS-like ansatz and by an additional zeta-valued constant $\rho$. For this normalization, we conjecture that the extended Steinmann relations and the coaction principle hold to all orders in the coupling. We describe an iterative algorithm for constructing the space of hexagon functions that respects both constraints. We highlight further simplifications that begin to occur in this space of functions at weight eight, and distill the implications of imposing the coaction principle to all orders. Finally, we explore the restricted spaces of transcendental functions and constants that appear in special kinematic configurations, which include polylogarithms involving square, cube, fourth and sixth roots of unity.}

\emailAdd{schuot@physics.mcgill.ca}\emailAdd{lance@slac.stanford.edu}\emailAdd{dulatf@slac.stanford.edu}\emailAdd{mvonhippel@nbi.ku.dk}\emailAdd{amcleod@nbi.ku.dk}\emailAdd{georgios.papathanasiou@desy.de}

\preprint{
\begin{flushright} DESY 19-062 \\ HU-EP-19/05 \\SLAC--PUB--17414
\end{flushright}
}

\begin{document}
\hypersetup{pageanchor=false}
\maketitle
\hypersetup{pageanchor=true}
\begin{fmffile}{feyndiags}

\section{Introduction}

Planar $\cN=4$ super-Yang-Mills (SYM) theory~\cite{Brink:1976bc,Gliozzi:1976qd} has proven to be an increasingly fruitful laboratory in which to explore the structure of quantum field theory and its intersection with contemporary mathematics. Part of the beauty of this theory is that it respects both a conformal~\cite{Mandelstam:1982cb,Brink:1982wv,Howe:1983sr} and a dual conformal symmetry~\cite{Drummond:2006rz,Bern:2006ew,Bern:2007ct,Alday:2007hr,Drummond:2008vq}, the latter of which is associated with a duality between its amplitudes and light-like polygonal Wilson loops~\cite{Alday:2007hr,Drummond:2007aua,Brandhuber:2007yx,Drummond:2007cf,Drummond:2007au,Alday:2008yw,Adamo:2011pv}. Strictly speaking, dual conformal symmetry is broken by the infrared divergences of these amplitudes, but their divergent structure is known to all orders in the form of the BDS ansatz~\cite{Bern:2005iz}. While the finite and dual-conformal-invariant functions that remain after dividing by the BDS ansatz are currently only known at specific loop orders and particle multiplicities, they are increasingly being recognized to exhibit many interesting geometric, algebraic, and motivic features. In this article, we expound on some of these surprising properties.

The BDS ansatz first receives a nontrivial correction in six-particle
kinematics~\cite{Bartels:2008ce,Bern:2008ap,Drummond:2008aq}.
This correction can be expressed as a linear combination of
dual superconformal invariants (encoding the helicity structure of the
amplitude), multiplied by transcendental functions of kinematic invariants (dual
conformally invariant cross ratios) that can be expanded in the coupling.
For six particles, both ingredients are well
understood~\cite{Drummond:2008vq,Dixon:2011pw}. In particular, the
transcendental functions that enter these amplitudes are composed of iterated
integrals over \emph{dlog} differential forms (or multiple
polylogarithms~\cite{Chen,G91b,Goncharov:1998kja,Remiddi:1999ew,Borwein:1999js,Moch:2001zr}) of uniform transcendental weight $2L$ at $L$ loops.
The branch cut structure of
these polylogarithmic functions is made manifest by considering their iterated
total differential, often expressed in the form of the
\emph{symbol}~\cite{Goncharov:2010jf,Duhr:2011zq}, which exposes the collection
of dlogs, or the \emph{symbol alphabet}, that contribute to each function. The alphabet of dlog forms relevant to six-particle scattering is (conjecturally) known~\cite{Goncharov:2010jf,Dixon:2011pw}, and has been observed (along with the alphabets entering higher-multiplicity scattering amplitudes) to have intriguing connections~\cite{,Golden:2013xva,Golden:2014xqa,Drummond:2014ffa,Golden:2014pua} to cluster algebras~\cite{1021.16017,1054.17024,GSV,FG03b}.

Given this knowledge of the transcendental functions entering the six-particle
amplitude, it is possible to construct an ansatz for it at any loop order. By
imposing symmetries and physical constraints (such as universal behavior in
singular limits) on this ansatz, the \emph{hexagon function bootstrap} program
has succeeded in identifying the complete amplitude at six points through six
loops, as well as the maximally helicity violating (MHV) amplitude at seven
loops~\cite{Dixon:2011pw,Dixon:2011nj,Dixon:2013eka,Dixon:2014voa,Dixon:2014xca,Dixon:2014iba,Dixon:2015iva,Caron-Huot:2016owq,Caron-Huot:2019vjl}.
The main computational challenge is constructing the space of functions
in which the ansatz lies; there is an overabundance of physical constraints.
Using input from the cluster algebra
structure of the space of kinematics, a \emph{heptagon bootstrap} has also been
carried out at seven points through four
loops~\cite{Drummond:2014ffa,Dixon:2016nkn,Drummond:2018caf}.

These bootstrap procedures can be carried out at two possible levels:
either at the level of the symbol, thus omitting any information about the
contour the dlog forms should be integrated over,
or at the level of fully integrated functions.
In order to capture the entire functional form of the amplitude, while
still retaining many of the simplifications afforded by the symbol,
it is possible to supplement the symbol with integration boundary data
using the full Hopf algebra structure of polylogarithms, which upgrades
the symbol to a
coaction~\cite{Gonch2,Brown:2011ik,Brown1102.1312,Duhr:2012fh,Chavez:2012kn}.
That is, by specifying the full coaction of the amplitude, which essentially
amounts to supplementing its symbol with certain boundary values,
all information about the amplitude can be encoded.

Stated more simply, any multi-variate function can be specified by giving
all of its first derivatives and its value at one point.  For multiple
polylogarithms, the first derivatives are expressible as a linear combination
involving a finite set of polylogarithms of one lower weight.
(The number of terms in the linear combination is equal to the number of
letters in the symbol alphabet.) These functions can in turn be specified
by their derivatives and values at the same point, and so on,
until one reaches weight-one functions, i.e.~logarithms.
However, the full coaction contains other components, which are not merely
(iterated) first derivatives.

The coaction is a specialized realization of a more general number-theoretical
structure that is concerned with \emph{motivic
periods}~\cite{Gonch3,FBThesis,Brown1102.1312,2015arXiv151206410B}.
On very general grounds, motivic periods are expected to be described by
a huge motivic Galois group.  When the periods are restricted to
correspond to a particular
class of amplitudes, a particular quotient of the motivic Galois group
can appear, called a \emph{cosmic Galois
group}~\cite{Cartier2001,2008arXiv0805.2568A,2008arXiv0805.2569A,Brown:2015fyf}.
Analogous to the algebraic Galois group that acts on the roots of polynomials,
the cosmic Galois group is conjectured to act on particular classes
of periods or amplitudes, exposing relations among them.
Different cosmic Galois groups can appear for different physical problems.
For instance, periods in $\phi^4$ theory are pure numbers, the coefficients of
the ultraviolet divergences for primitively divergent graphs.
They have long been known to have interesting number-theoretic
properties~\cite{Broadhurst:1996kc}.  More recently it was
observed~\cite{Schnetz:2013hqa,Panzer:2016snt}
that $\phi^4$ periods show a certain stability
under a cosmic Galois group; namely, 
a certain component of the coaction of higher-loop $\phi^4$ periods is
composed \emph{entirely} of only lower-loop $\phi^4$ periods.
This so-called \emph{coaction principle} was proven for certain graphs~\cite{Brown:2015fyf} by embedding the phenomenon into the larger conjectural framework of cosmic Galois theory. The coaction principle was further verified in $\phi^4$ periods up to 11 loops~\cite{Panzer:2016snt}, and has also been observed to hold for the polylogarithmic part of the anomalous magnetic moment of the electron through four loops~\cite{Laporta:2017okg,Schnetz:2017bko}. Only certain numbers appear at lower loops, and the coaction principle makes predictions restricting the possible higher loop numbers. In string perturbation theory, similar structures have also been observed, connecting not different loop orders but rather different orders in the $\alpha'$ expansion of tree-level string amplitudes~\cite{Schlotterer:2012ny}.

In this paper, together with a companion paper~\cite{Caron-Huot:2019vjl}, we provide further evidence for the existence of a coaction principle in quantum field theory by analyzing the six-point amplitude in planar $\cN=4$ SYM.
To do so, we first characterize the minimal space of Steinmann hexagon functions needed to express the six-point amplitude through seven loops, as a subspace of the space ${\cal G}$ of generalized polylogarithms built from the hexagon symbol alphabet. This space can be decomposed in transcendental weight as
\begin{equation}
{\cal G}\ =\ \bigoplus_{n=0}^\infty \, {\cal G}_n \, ,
\end{equation}
where ${\cal G}_n$ denotes the space of weight-$n$ functions built from the hexagon symbol alphabet by carrying out $n$ iterated integrations. More precisely, we study the spaces of polylogarithms that appear in the (iterated) derivatives of the amplitude at successive loop orders. As hinted at above, this is most conveniently carried out using the coaction map
\be
\Delta({\cal G}) = {\cal G} \otimes {\cal G}^{\rm \mathfrak{dR}} \, ,
\label{DeltaG}
\ee
which sends (motivic) polylogarithms in $\mathcal{G}$ into into a tensor space of the original space $\mathcal{G}$ times a new de Rham space $\mathcal{G}^{\mathfrak{dR}}$. While functions in $\mathcal{G}$ can be thought of as a pairing between a differential form and a cycle (or integration contour), objects in $\mathcal{G}^{\mathfrak{dR}}$ should be thought of as pairings between differential forms and their associated duals.\footnote{We thank Claude Duhr for illuminating discussions on this topic.} (A familiar example of this pairing is provided by closed string amplitudes~\cite{Brown:2018omk}.) Concretely, this means that the objects in $\mathcal{G}^{\mathfrak{dR}}$ carry no information about the original contour of integration. 
The objects appearing in the left entry of the tensor product in
\eqn{DeltaG} can be seen as the transcendental part of the total
derivative of the object on the left-hand side, in the sense that the derivative
$\textrm{d}\mathcal{G}$ of an iterated integral obeys
\be
\Delta(\textrm{d}\mathcal{G})
= (\mathrm{id}\otimes\textrm{d})\Delta(\mathcal{G})\,,
\label{eq:DeltadG}
\ee
i.e.~it acts only on the last entry of the tensor product.

The coaction is coassociative, and therefore we can again apply the coaction to functions in the first factor of~\eqref{DeltaG}. In particular, we can map the amplitude to an object in ${\cal G} \otimes {\cal G}^{\rm \mathfrak{dR}} \otimes \cdots \otimes {\cal G}^{\rm \mathfrak{dR}}$ in which only logarithms (or rather, their de Rham avatars) appear in all but the first tensor factor.
The $L$-loop six-point amplitude provides six different transcendental
functions at weight $2L$; one is associated with the MHV amplitude
and five are associated with different components of the next-to-MHV (NMHV)
amplitude.  We would like to study the space of lower-weight functions
that can be generated from these weight-$2L$ functions.
In particular, we consider the functions appearing in the left-most
entry of the tensor product obtained from iterated application
of the coaction $\Delta$. Concretely, we can consider the $k$-fold iteration
of the coaction for ${\cal G}^{\rm \mathfrak{dR}}$ always of weight one,
which allows us to associate a set
of weight-$(2L-k)$ functions to the original weight-$2L$ functions.
Stated more simply, we construct the span of all the weight-$(2L-1)$ functions
appearing in the derivative of the amplitude, then compute all of their
derivatives and construct the span again, and repeat $k$ times.
We observe that the dimension of the weight-$(2L-k)$ function space
generated in this fashion increases with $k$ until it saturates,
usually around $k=L$.

The space $\Hhex \subset {\cal G}$ that we construct in this way
obeys a coaction principle, which we will explain further in
section~\ref{sec:cgg}, but which is encapsulated by the statement that
\be
\Delta \Hhex \subset \Hhex \otimes {\cal K}^\pi \,,
\label{eq:coaction_principle_intro}
\ee
where ${\cal K}^\pi$ is unimportant for now.  The important statement
in \eqn{eq:coaction_principle_intro} is that the left part of the coaction
on any element of $\Hhex$ is always in $\Hhex$, not just in ${\cal G}$.
Part of this statement is well known to physicists.  At symbol level,
when the left part of the coaction has weight one,
\eqn{eq:coaction_principle_intro} just says that for a given scattering
amplitude, to all loop orders, the first entry of its symbol can be
consistently restricted to a subset of the symbol alphabet, corresponding
to the location of physical branch cuts~\cite{Gaiotto:2011dt}. 
Furthermore, because derivatives commute with taking branch cuts,
as reflected in \eqn{eq:DeltadG}, the branch cut conditions apply to
all the functions obtained by taking derivatives of the loop
amplitudes, i.e.~they apply to all of $\Hhex$.
The same statements hold at function level, and this is the
essence of the hexagon function bootstrap as implemented
in ref.~\cite{Dixon:2013eka},
to restrict ${\cal G}$ to a subspace having good branch cuts.
The space $\Hhex$, like ${\cal G}$, has a decomposition,
\begin{equation}
\Hhex\ =\ \bigoplus_{n=0}^\infty \, \Hhex_n \,,
\end{equation}
i.e.~a grading by the weight $n$.

It was later realized that (for amplitudes normalized by
the BDS-like ansatz~\cite{Alday:2009dv,Yang:2010as})
there was also a consistent restriction on
the first {\it two} entries~\cite{Caron-Huot:2016owq}. 
This restriction, known as the Steinmann
relations~\cite{Steinmann,Steinmann2,Cahill:1973qp},
enforces the compatibility of branch cuts in different channels.
Again, because of the commutativity of derivatives and branch cuts,
these conditions automatically apply to all functions in $\Hhex$.

However, even the Steinmann restrictions are insufficient to
account for the number of functions in $\Hhex$.  For example, at weight
two they would permit a constant, the Riemann zeta value $\zeta_2 = \pi^2/6$,
to be a member of $\Hhex$.  It has no branch cuts, so it automatically
satisfies all branch-cut restrictions.  But when the derivatives
of the amplitudes are computed, $\zeta_2$ does not appear as
an independent element.  Neither does $\zeta_3$, whereas $\zeta_4$ does.
Our goal in this paper is to identify the {\it minimal} space
of functions $\Hhex$ which can contain the amplitudes and all their
derivatives, and to verify that \eqn{eq:coaction_principle_intro}
holds as generally as possible, not only for the full functions,
but also for constants that appear when the functions are evaluated
at specific kinematic points.

As was also mentioned in the companion paper~\cite{Caron-Huot:2019vjl},
\eqn{eq:coaction_principle_intro} is {\it not} obeyed for
the BDS-like normalized amplitude, but the situation can be
remedied simply by dividing the amplitudes
by a kinematical constant, $\rho$, which
depends on the coupling but at each order is a multiple zeta value.
At present, $\rho$ needs to be determined at each loop order,
and through seven loops, only Riemann zeta values appear in it.
Because it is a constant, $\rho$ does not affect the Steinmann relations.
The six-point amplitudes, normalized by the product of $\rho$ and
the BDS-like ansatz, and all their derivatives, are what we use to define
the space $\Hhex$.

Having thus identified the space $\Hhex$, we can search for any systematic constraints that it obeys to all orders.  One constraint is a generalization of the Steinmann relations.  While the Steinmann relations were originally formulated as constraints on the first two discontinuities of any amplitude, we observe that they are obeyed to all depths in the symbol of functions in $\Hhex$. That is, instead of just imposing restrictions on the first two entries of the symbol, these \emph{extended Steinmann relations} impose restrictions on \emph{all adjacent pairs} of symbol entries. There is a physical argument for why one should also impose the extended Steinmann relations. Namely, the Steinmann relations should hold on any Riemann sheet.  Moving from one sheet to another involves shifting functions by their discontinuities, and then by their discontinuities' discontinuities, and so on for generic Riemann sheets. At the level of the symbol, these operations correspond to removing successive initial entries of the symbol.  Thus they convert a condition between any pair of adjacent entries into the same one between the first two entries. The extended Steinmann relations can also be understood in the context of cluster algebras as the \emph{cluster adjacency} of the (appropriately normalized) amplitude~\cite{Drummond:2017ssj,Drummond:2018dfd}, which imply the extended Steinmann relations at all particle multiplicity~\cite{Golden:2019kks}. 

As mentioned earlier, there are also constraints on the members of $\Hhex$ that are transcendental {\it constants}, functions that are totally independent of the kinematics.  On general grounds, these constants are expected to be multiple zeta values (MZVs).  Through weight 12, there are 47 such MZVs.  However, the
only ones that we need to include as independent elements of $\Hhex$ are the
five that are even powers of $\pi$:
\be
\zeta_4 \,,\ \zeta_6 \,,\ \zeta_8 \,,\ \zeta_{10} \,,\ \zeta_{12} \,,\ \ldots.
\label{eq:indepzetas}
\ee
(Recall that $\zeta_2$ is not independent.)
Further constraints are also found to apply to the span of the
transcendental constants that appear as integration constants in this space.
We fix the integration constants at a special, symmetric point in the space
of kinematics in the bulk of the Euclidean region, called ``$(1,1,1)$'',
where the three kinematical variables (cross ratios) become unity.
At this point, all the functions in $\Hhex$ evaluate to MZVs, but only
particular linear combinations appear.  Because only particular combinations
appear, there is a nontrivial coaction principle at this point,
\be
\boxed{
\Delta \Hhex(1,1,1) \subset \Hhex(1,1,1) \otimes {\cal K}^\pi(1,1,1) \,. }
\label{eq:coaction_principle_111_intro}
\ee
If we had not divided by $\rho$, this principle would not be obeyed,
as explained in ref.~\cite{Caron-Huot:2019vjl} for the case of $(\zeta_3)^2$.
Thanks to $\rho$, we find that it is obeyed.  It may be that
\eqn{eq:coaction_principle_111_intro} is guaranteed given
\eqn{eq:coaction_principle_intro}, but in practice we can
check \eqn{eq:coaction_principle_111_intro} to much higher weight
than we can verify all the components of \eqn{eq:coaction_principle_intro}.

Although we have given a ``top-down'' definition of $\Hhex$, where
we compute loop amplitudes and then take their derivatives, there is
also a ``bottom-up'' approach, where we build the function space iteratively
in the weight.  We need the bottom-up approach past weight 7, at which point
we do not yet have enough derivatives to span the full space.  On the other
hand, we do have enough information about the independent constants
and the constants at $(1,1,1)$, to be able to construct the full 
function space $\Hhex$ through weight 11 (weight 12 up to a small ambiguity).

The constraints~\eqref{eq:indepzetas} on the independent constants,
in combination with the extended Steinmann relations, greatly reduce
the size of $\Hhex$, relative even to the earlier Steinmann hexagon
space~\cite{Caron-Huot:2016owq}.  The smaller size has made it possible to
bootstrap the MHV amplitude through seven loops and the NMHV
amplitude through six loops~\cite{Caron-Huot:2019vjl}.

We expect the coaction principle to hold in general kinematics.  However,
it is nontrivial to compute all components of the coaction for
general kinematics.  For a weight-$n$ function in $\Hhex$, the components of
the coaction with weight $\{n-k,1,\ldots,1\}$,
constructed by taking $k$ derivatives, give a weight $n-k$ function
that is in $\Hhex$ by construction.  However, the weight $\{n-k,k\}$
component of the coaction could contain a constant $\zeta_k$ in the second
entry, which would not be captured by the weight $\{n-k,1,\ldots,1\}$
component.  In order to investigate whether the coaction principle
holds for $\{n-k,k\}$ components for generic $k$, beyond the point $(1,1,1)$,
we study further kinematical points.  At many of these points,
transcendental constants beyond MZVs appear, such as alternating sums and
multiple polylogarithms evaluated at higher roots of unity.  To study the
coaction at these points, it is especially useful to work in terms of an
$f$-alphabet, which makes the coaction structure of these constants
manifest~\cite{Brown:2011ik,HyperlogProcedures}.  We also explore
particular dimension-one limits, i.e.~lines through the three-dimensional
space of cross ratios, in which the symbol alphabet simplifies to just
a few letters, and all functions in $\Hhex$ can be expressed as simpler
polylogarithms, usually harmonic polylogarithms~\cite{Remiddi:1999ew}. 
In all such limits, we find that the coaction principle holds.

The remainder of this paper is organized as follows: in section~\ref{sec:hex}, we set the stage for our discussion of the hexagon function space by describing the kinematical setup and defining the analytical properties of the space. In section~\ref{sec:stein} we discuss the extended Steinmann relations and show the restrictions they impose on the space of hexagon functions. Afterwards, in section~\ref{sec:construction} we show how the space of hexagon functions can be constructed in practice, including the determination of the constant boundary terms that are needed to promote the symbol expression to a full function. Equipped with a concrete realization of the function space, we can study the implications of the coaction principle and cosmic Galois theory on this space, which we describe in section~\ref{sec:cgg}. In section~\ref{sec:saturation} we focus on our top-down definition of $\Hhex$, examining when the space of functions that appears in the amplitude saturates for each weight. Section~\ref{sec:hexagon_limits} investigates the implications of the coaction principle on various lines and points within $\Hhex$. We conclude in section~\ref{sec:conclusions}.
Two appendices contain the values of the amplitudes at $(1,1,1)$ in the $f$-basis (\ref{appendix:fbasis}) and some empirical longer-range restrictions on symbol entries (\ref{appendix:longrange}). An ancillary file {\tt ftoMZV.txt} provides the conversion from the $f$-alphabet to MZVs through weight 14.

%%%%%%%%%%%%%%%%%%%%%%%%%%%%%%%%%%%%%%%%%%%%%%%%%%%

\section{Analytic Properties of the Six-Particle Amplitude}\label{sec:AnalyticProperties}
\label{sec:hex}

\subsection{Normalization and kinematic dependence}\label{subsec:kinematics}

The kinematic dependence of an amplitude in planar ${\cal N} = 4$ SYM is strongly constrained by dual conformal symmetry~\cite{Bern:2005iz,Drummond:2007au,Bern:2008ap,Drummond:2008aq,Drummond:2006rz,Bern:2006ew,Bern:2007ct,Alday:2007hr,Drummond:2008vq}. After normalizing by the BDS ansatz $\mathcal{A}^{\rm BDS}_n$, which accounts for the infrared divergences of the amplitude and an associated dual-conformal anomaly, the amplitude becomes finite and its kinematic dependence is restricted to dual-conformal-invariant cross ratios. Using ${\cal N} = 4$ supersymmetry, amplitudes with different external particles can be combined into a single superamplitude $\mathcal{A}_n$.  The superamplitude can be further factorized into an exponentiated \emph{remainder function} and an expansion $\cP_n$ in \emph{ratio functions} encoding the ratio of the N$^k$MHV superamplitude to the MHV one, as
\be \label{eq:BDS_regularized}
\mathcal{A}_n=\mathcal{A}^{\rm BDS}_n \times\exp({\cal R}_n)\times \cP_n\,.
\ee
The remainder function thus contains all nontrivial information about the MHV amplitude, and it is a bosonic, pure transcendental function of dual-conformally-invariant cross ratios. Restricting from now on to multiplicity $n=6$, which will be the focus of this article, only three such cross ratios can be formed, and they can be chosen to be
\be \label{uvw_def}
u = \frac{s_{12} s_{45}}{s_{123} s_{345}}\,, 
\qquad v = \frac{s_{23} s_{56}}{s_{234} s_{123}}\,, \qquad
w = \frac{s_{34} s_{61}}{s_{345} s_{234}}\,,
\ee
where $s_{i\dots j} \equiv (p_i + \dots + p_j)^2$ are Mandelstam invariants. Beyond MHV, the only other inequivalent helicity configuration is NMHV, for which the ratio function reads
\begin{align}
&\cP_{\rm NMHV}\ =\ \frac{1}{2}\Bigl[
 [(1) + (4)] V(u,v,w) + [(2) + (5)] V(v,w,u) + [(3) + (6)] V(w,u,v) 
\nonumber \\
&\hskip2.2cm 
+ [(1) - (4)] \tilde{V}(u,v,w) - [(2)-(5)] \tilde{V}(v,w,u)
  + [(3) - (6)] \tilde{V}(w,u,v) \Bigr] \,. \ \ \ 
\label{PVform}
\end{align}
In the latter equation, $V$ and $\tilde{V}$ are pure functions similar to $\cR_6$. They are accompanied by dual superconformal $R$-invariants denoted by $(f)$~\cite{Hodges:2009hk,Mason:2009qx}, which contain Grassmann variables and rational dependence on the kinematical variables. The precise form of the $R$-invariants will not be important for our purposes, but it may be found for example in ref.~\cite{Elvang:2013cua} or our companion paper~\cite{Caron-Huot:2019vjl}.

As we have reviewed so far, the computation of the six-particle amplitude of any helicity in ${\cal N} = 4$ SYM boils down to the determination of the functions $\cR_6, V$ and $\tilde{V}$, given the known form of the $R$-invariants $(f)$ and the BDS ansatz $\mathcal{A}^{\rm BDS}_6$. It is important to bear in mind, however, that the factorization \eqref{eq:BDS_regularized} is not unique. Apart from the infrared-divergent part, there is still freedom in choosing the finite piece that enters in the first, normalization factor. A main thesis of this article is that it is meaningful to \emph{tune} the definition of this normalization factor, such that the remaining finite, normalized amplitude becomes simpler to compute, and manifests certain important physical and mathematical properties. 

This strategy has already proven fruitful once in the past when
considering the causal properties of amplitudes.
The Steinmann relations~\cite{Steinmann,Steinmann2,Cahill:1973qp} govern
the consistency of multiple discontinuities in overlapping channels,
in particular those involving different three-particle invariants.
The BDS ansatz violates these conditions~\cite{Bartels:2008ce},
and therefore so does the amplitude normalized by the BDS ansatz.
However, the unique, dual conformal finite piece of $\mathcal{A}^{\rm BDS}_6$
that depends on three-particle invariants can be removed from the BDS ansatz,
yielding the so-called BDS-like ansatz~\cite{Alday:2009dv,Yang:2010as}.
When the amplitude is normalized by this latter ansatz, it obeys the
Steinmann relations~\cite{Caron-Huot:2016owq} (see also
ref.~\cite{Dixon:2016nkn}) which greatly reduces the size of the space of
functions to which it belongs and thus facilitates its determination,
as we will review in subsection~\ref{subsec:Steinmann}.
The part of the BDS ansatz that must be removed is
\be
\exp\biggl[ \frac{1}{4} \Gcusp \EE^{(1)} \biggr] \,,
\label{BDStoBDSlike}
\ee
where
\be
\EE^{(1)}(u,v,w) = \Li_2\Bigl(1-\frac{1}{u}\Bigr)
+ \Li_2\Bigl(1-\frac{1}{v}\Bigr) + \Li_2\Bigl(1-\frac{1}{w}\Bigr)\,,
\label{E1}
\ee
and $\Gcusp$ is the cusp anomalous dimension
for planar $\mathcal{N}=4$ SYM~\cite{Beisert:2006ez}.
The kinematic dependence of the factor~\eqref{BDStoBDSlike} is fixed
by the requirement that the Steinmann relations are preserved. However,
it could still be multiplied by a constant.

Indeed, we will see that it is advantageous to further redefine our normalization by a coupling-dependent constant $\rho$, such that the amplitude and its iterated derivatives respect a coaction principle. We denote the new normalization as ``cosmic'' to indicate invariance of the associated function space under a cosmic Galois group~\cite{Cartier2001,2008arXiv0805.2568A,2008arXiv0805.2569A,Brown:2015fyf}. All in all, the cosmically normalized functions $\cE$, specifying the MHV amplitude, as well as $E$ and $\tilde E$, associated with the NMHV one, will be related to their BDS-normalized analogs by
\be
\EE = \frac{{\cal A}_6}{\rho\, {\cal A}_6^{\rm BDS-like}}
    = \frac{1}{\rho} \, 
\exp\biggl[ \frac{1}{4} \Gcusp \EE^{(1)} + {\cal R}_6 \biggr]\,, \qquad E = \EE \times V, \qquad
\Et = \EE \times \tilde{V}\,.
\label{EXMHVtoR6}
\ee
We will quote the value of $\rho$ through seven loops in section \ref{sec:cgg}, see in particular eq.~\eqref{rho}, after describing the coaction principle giving rise to it. In practice we determine $\rho$ order by order in perturbation theory, in parallel with the amplitude; it forms part of the ansatz we use in order to identify the amplitude from within our minimal space of polylogarithmic functions $\Hhex$, with the procedure detailed in our companion paper~\cite{Caron-Huot:2019vjl}.

In the remainder of this section, we will discuss the class of functions the cosmically normalized amplitude (coefficients) $\EE$, $E$ and $\Et$ belong to, and their analytic properties.\footnote{To avoid confusion, note that in ref.~\cite{Caron-Huot:2016owq} the same notation was used for the BDS-like normalized amplitude coefficients, which are obtained from \eqref{EXMHVtoR6} after replacing $\rho\to1$.  At the level of the symbol (defined in the next subsection), the two normalizations are identical, because the symbol of $\rho$ is equal to unity.}

%%%%%%%%%%%%%%%%%%%%%%%%%%%%%%%%%%
\subsection{Multiple polylogarithms, coproducts and symbols}

For the $n$-particle amplitude in planar ${\cal N}=4$ SYM, the transcendental
functions entering the remainder function and the NMHV ratio function
(and hence also $\EE$, $E$ and $\Et$), are
expected to be \emph{multiple polylogarithms} (MPL) of weight $2L$ at any loop order $L$~\cite{ArkaniHamed:2012nw}. A function $F$ is defined to be an MPL
of weight $n$ if its total differential obeys
\be 
dF = \sum_{\phi_\beta\in{\Phi}} F^{\phi_\beta} d\ln \phi_\beta\,,
\label{dFPhi}
\ee
such that $F^{\phi_\alpha}$ is an MPL of weight $n-1$, satisfying
\be
dF^{\phi_\beta} = \sum_{\phi_\alpha\in{\Phi}} F^{\phi_\alpha,\phi_\beta} d\ln \phi_\alpha\label{dFPhiCop}\,,
\ee
and so on, with the recursive definition terminating with the usual logarithms
on the left-hand side at weight one, and rational numbers as coefficients of the
total differentials on the right-hand side corresponding to weight zero. The set
$\Phi$ of arguments of the dlogs is called the \emph{symbol alphabet}.
It encodes the positions of the possible branch-points of the
transcendental functions.
This iterative structure forms part of the Hopf algebra of MPLs. In
particular the \emph{coaction} operation $\Delta$ (sometimes loosely
referred to as a \emph{coproduct}), maps an MPL
of weight $n$ to linear combinations of pairs of MPLs with weight $\{n-k,k\}$
for $k=0,1,\ldots n$.

The $\{n-1,1\}$ component of $\Delta$ is essentially equivalent to the total differential \eqref{dFPhi}, and can be realized straightforwardly as
\be
\Delta_{n-1,1} F = \sum_{\phi_{\beta} \in \Phi} F^{\phi_\beta}\otimes
\big[\ln\phi_{\beta}\mod\,(i\pi)\big]\,.
\ee
Recall that in the general definition of the coaction, cf.~\eqn{DeltaG},
the second factor is an element of $\mathcal{G}^{\mathfrak{dR}}$ and thus
agnostic of the contour of integration of the original polylogarithm. This means
in particular, that the second entry of the coaction needs to be invariant under
analytic continuation, or shifts of the integration contour around poles. 
For multiple polylogarithms, all monodromies around poles are
proportional to $(i\pi)$.  Thus the required invariance can be realized
by modding the second entry of the coaction by $(i\pi)$.
In the following we will tacitly assume that the second entry of the coaction
is modulo monodromies, and we will suppress the explicit notation.

The coaction may be repeatedly applied to either the first or the second factor
of the pair, yielding a further decomposition. As a result of the
coassociativity of the coaction there is a unique decomposition of an MPL of weight $n$ into subspaces of MPLs with weight $\{k_1,\ldots,k_m\}$, such that $\sum_{i=1}^m k_i=n$. Denoting the projection of the coaction on each of these subspaces by $\Delta_{k_1,\ldots,k_m}$, the previous equations \eqref{dFPhi}--\eqref{dFPhiCop} may be rewritten as\footnote{In section \ref{sec:cgg} we will provide the general form of the coaction on MPLs, and provide more information on the relatively minor distinction between the latter and the coproduct.}
\begin{align}
\Delta_{n-1,1}F &= \sum_{\phi_\beta\in{\Phi}} F^{\phi_\beta} \otimes \ln \phi_\beta\,,
\label{Delta1}\\
\Delta_{n-2,1,1}F &= \sum_{\phi_\alpha,\phi_\beta\in{\Phi}} F^{\phi_\alpha,\phi_\beta} \otimes \ln \phi_\alpha \otimes \ln \phi_\beta\,.\label{Delta11}
\end{align}
We will colloquially refer to the leftmost factors $F^{\phi_\beta}, F^{\phi_\alpha,\phi_\beta}$ as the single and double coproducts of the function $F$.  Note that the relations \eqref{Delta1}--\eqref{Delta11} also hold when the leftmost factors are weight zero, i.e.~rational numbers.  Furthermore, maximally iterating the procedure we just described defines the  \emph{symbol},
\be\label{symbol}
S[F]=\Delta_{\fwboxL{27pt}{{\underbrace{1,\ldots,1}_{n\,\,\text{times}}}}}F=\sum_{\phi_{\alpha_1},\ldots,\phi_{\alpha_n}} F^{\phi_{\alpha_1},\ldots,\phi_{\alpha_n}}\,  \left[\ln {\phi_{\alpha_1}} \otimes\cdots \otimes\ln \phi_{\alpha_n}\right]\,,
\ee
where one typically also simplifies the notation by replacing $\ln \phi_{\alpha_i}\to \phi_{\alpha_i}$ for compactness. 

The \emph{symbol letters} $\phi_\alpha$ are algebraic functions of the variables that $F$ depends on. Particularly for the six-particle amplitude, there exist three independent variables, which may be chosen to be the cross ratios \eqref{uvw_def}, whereas the set of symbol letters or \emph{alphabet} is
\be
\Phi\to \mathcal{S} = \{u, v, w, 1-u ,1-v, 1-w, y_u, y_v, y_w \}\ , \label{eq:hex_letters}
\ee
with
\be
y_u = \frac{u-z_+}{u-z_-}\,, \qquad y_v = \frac{v-z_+}{v-z_-}\,, 
\qquad y_w = \frac{w - z_+}{w - z_-}\,,
\label{yfromu}
\ee
\be
z_\pm = \frac{1}{2}\Bigl[-1+u+v+w \pm \sqrt{\Delta}\Bigr]\,, 
\qquad \Delta = (1-u-v-w)^2 - 4 uvw\,.
\label{z_definition}
\ee
Parity acts as an inversion $y_i\to1/y_i$ on the variables $y_i\in\{y_u,y_v,y_w\}$, or equivalently it sends \mbox{$\sqrt{\Delta}\to-\sqrt{\Delta}$}, while leaving the cross ratios $u$, $v$, and $w$ invariant. Consequently, each point in $(u,v,w)$ space corresponds to two points in $(y_u,y_v,y_w)$ space, with parity-even functions taking the same value at both points, and parity-odd functions changing sign when going from one point to the other. In other words, while even functions are well-defined in cross-ratio space, odd functions are only defined up to a common overall sign.\footnote{For this reason, it may some times be more convenient to use another set of three independent variables, in which all letters become rational, such as the `$y$' variables, or cluster ${\mathcal X}$-coordinates \cite{Golden:2013xva,Parker:2015cia}.}

Given a symbol alphabet, any set with the same size, consisting of multiplicatively independent combinations of its letters, is also equivalent: it simply amounts to a linear change of basis in the equations \eqref{Delta1}--\eqref{symbol}. Taking advantage of this freedom, apart from $\cal S$ we will also define and make use of the following equivalent alphabet,
\be
\Phi\to \mathcal{S}' = \{a, b, c, m_u , m_v, m_w, y_u, y_v, y_w \} \, ,
\label{eq:new_hex_letters}
\ee
where
\begin{align}\label{eq:abc}
a = \frac{u}{v w}, \quad b = \frac{v}{u w}, \quad c = \frac{w}{u v},
\end{align}
\vspace{-.6cm}
\begin{align}
m_u = \frac{1-u}{u}, \quad m_v = \frac{1-v}{v}, \quad m_w = \frac{1-w}{w}.
\end{align}
As we will see later in this section, $\mathcal{S}'$ has the virtue of exposing important analytic properties of the (properly normalized) amplitude in the most transparent fashion.

Before closing this subsection, let us also record the form of the new letters in terms of the $y$-variables,
\begin{align}
a = \frac{y_u (1-y_v y_w)^2}{(1-y_u)^2 y_v y_w}, \quad
b = \frac{y_v (1-y_u y_w)^2}{y_u (1-y_v)^2 y_w}, \quad
c = \frac{y_w (1-y_u y_v)^2}{y_u y_v (1-y_w)^2},
\end{align}
\vspace{-.6cm}
\begin{align}
m_u = \frac{(1-y_u)(1-y_uy_vy_w)}{y_u(1-y_v)(1-y_w)}, \quad
m_v = \frac{(1-y_v)(1-y_uy_vy_w)}{y_v(1-y_w)(1-y_u)}, \quad
m_w = \frac{(1-y_w)(1-y_uy_vy_w)}{y_w(1-y_u)(1-y_v)},\nonumber
\end{align}
which illustrates explicitly how using $(y_u,y_v,y_w)$ as independent variables
rationalizes the alphabet.

\subsection{Integrability conditions}\label{subsec:integrability}

In the previous subsection we specified the alphabet of a particular class of
MPLs. However, not every word we can form from this alphabet corresponds
to a function. We need to integrate a word of differential forms,
drawn from our alphabet, along an integration contour (see eqn.~\eqref{eq:I_def}).
In general, the value of the integral will depend on the contour. Only
certain words can be lifted to functions that are independent of the details of
the contour but only depend on the endpoints (and the homotopy class of the
contour). The conditions for such homotopy
invariant words are that the double derivatives of $F$
with respect to two different independent variables should
commute, $d^2F=0$, or more explicitly
\be
\frac{\partial^2F}{\partial u_i\partial u_j}\ =\
\frac{\partial^2F}{\partial u_j\partial u_i} \,, \qquad i \neq j,
\label{eq:commdoublederiv}
\ee
where $u_1=u$, $u_2=v$, $u_3=w$.
This condition, when computed using eqs.~\eqref{dFPhi} and~\eqref{dFPhiCop}, induces linear relations between the double coproducts $F^{\phi_\alpha,\phi_\beta}$, known as the $\{n-2,1,1\}$ \emph{integrability conditions}. 

In particular, for the hexagon functions relevant for the six-particle amplitude in planar $\mathcal{N}=4$ SYM, the kinematic dependence of the nine-letter alphabet yields 26 linear equations between the 81 double coproducts. Integrability conditions only involve the antisymmetric combinations of double coproducts, which we denote by
\be
F^{[x,y]} \equiv F^{x,y} - F^{y,x} \,.
\label{eq:Fxycomm}
\ee
The hexagon function integrability conditions can be conveniently expressed
in the alphabet $\mathcal{S}'$, defined in \eqn{eq:new_hex_letters}, as
\begin{align}
 F^{[a,b]} &=0\,,\label{FabIntegrabilityFirst}\\
 F^{[a,m_u]} &=0\,, \\
 F^{[a,y_u]} &=0\,, \\
 F^{[a,y_v]}-F^{[a,y_w]} &=0\,,\\
 F^{[m_u,y_v]}-F^{[m_u,y_w]} &=0\,,
\end{align}
plus their two $a\to b\to c$ cyclic permutations,
\begin{align}
 F^{[m_u,m_v]}+F^{[m_u,m_w]} &=0\,, \\
F^{[m_w,a]}+F^{[b,m_w]}+F^{[m_u,m_w]}+F^{[y_u,y_v]} &=0\,,\label{FabIntegrabilityMiddle}
\end{align}
plus a single $a\to b\to c$ cyclic permutation, and finally
\begin{align}
 F^{[b,y_u]}+F^{[c,y_u]}+F^{[m_u,y_u]} &=0\,, \\
  F^{[a,y_v]}+F^{[c,y_u]}+F^{[m_v,y_v]} &=0\,,\\
 F^{[a,y_v]}+F^{[b,y_u]}+F^{[m_w,y_w]} &=0\,, \\
 F^{[b,y_u]}-F^{[c,y_u]}-F^{[m_v,y_u]}+F^{[m_w,y_u]} &=0\,, \\
 F^{[a,y_v]}-F^{[c,y_u]}-F^{[m_u,y_v]}+F^{[m_w,y_u]} &=0\,, \\
\hspace{-8pt} F^{[m_v,a]}+F^{[c,m_v]}+F^{[m_u,m_v]}+F^{[y_u,y_w]} &=0\,, \\
\hspace{-8pt} F^{[m_u,m_v]}-F^{[y_u,y_v]}+F^{[y_u,y_w]}-F^{[y_v,y_w]} &=0\,.\label{FabIntegrabilityLast}
\end{align}
For example, starting with the nine logarithms at weight one,
\eqn{eq:new_hex_letters}, we can form an
81-dimensional ansatz for the symbol of weight two functions,
cf.~\eqn{Delta11}. Solving the twenty-six integrability equations, we find
a 55-dimensional basis for the most general space of weight-two MPLs built from
the hexagon alphabet.  The integrability equations can be solved iteratively for all adjacent pairs of entries, and the resulting space of MPLs is denoted by 
${\cal G}$.

\subsection{Physical singularities and the Steinmann relations}\label{subsec:Steinmann}

While the six-particle amplitude certainly lies within ${\cal G}$, it turns out that it occupies a much smaller subspace thereof, due to additional analytic properties. The most elementary such property is a consequence of locality, known as the \emph{first-entry condition}. It states that in order for color-ordered planar amplitudes (of any multiplicity) in massless gauge theories to have physical singularities, the first entry of their symbol must necessarily be a Mandelstam invariant made of consecutive external momenta \cite{Gaiotto:2011dt}. If we additionally have dual conformal invariance, as is the case with $\cN=4$ SYM, this condition in particular picks out the cross ratios \eqref{uvw_def}, or equivalently the letters $a,b,c$ of the alphabet \eqref{eq:new_hex_letters}. With this restriction, it is evident that the subspace of MPLs in which the amplitude and its derivatives/coproducts live will just contain the three logarithms formed by these letters at weight one:
\be
\Hhex_1\ =\ \{ \ln a,\ \ln b,\ \ln c\}\ \equiv\ \{ \ln a_i \} \,.
\label{eq:wt1}
\ee
At weight two, the first-entry and integrability conditions allow only 9 of
the 55 most general MPLs with this alphabet at weight two, plus the constant
$\zeta_2$,
\be
\biggl\{
\Li_2\left(1-\frac{1}{u_i}\right),\ \ln^2a_i,\ \ln a_i \ln a_{i+1},\
\zeta_2 \biggr\} \,,
\qquad i=1,2,3,
\label{eq:wt2noSteinmann}
\ee
for a total of 10 weight two functions.

The next analytic constraints we will exploit are the Steinmann relations~\cite{Steinmann,Steinmann2,Cahill:1973qp}, which demand that the double discontinuities of any amplitude vanish when taken in overlapping channels. Focusing in particular on three-particle Mandelstam invariants, for the six-particle amplitude the Steinmann relations forbid the following overlapping discontinuities,
\be
\label{eq:disc_SteinmannA}
  \Disc_{s_{234}}\left(\Disc_{s_{345}}\left({\cal A}_6\right) \right)
= \Disc_{s_{345}} \left(\Disc_{s_{123}} \left({\cal A}_6\right) \right)
= \Disc_{s_{234}} \left(\Disc_{s_{123}} \left({\cal A}_6\right) \right)=0\,.
\ee
As already remarked in subsection \ref{subsec:kinematics}, these conditions carry over to the BDS-like or cosmically normalized amplitude defined in this paper, since in both cases the infrared-divergent normalization factor by which we divide $\cA_6$ has no dependence on three-particle invariants, and thus commutes with the discontinuities in \eqref{eq:disc_SteinmannA}. In contrast, $\mathcal{A}_6^{\rm BDS}$ does depend on three-particle invariants, therefore the BDS-normalized amplitudes (and also the functions $\cR_6, V$ and $\tilde{V}$) will generically have nonvanishing double discontinuities that only cancel out in the product \eqref{eq:BDS_regularized}.

At this point we can justify our choice of alternative alphabet $\cS'$ in \eqn{eq:new_hex_letters}:
each of the letters $\{a, b, c\}$ depends on only a single three-particle
Mandelstam invariant.  For example, $a$ contains only $s_{234}$ (and a number of two-particle invariants).  Therefore,  \eqn{eq:disc_SteinmannA} translates directly into the following simple conditions on the functions $F \equiv \EE,\,E,\,\Et$, defined in \eqref{EXMHVtoR6}, and as remarked earlier on all their
derivatives:
\be\label{eq:disc_Steinmann}
  \Disc_{a}\left(\Disc_{b}\left(F\right) \right)
= \Disc_{b} \left(\Disc_{c} \left(F\right) \right)
= \Disc_{a} \left(\Disc_{c} \left(F\right) \right)=0\,.
\ee
At the level of the symbol, taking a discontinuity around a given letter is
particularly simple: if the first entry of a term in the symbol is the letter
under consideration we clip it off and retain the remaining tail
(or de Rham part) of the symbol, otherwise
we discard the term. This means that we can recast the Steinmann relation in
\eqn{eq:disc_Steinmann} in the coproduct notation of \eqn{Delta11} as
\be\label{eq:FabSteinw2}
F^{a,b}=0\,,\quad\text{if $F$ is a function of weight two}\,,
\ee
plus two more cyclic permutations. We have not included the equations where the order of letters is reversed, as it can be easily checked that eqs.~\eqref{FabIntegrabilityFirst} and \eqref{eq:FabSteinw2} (as well as their cyclic permutations) automatically imply them.  Imposing \eqn{eq:FabSteinw2} in the most general space of MPLs with the alphabet \eqref{eq:new_hex_letters} takes us to a 52-dimensional subspace.

Finally, combining the last formula with the first-entry condition and integrability (plus certain beyond-the-symbol physical branch cut conditions we will review in subsection \ref{sec:symbols_to_functions}), defines what have been previously coined as the Steinmann Hexagon Functions~\cite{Caron-Huot:2016owq}.
The weight-one part of this space is still given by \eqn{eq:wt1}, but the
weight-two part is trimmed from the 10 functions in \eqn{eq:wt2noSteinmann}
down to seven:
\be
\biggl\{
\Li_2\left(1-\frac{1}{u_i}\right),\ \ln^2a_i,\ \zeta_2 \biggr\} \,,
\qquad i=1,2,3.
\label{eq:wt2Steinmann}
\ee
The reduction in the size of the space, compared with not imposing the Steinmann relation \eqref{eq:FabSteinw2}, is even more drastic at higher weight. Perhaps more importantly, it is possible to generalize this condition, with far-reaching consequences that we will now move on to discuss.

%%%%%%%%%%%%%%%%%%%%%%%%%%%%%%%%%%%%%%%%%%%%%%%%%%%%%%%%%%%%%

\section{The Extended Steinmann Relations}
\label{sec:stein}

While the first-entry condition and Steinmann relations restrict which
letters can appear in the two leftmost symbol entries of the six-point
amplitude, there are additional restrictions on the symbol entries
appearing at all further depths in the symbol. These restrictions arise
when we construct the higher-weight spaces iteratively in the weight
(see section~\ref{sec:construction}), by imposing the first
two entry conditions and integrability. Out of the 55 integrable weight-two
symbols, only 43 linear combinations of adjacent symbol entries actually
appear in the space of Steinmann hexagon functions.\footnote{Here
    we consider pairs of adjacent symbol entries in the \emph{middle} of the
    symbol, i.e.~not the first two entries, which are further
    restricted by the first entry condition, nor the last two entries,
    which for the amplitude are constrained by dual superconformal
    symmetry~\cite{CaronHuot:2011ky}.}
In other words, the branch-cut condition, integrability condition and Steinmann relations
jointly imply an additional 12 equations between double coproducts, on top of \eqref{FabIntegrabilityFirst}--\eqref{FabIntegrabilityLast} and  \eqref{eq:FabSteinw2}, which prohibit an equal number of integrable pairs of adjacent letters from appearing at any depth in the symbol. These equations may be written as
\begin{align}
F^{a,m_u} = 0, \qquad F^{a,y_v} = F^{a,y_w},& \nonumber\\
F^{m_u, y_v} + F^{y_v, m_w} = F^{m_u, y_u} + F^{y_w, m_w} ,& 
\label{eq:adjlettconstraint}\\
F^{m_v, m_u} + F^{y_u, y_v} + F^{y_w, y_w} = F^{y_u, y_w} + F^{y_w, y_v} ,&
\nonumber
\end{align}
plus cyclic permutations.

This simplification is only part of the story.  The space of adjacent symbol
entries appearing in the six-point BDS-like normalized amplitude itself is
yet smaller. To observe this, we consider (at symbol level)
the $L$-loop amplitudes,
and all components of the coaction $\Delta$ on them which take the form
$\Delta_{w_1,1,1,w_2}$, for any nonnegative integers $w_1$, $w_2$ satisfying
$w_1+w_2=2L-2$. The linear combination of adjacent symbol letters in the
weight-one slots, appearing between each independent pair of functions $f,g$ in
the $w_1,w_2$ slots, respectively, represents an independent weight-two symbol.
(See \eqn{eq:11notation} and the text below it for more details.)
We determine the span of \emph{all} weight-two symbols in these amplitudes
at a given loop order by simultaneously considering all allowed values
of $w_1$ and $w_2$.

Carrying out this analysis on all previously available results up to
five loops~\cite{Caron-Huot:2016owq}, it is found that only 40 linear
combinations of adjacent symbol entries actually appear in the
amplitude~\cite{Caron-Huot:2018dsv}. 
The three additional pairs of adjacent symbol entries that are present
in the Steinmann hexagon space we have presented so far,
but are absent in the amplitude, are\footnote{These results were initially reported at {\it Amplitudes 2017}, in a talk by one of the authors~\cite{YorgosAmps17}.}
\begin{align} 
\cancel{\ldots \otimes a\otimes b \otimes \ldots}, \quad
\cancel{\ldots \otimes b\otimes c \otimes \ldots}, \quad
\cancel{\ldots \otimes c\otimes a \otimes \ldots}\ . \label{eq:ExtSteinmannabc}
\end{align}
In other words, the amplitudes reside in a space smaller than previously thought, with the double coproduct \eqref{Delta11} of every function $F$ within this space obeying the extra condition
\be\label{eq:FabStein}
\boxed{F^{a,b}=0}\,,
\ee
plus cyclic permutations.  Comparison with \eqn{eq:FabSteinw2} reveals that this condition is precisely the application of the Steinmann relations to \emph{all} depths in the symbol, and we thus refer to \eqn{eq:FabStein} as the \emph{extended Steinmann relations}.

The extended Steinmann relations form an integral part of the refined hexagon function space $\Hhex$ that we will define in the upcoming sections, but as we can see already at the level of the symbol in Table \ref{tab:SymbolDim}, at weight 10 and above it leads to a more than 50\% reduction in the size of the space in which the six-particle amplitude needs to be identified. The extended Steinmann dimensions at symbol level agree with ref.~\cite{DFGPrivate}. As mentioned in the introduction, the extended Steinmann relations appear to follow from the physical requirement that the ordinary Steinmann relations hold not only in the Euclidean region, but also on any Riemann sheet.

%%%%%%%%%%%%%%%%%%%%%%%%%%%%%%%5
\renewcommand{\arraystretch}{1.25}
\begin{table}[!t]
\centering
\begin{tabular}[t]{l c c c c c c c c c c c c c c}
\hline\hline
weight $n$
& 0 & 1 & 2 & 3 & 4 &  5 &  6 &  7 &  8 &  9 & 10 & 11 & 12 & 13 \\
\hline\hline
First entry
& 1& 3 & 9 & 26 & 75 & 218 & 643 & 1929 & 5897 & ? & ? & ? & ? & ? 
\\\hline
Steinmann
& 1 & 3 & 6 & 13 & 29 & 63 & 134 & 277 & 562 & 1117 & 2192 & 4263 & 8240 & ?
\\\hline
Ext.~Stein.
& 1 & 3 & 6 & 13 & 26 & 51 & 98 & 184 & 340 & 613 & 1085 & 1887 & 3224 & 5431 
\\\hline\hline
\end{tabular}
\caption{The dimensions of the hexagon, Steinmann hexagon, and extended Steinmann hexagon spaces at symbol level.}
\label{tab:SymbolDim}
\end{table}
%%%%%%%%%%%%%%%%%%%%%%%%%%%%

To express this allowed 40-dimensional space of adjacent symbol entries~\cite{Caron-Huot:2018dsv}, we adopt the notation
\be\label{eq:11notation}
f_i \otimes \ln x \otimes \ln y \otimes g_j \quad \Rightarrow \quad [x,y]\,,
\ee
so that a sum of $[x,y]$ denotes symbols of the form
\ba
&f_i \otimes \ln x \otimes \ln y \otimes g_j + f_i \otimes \ln z \otimes \ln w \otimes g_j  \quad \Rightarrow  \quad [x,y]+[z,w]\,.
\ea
We emphasize that weight-two symbols should only be isolated in this way when each term appears between the same functions $f_i$ and $g_j$, which should themselves be linearly independent from the other functions appearing in the first and last coproduct entries.  Also note that the bracket notation here, unlike in \eqn{eq:Fxycomm}, does {\it not} imply any commutator or antisymmetrization. To denote cyclic classes, we write $a_i\in\{a,b,c\}$, $m_i\in\{m_u,m_v,m_w\}$, and $y_i\in\{y_u,y_v,y_w\}$, where $i\neq j\neq k$. In this notation, the 16 allowed odd pairs are
\bea
[a_i,y_i]+[y_i,a_i],\nonumber\\ 
{[}a_i,y_j y_k]+[y_j y_k,a_i],\nonumber\\ 
{[}m_j/m_k,y_i]+[y_i,m_j/m_k],\label{oddintegpairs}\\ 
{[}m_i,y_u y_v y_w]+[y_u y_v y_w,m_i],\nonumber\\ 
{[}a_i m_i,y_j y_k]-[m_j, y_j]-[m_k,y_k]-[y_j y_k,a_i m_i]+[y_j,m_j]+[y_k,m_k],\nonumber\\ 
{[}m_u,y_v y_w]+[m_v,y_u y_w]+[m_w,y_u y_v]-[y_v y_w,m_u]-[y_u y_w,m_v]-[y_u y_v,m_w],\nonumber
\eea
while the 24 allowed even pairs are
\bea
[a_i,a_i],\nonumber\\
{[}m_i,m_i],\nonumber\\
{[}a_i,m_j]+[m_j,a_i],\quad [a_i a_j,m_k],\nonumber\\
{[}m_j,m_k]+[m_k,m_j]-[y_i,y_i],\label{evenintegpairs} \\
{[}a_i,m_j m_k]+[y_i,y_u y_v y_w],\nonumber\\
{[}y_u,y_u^2 y_v y_w]+[y_u^2 y_v y_w,y_u],\quad [y_v,y_u y_v^2 y_w]+[y_u y_v^2 y_w,y_v],\nonumber\\
{[}a,m_v]+[m_u,m_v]-[m_w,b]+[m_w,m_u]-[m_w,m_v]+[y_v,y_w].\nonumber
\eea
The adjacent symbol entries of the double pentaladder integrals (which contribute to the six-point amplitude at all loop orders) are also contained within this space~\cite{Caron-Huot:2018dsv}.

Let us reiterate that the 15 constraints embodied by eqs.~\eqref{oddintegpairs}
and \eqref{evenintegpairs}, which reduce the allowed adjacent pairs from
55 to 40, are empirically consequences of just the three
constraints~\eqref{eq:FabStein}
together with the first-entry condition.
While we have not been able to prove this connection analytically,
we have verified that it holds at least to weight 13 at symbol level,
and weight 11 at function level.

It is interesting that the combination of first-entry, integrability and Steinmann conditions have a ``nonlocal'' effect anywhere in the symbol, which may be equivalently recast in terms of the local equations \eqref{eq:adjlettconstraint}, as we observed at the beginning of this subsection. Such a local restriction can alternatively be accomplished using cluster
adjacency~\cite{Drummond:2017ssj,Drummond:2018caf}, which
is often phrased in terms of non-dual-conformally-invariant four brackets. While we do not need to impose the equations \eqref{eq:adjlettconstraint} when constructing our minimal space $\Hhex$ recursively in the weight, as they follow for free (empirically) given~\eqn{eq:FabStein}, we do have to impose them when relaxing the first entry condition, in order to study the full space of symbols that is expected to appear in any middle $w$ entries of the BDS-like or cosmically normalized amplitudes at arbitrary loop order.

Explicitly constructing this space, we find that its dimension is $\{9, 40, 140, 432, 1233, 3340\}$ at weights $w=1,2,3,4,5,6$. These dimensions coincide with an analysis of the cluster-adjacency condition~\cite{Drummond:2017ssj,DFGPrivate}. For comparison, $\{9, 55, 285, 1351\}$ analogous non-Steinmann satisfying symbols were reported for $w=1,2,3,4$ in eq.~(3.2) of ref.~\cite{Harrington:2015bdt}. The five-loop amplitudes saturate the 140-dimensional weight-three space but not the 432-dimensional weight-four space. The six-loop amplitudes saturate this latter space, but not the 1233-dimensional weight-five space.

Notice that \eqns{eq:adjlettconstraint}{eq:FabStein} only allow the 
symbol letters $a$, $b$, and $c$ to appear adjacent to the letters
\begin{align} 
\mathcal{S}_{\text{a}} &= \{a,m_v,m_w,y_u,y_v y_w\}, \nonumber \\
\mathcal{S}_{\text{b}} &= \{b,m_w,m_u,y_v,y_w y_u\},
\label{eq:adjacent_letters_constraint} \\
\mathcal{S}_{\text{c}} &= \{c,m_u,m_v,y_w,y_u y_v\}. \nonumber
\end{align} 
It would be nice to develop a physical intuition for what the restrictions such as~\eqn{eq:adjacent_letters_constraint} are enforcing. To do so, we search for analogous restrictions on pairs of letters that are not adjacent but at larger separation in the symbol. We consider first next-to-adjacent symbol entries. While all nine hexagon letters~\eqref{eq:new_hex_letters} can be next-to-adjacent to all other hexagon letters, a different type of restriction still occurs. In particular, only special linear combinations of letters appear \emph{between} letters that are not allowed to be adjacent by the constraint~\eqref{eq:adjacent_letters_constraint}. 

Consider for instance the 140-dimensional space of weight-three symbols obeying the constraints~\eqref{eq:adjlettconstraint} and \eqref{eq:FabStein} but not the first-entry condition. Any symbol letter that appears between $a$ and $b$ must reside in both $\mathcal{S}_{\text{a}}$ and $\mathcal{S}_{\text{b}}$. There are only two possible letters, $m_w$ and $y_u y_v y_w$. However, the term $a \otimes y_u y_v y_w \otimes b$ never appears in any integrable symbol, leaving just a single term of this form, $a \otimes m_w \otimes b$. Intriguingly, there also exists a clear physical difference between these two terms. Consider the kinematic limit where both discontinuities in $a \sim s_{234}$ and $b \sim s_{345}$ are simultaneously accessible.  As these variables go to zero, $w = 1/\sqrt{ab} \to \infty$ and so $m_w = (1-w)/w$ approaches a constant, and the symbol $a \otimes m_w \otimes b$ vanishes. 
On the other hand, $y_u y_v y_w \to w/(uv) \to \infty$ as $w \to \infty$ with $u,v$ fixed, so the symbol $a \otimes y_u y_v y_w \otimes b$ remains nonzero in the region probed by the Steinmann relations for the overlapping channels $s_{234}$ and $s_{345}$. (See also the discussion in appendix~\ref{sec:long_range_symb_restrictions}.) Perhaps the vanishing of $a \otimes m_w \otimes b$ in this region is suppressing a subleading overlapping branch-cut singularity, thus explaining why this combination can appear, and not $a \otimes y_u y_v y_w \otimes b$. In fact, this interpretation can be extended to higher depths in the symbol, and to sequences of iterated discontinuities between any pair of symbol letters that are restricted by~\eqn{eq:adjacent_letters_constraint}, as we show in appendix~\ref{sec:long_range_symb_restrictions}.

While not the focus of this article, for the amplitude with $n=7$ particles the usual Steinmann relations \cite{Dixon:2016nkn} may also be extended to apply anywhere in the symbol. Intriguingly, for both $n=6,7$ it has been found that the space of integrable symbols with physical branch cuts respecting them is also uniquely picked out by the principle of ``cluster adjacency''~\cite{Drummond:2017ssj,Drummond:2018dfd}. This principle states that symbol letters can only appear next to each other when they also appear together in a cluster of Gr(4,$n$). (See refs.~\cite{ArkaniHamed:2012nw} and \cite{Golden:2013xva} for more background on how cluster algebras appear in the integrand and kinematic space of planar ${\cal N}=4$ SYM amplitudes, respectively.)  This condition has also helped in determining the four-loop NMHV seven-particle amplitude \cite{Drummond:2018caf}. Like the extended Steinmann relations, cluster adjacency gives rise to a set of constraints that are expected to be obeyed by all BDS-like normalized amplitudes. (Cluster algebras also encode information about which symbol letters are allowed to appear in the amplitude at larger separations~\cite{Drummond:2018dfd}.) While no BDS-like ansatz can be formed when $n$ is a multiple of four~\cite{Alday:2010vh,Yang:2010az}, generalized BDS normalizations can be formed that make the Steinmann relations manifest for any number of particles~\cite{Golden:2018gtk}, in which cluster adjacency can also be shown to hold~\cite{Golden:2019kks}. As shown in the latter reference, cluster adjacency implies the extended Steinmann relations at all $n$, however it is not yet known whether these two conditions are equivalent in integrable symbols that have physical branch cuts more generally.

\section{Constructing \texorpdfstring{$\Hhex$}{Hhex}} \label{sec:constructing_the_space}
\label{sec:construction}

In this section, we describe our general
procedure for building the function space
$\Hhex$ relevant for six-particle scattering in $\mathcal{N}=4$ SYM up to
weight 12 (as well as to weight 13 at symbol level, and to weight 14 for MHV final entries).
We incorporate in particular the extended Steinmann relations, the evidence for
which we described in the previous section.  At function level, we have
to maintain the proper branch cuts and Steinmann relations, and these
conditions fix certain zeta values~\cite{Dixon:2013eka,Caron-Huot:2016owq}.
Here we impose another restriction on $\Hhex$:
we only include constant functions (MZVs) as independent elements
of the function space when we are forced to.  We will find that very few
such independent constants are required.  Another surprising aspect
is that certain symbols that pass all symbol-level conditions cannot
be completed to functions passing all the zeta-valued conditions,
starting at weight eight.  We will defer the latter details until
section~\ref{sec:saturationfull}, after discussing
the coaction principle.

The function space $\Hhex$
was an essential ingredient in the determination of the six-loop NMHV and
seven-loop MHV six-particle amplitudes in a companion
paper \cite{Caron-Huot:2019vjl}.
It also provides an important testing ground for
elucidating the precise form of a coaction principle on this space,
to be discussed in the next section.

In addition to imposing the extended Steinmann relations and zeta-valued restrictions just mentioned, there are two new technical aspects of our approach to constructing $\Hhex$. First, instead of the original symbol alphabet \eqref{eq:hex_letters}, we use the multiplicatively equivalent alphabet \eqref{eq:new_hex_letters}, which maximally simplifies the (extended) Steinmann relations, as well as the MHV final-entry condition.  (In these respects, it is similar to the choice of alphabet for the seven-particle amplitude bootstrap \cite{Drummond:2014ffa}.) Second, we adopt the method described in refs.~\cite{Dixon:2013eka,Dixon:2016nkn}, also building on the latter reference, for representing and constructing integrable symbols, and functions, in terms of sparse tensors with purely numeric, integer entries. These new aspects drastically reduce the complexity of the linear systems one has to solve in the process of building the function space, thus allowing one to push the latter to higher weights.

\subsection{Representing coproducts efficiently}
As we saw in sections \ref{sec:AnalyticProperties} and \ref{sec:stein}, the simplest space containing the six-particle amplitude consists of MPLs with alphabet \eqref{eq:new_hex_letters} whose first symbol entry contains the letters \eqref{eq:abc} and whose 81 double coproducts \eqref{Delta11} obey the 26+3 integrability relations~\eqref{FabIntegrabilityFirst}--\eqref{FabIntegrabilityLast} plus extended Steinmann relations \eqref{eq:FabStein}. More generally, once we have specified our set of symbol letters $\Phi$ with size $|\Phi|$, then any set of $l$ linearly independent equations on the double coproducts of the functions we wish to construct is fully encoded in a
$l \times |\Phi|\times |\Phi|$ tensor $D$,
\be
\sum_{\alpha,\beta=1}^{|\Phi|} D_{m\alpha\beta} \, F^{\phi_{\alpha},\phi_{\beta}}=0\,,\qquad m=1,2,\ldots,l\,.\label{DoubleCopMatrix}
\ee
In a similar vein, if we have a basis 
\be
F^{(n)}_{i_n}\,,\qquad i_n=1,2,\ldots, d_n\,,
\ee
of multiple polylogarithms obeying any given set of constraints of the form \eqref{DoubleCopMatrix} at weight $n$, then the $\{n-1,1\}$ coproduct component of each basis element may be represented as a $d_n\times d_{n-1}\times |\Phi|$ tensor $T$,
\be\label{tensor_coproduct_single}
\Delta_{n-1,1}F^{(n)}_{i_n}=\sum_{i_{n-1},\alpha} T^\alpha_{i_n,i_{n-1}}F^{(n-1)}_{i_{n-1}}\otimes \ln \phi_\alpha\,,
\ee
once we have also specified the corresponding basis $F^{(n-1)}_{i_{n-1}}$ at one weight fewer, in addition to the alphabet $\Phi$. This representation of the relevant function space in terms of matrices and tensors is extremely efficient~\cite{Dixon:2016nkn}, owing to the fact that the entries of $T$ are simply rational numbers, and $T$ is usually very sparse.

We may generalize the above representation to any $\{n-k,1,\ldots,1\}$ coproduct,
\be\label{Coproduct_nk}
\Delta_{n-k,\fwboxL{27pt}{{\underbrace{1,\ldots,1}_{k\,\,\text{times}}}}}F^{(n)}_{i_n}=\sum_{i_{n-k},\alpha_1,\ldots,\alpha_k} T^{\alpha_1,\ldots,\alpha_k}_{i_n,i_{n-k}}F^{(n-k)}_{i_{n-k}}\otimes \ln\phi_{\alpha_1}\otimes\cdots\otimes\ln\phi_{\alpha_k}\,,
\ee
with
\be\label{Tnk}
T^{\alpha_1,\ldots,\alpha_k}_{i_n,i_{n-k}}=\sum_{i_{n-1},\ldots,i_{n-k+1}}T^{\alpha_k}_{i_n,i_{n-1}}T^{\alpha_{k-1}}_{i_{n-1},i_{n-2}}\cdots T^{\alpha_{1}}_{i_{n-k+1},i_{n-k}}\,,
\ee
which is also valid for $k=1$ provided no summation is implied in that case. Finally, we may extend this notation to the case where $k=n$, for which 
there exists a single index $i_0=1$.  So for example at weight one,
with $k=n=1$, then $T^{\alpha}_{i_{1},1}$ essentially becomes a matrix rather than a tensor, and without loss of generality we can also set set $F_{1}^{(0)}=1$ for the basis element multiplying it, since the latter is now just a rational number. For example in the ordered alphabet \eqref{eq:new_hex_letters}, we choose the weight-one extended Steinmann hexagon functions~\eqref{eq:wt1}, and thus their corresponding matrix representation, as
\be\label{eq:w1hex}
F^{(1)}_1=\ln a\,,\,\,F^{(1)}_2= \ln b\,,\,\,F^{(1)}_3=\ln c\,
\qquad\Leftrightarrow\qquad
T^{\alpha}_{i_{1},1}=\left(
\begin{array}{ccccccccc}
 1 & 0 & 0 & 0 & 0 & 0 & 0 & 0 & 0 \\
 0 & 1 & 0 & 0 & 0 & 0 & 0 & 0 & 0 \\
 0 & 0 & 1 & 0 & 0 & 0 & 0 & 0 & 0 \\
\end{array}
\right)\,,
\ee
with rows labeled by $i_1$ and columns by $\alpha$.

\subsection{Constructing integrable symbols via tensors}
Let us now describe how we iteratively construct the space of symbols of a given alphabet, subject to the integrability conditions as well as any other linear constraints on their double coproducts such as the extended Steinmann relations. Suppose we already have a basis of such symbols $F^{(n)}_{i_{n}}$ at weight $n$. Then, the $\{n,1\}$ coproduct of any function $F$ of the same alphabet at weight $n+1$ lies in the tensor product space with elements
\be\label{tensor_product_n1}
F^{(n)}_{i_{n}}\otimes \ln \phi_\beta\,\,,\,
\, \forall \, i_n,\beta\,.
\ee
We can thus form an ansatz,
\be\label{n1_ansatz}
\Delta_{n,1} F=\sum_{j,\gamma,i_{n},\beta}  c_{j\gamma}L_{j i_{n}}^{\gamma\beta}F^{(n)}_{i_{n}}\otimes \ln \phi_\beta\,,
\ee
where the `$c$'s are yet-to-be determined coefficients, and $L$ is a known tensor, which in the most generic case can be chosen as
\be
L_{j i_{n}}^{\gamma\beta}=\delta_{ji_{n}}\delta^{\gamma\beta}\,,
\ee 
corresponding to the largest possible ansatz with $d_n\times |\Phi|$ variables, namely the case where we attach an independent unknown coefficient to each element of the tensor product space \eqref{tensor_product_n1}. 

In order to reduce the initial size of our ansatz, we may however make more restricted choices exploiting any additional property or symmetry of the function space. For example, if we wish to restrict ourselves to the weight-$(n+1)$ hexagon function space with MHV final entries ($\EE^a=\EE^b=\EE^c=0$), we may choose
\be\label{LMHV}
L_{j i_{n}}^{\gamma\beta}=\delta_{ji_{n}}L_{\text{MHV}}^{\gamma\beta}\,,\quad L_{\text{MHV}}^{\gamma\beta}=\left(
\begin{array}{ccccccccc}
 0 & 0 & 0 & 1 & 0 & 0 & 0 & 0 & 0 \\
 0 & 0 & 0 & 0 & 1 & 0 & 0 & 0 & 0 \\
 0 & 0 & 0 & 0 & 0 & 1 & 0 & 0 & 0 \\
 0 & 0 & 0 & 0 & 0 & 0 & 1 & 0 & 0 \\
 0 & 0 & 0 & 0 & 0 & 0 & 0 & 1 & 0 \\
 0 & 0 & 0 & 0 & 0 & 0 & 0 & 0 & 1 \\
\end{array}
\right)\,.
\ee
Similarly, we can choose the tensor $L$ so as to construct and solve ans\"atze for the parity-even and -odd functions separately. Indeed, we have found it advantageous to construct our extended Steinmann hexagon symbol space in this manner, as it leads to smaller and simpler systems of equations.

Once we have built an ansatz of the form \eqref{n1_ansatz} at weight $n+1$, the next step is to enforce the appropriate conditions \eqref{DoubleCopMatrix} on its $\{n-1,1,1\}$ coproduct components. By virtue of eqs.~\eqref{tensor_coproduct_single} and \eqref{n1_ansatz} we may show, analogously to eqs.~\eqref{Coproduct_nk}--\eqref{Tnk}, that the double coproduct of our ansatz for the function $F$ will be
\be
F^{\phi_{\alpha},\phi_{\beta}}=\sum_{j,\gamma,i_{n},i_{n-1}}  c_{j\gamma}L_{j i_{n}}^{\gamma\beta}T^\alpha_{i_n,i_{n-1}}F^{(n-1)}_{i_{n-1}}\,.
\ee 
Given that $F^{(n-1)}_{i_{n-1}}$ is a basis of independent functions, the equations \eqref{DoubleCopMatrix} will have to hold separately for each of their coefficients in the above equation. In this manner, we arrive at the following system of linear equations for the unknowns $c_{j\gamma}$,
\be
\sum_{j,\gamma} M_{(m i_{n-1})(j\gamma)}c_{(j\gamma)}=0\,,
\ee
where $(m i_{n-1})=(11),\ldots,(1d_{n-1}),(21),\ldots,(2d_{n-1}),\ldots,(l d_{n-1})$ denotes a combined index, similarly for $(j\gamma)$, and finally the elements of the matrix $M$ are given by
\be\label{Mmatrix}
M_{(m i_{n-1})(j\gamma)}\equiv\sum_{\alpha,\beta,i_n}D_{m\alpha\beta} T^\alpha_{i_n,i_{n-1}}L_{j i_{n}}^{\gamma\beta}\,.
\ee
In summary, starting from a basis of symbols \eqref{tensor_coproduct_single} at weight $n$, obeying conditions of the form \eqref{DoubleCopMatrix} on their double coproducts, we may construct a basis with the same properties at weight $n+1$, by determining the right kernel, or nullspace, of the matrix $M$ in \eqn{Mmatrix}, with the known tensor $L$ encoding optional additional restrictions on our initial ansatz \eqref{n1_ansatz}, for example such as in \eqn{LMHV} for specific final entries in the case of hexagon functions. Letting $N_{(j\gamma)i_{n+1}}$ denote the elements of the matrix whose columns correspond to different basis vectors on the nullspace of $M$, $M\cdot N=0$, the new basis of symbols at weight $n+1$ will be explicitly given by
\be
\Delta_{n,1}F^{(n+1)}_{i_{n+1}}=\sum_{i_{n},\alpha} T^\alpha_{i_{n+1},i_{n}}F^{(n)}_{i_{n}}\otimes \ln \phi_\alpha\,,\quad\text{with}\,\,T^\alpha_{i_{n+1},i_{n}}=\sum_{j,\gamma}N_{(j\gamma)i_{n+1}}L_{j i_{n}}^{\gamma\alpha}\,.
\ee
The procedure we have described can be applied to the construction of general integrable symbols subject to additional analytic constraints, with the ``data'' characterizing each specific realization being the particular choices of alphabet $\Phi$, weight-1 functions $T_{i_11}^\alpha$, $\{n-1,1,1\}$ coproduct conditions $D_{m\alpha\beta}$, as well as optional restrictions to particular subspaces~$L_{j i_{n}}^{\gamma\alpha}$. The application we have in mind here is of course to extended Steinmann hexagon symbols, for which we reiterate that we have chosen the alphabet \eqref{eq:new_hex_letters}, weight-1 functions \eqref{eq:w1hex}, double coproduct conditions that may be inferred from eqs.~\eqref{FabIntegrabilityFirst}--\eqref{FabIntegrabilityLast} and \eqref{eq:FabStein}, and separate ans\"atze for the parity even and odd subspaces respectively.

Before closing this section, let us also briefly comment on our strategy for
tackling the most computationally challenging step in the construction of our
extended Steinmann hexagon function space, the computation of the nullspace of
the matrix $M$ in \eqref{Mmatrix}. The main idea, advocated in
ref.~\cite{Drummond:2014ffa},
is to choose the constituents of the matrix $M$ such
that they only have integer entries. On the one hand, this allows one to bound
the size of the entries of $M$ at intermediate stages of its Gaussian
elimination, thereby reducing the runtime and intermediate storage required. On
the other hand, it gives the opportunity to apply the Lenstra-Lenstra-Lov\`asz
algorithm to further improve the sparsity and/or entry size of the final
expression for the nullspace matrix $N$, and thus facilitate the repetition of
the procedure at higher weight. In this manner, standard symbolic software such
as \texttt{Maple} and \texttt{Mathematica} was sufficient for going up 
to weight 11.  Beyond this point, more specialized tools were required,
such as \texttt{SageMath} \cite{SageMath} at weight 12,
\texttt{SpaSM} at weight 13, and custom \texttt{C++} code at weight 14
with MHV final entries that exploits finite
field techniques for solving the linear systems, avoiding the generation of
complicated rational numbers in intermediate steps. 

\subsection{Promoting symbols to functions} \label{sec:symbols_to_functions}

A basis of symbols can be iteratively promoted to a basis of functions.  There are two separate aspects to this promotion.  One aspect is to associate, if possible, each non-vanishing symbol with a unique function that satisfies function-level conditions corresponding to those imposed already at symbol level.  The second aspect is to allow for functions that vanish entirely at symbol level.  We will only add such functions when we determine that a particular constant zeta value must be included as an independent element of the function space.  As mentioned in the introduction and in section~\ref{sec:intconsts}, for $\Hhex$ the first time this happens is for $\zeta_4$.  This independent zeta value then spawns a set of allowed functions at weight $n$ of the form
\be
\zeta_4 \, F^{(n-4)}_{i_{n-4}} \,, 
\qquad i_{n-4} = 1,2,\ldots,d_{n-4} \,.
\label{clonezeta4}
\ee
In other words, the second aspect of the function-level construction is rather trivial, because we just need to clone the function space from four weights lower, and it will automatically obey all function-level conditions. In the rest of this section, therefore, we will focus on the first aspect, associating consistent functions iteratively with non-vanishing symbols.

At each weight, the $\Delta_{n-1,1}$ coproduct component encodes the total derivative of each function, which can be integrated into multiple polylogarithms once the symbols appearing in the weight $n-1$ entry have been upgraded to functions. In the case of hexagon functions, there is a natural kinematic point at which to set the integration constant---the point where all three cross ratios $u$, $v$, and $w$ are 1, on the Euclidean sheet, which we refer to as $(1,1,1)$. The physical branch cut condition guarantees that hexagon functions are finite and smooth at this point (whereas they can develop logarithmic singularities when one of the cross ratios vanishes). Moreover, it has been observed that the six-point amplitude and its coproducts only involve multiple zeta values at this point, providing a natural restriction on the types of boundary data that must be considered here.   

Steinmann Hexagon functions in fact require the appearance of multiple zeta values in their coproduct entries in order to remain consistent with the branch cut condition. This is due to the existence of kinematic limits where the derivatives of these functions have the potential to become singular---namely, where the symbol letters in their last entry vanish. To avoid these singularities, the lower-weight functions appearing in front of them in the coproduct must vanish in this potentially singular limit. Intuitively, this is just the statement that in the limit that any hexagon symbol letter $\phi_\alpha$ other than $a$, $b$, or $c$ vanishes, hexagon functions must be free of coproduct terms such as $\zeta_{n-1} \otimes \ln \phi_\alpha$ (or more generally, free of any weight $n-1$ function that doesn't vanish in the $\phi_\alpha \rightarrow 0$ limit). 

This manifestation of the branch cut condition at higher weight does not, as one might na\"ively expect, amount to the requirement that $F^{1-u_i} \rightarrow 0$ as $u_i \rightarrow 1$ and $F^{y_i} \rightarrow 0$ as $y_i \rightarrow 0$. In general, these coproduct entries get mixed together in kinematic limits, allowing for more complicated cancellations to take care of unphysical singularities. For instance, in the limit that $w \rightarrow 1$, the $y_i$ letters become
\begin{equation}
y_u \rightarrow (1-w) \frac{u (1-v)}{(u-v)^2}, \quad y_v \rightarrow \frac{1}{(1-w)} \frac{(u-v)^2}{v(1-u)}, \quad y_w \rightarrow \frac{1-u}{1-v}.
\label{weq1yi}
\end{equation}
Thus, the coproduct entry $F^{1-w}$ will get mixed with the functions $F^{y_u}$ and $F^{y_v}$, and it is sufficient to require that
\begin{equation} \label{eq:coproduct_entry_branch_cut_condition}
\Big[ F^{1-w} + F^{y_u} - F^{y_v} \Big]_{w\rightarrow1} = 0.
\end{equation}
In general, this relation only requires the addition of zeta-valued constants to these coproduct entries.  It can be imposed anywhere on the $w=1$ surface.

If $F$ is a parity-even function, then in ~\eqn{eq:coproduct_entry_branch_cut_condition} zeta values can only be added to $F^{1-w}$, and it is convenient to impose this condition directly at the point $u=v=w=1$, which is located on the surface $\Delta(u,v,w)=0$ where all parity-odd functions $F^{y_i}$ vanish.  Thus we require, considering also the cyclic images of~\eqref{eq:coproduct_entry_branch_cut_condition},
\begin{equation} \label{branchcutevencondition}
F^{1-u_i}(1,1,1) = 0, \qquad F~\rm{parity~even}.
\end{equation}
Since this condition is homogeneous, it can only force functions to vanish at $(1,1,1)$, i.e.~set potential coefficients of MZVs to zero.

If $F$ is a parity-odd function, then zeta values can only be added to the coproduct entries $F^{y_i}$. However, the condition~\eqref{eq:coproduct_entry_branch_cut_condition} is not sufficient to determine these zeta-valued contributions, since only differences of these coproduct entries appear. Instead, they can be determined on the surface where one of the $y_i$ variables becomes unity, which is also part of the parity-odd vanishing surface $\Delta(u,v,w)=0$. In this limit, the derivatives with respect to the other two variables $\partial/\partial y_{j\neq i}$ become proportional to $F^{y_{j\neq i}}/y_{j\neq i}$. It therefore suffices to require that
\begin{equation} \label{eq:coproduct_entry_branch_cut_condition_odd}
F^{y_v} \Big|_{y_u \rightarrow 1} = 0 \ ,
\end{equation} 
as well as all $S_3$ permutations of this condition when $F$ is parity odd.

A convenient place to impose \eqn{eq:coproduct_entry_branch_cut_condition_odd} is on the line $(u,u,1)$ in the limit that $u=v \to 0$.  In this limit, from \eqn{weq1yi}, $y_w \to 1$ while $y_u$ and $y_v$ can remain different from 1.  Thus we can impose
\begin{equation} \label{branchcutoddcondition}
F^{y_u}(u,u,1)|_{u\to0} = F^{y_v}(u,u,1)|_{u\to0} = 0,
\qquad F~\rm{parity~odd},
\end{equation}
as well as the cyclically related constraints.
On the line $(u,u,1)$, all hexagon functions collapse to   
harmonic polylogarithms (HPLs)~\cite{Remiddi:1999ew} $H_{\vec{w}}(u)$ with indices $w_i \in \{0,1\}$.  The constraint~\eqref{branchcutoddcondition} sets
the coefficient of all independent zeta values to zero, but this does not imply that the value of these coproduct entries vanishes at the point $(1,1,1)$. Rather, the
functions $F^{y_u}$ and $F^{y_v}$ can still generate nonzero zeta-valued contributions when integrated along the line back to $(1,1,1)$ (as can be seen in identities relating HPLs with argument $u$ to HPLs with argument $1-u$). Thus, in general, nonzero coefficients are induced for MZVs appearing in 
$F^{y_i}(1,1,1)$.

The conditions~\eqref{eq:coproduct_entry_branch_cut_condition} and~\eqref{eq:coproduct_entry_branch_cut_condition_odd} must first be imposed for weight 2 functions $F$, where the coproduct entries $F^{1-u_i}$ are nonzero. However, at this weight all $F^{y_i} = 0$, reducing~\eqn{eq:coproduct_entry_branch_cut_condition} to the condition~\eqref{branchcutevencondition}, which is automatically satisfied since all $\ln a_i$ vanish at $(1,1,1)$. At weight 3, this condition becomes nontrivial for the first time in a parity odd function, which can be identified as the one-loop six-dimensional hexagon integral ${\tilde \Phi}_6$~\cite{DelDuca:2011ne,Dixon:2011ng}. From eq.~(B.8) of ref.~\cite{Dixon:2013eka}, its $y_u$ coproduct is
\begin{align}
{\tilde \Phi}_6^{y_u} &= - \sum_{i=1}^3 \Li_2(1-u_i)
- \ln v \ln w + 2 \zeta_2 \label{PhitildeyuA}\\
&= \sum_{i=1}^3 \Li_2\left(1-\frac{1}{u_i}\right)
+ \frac{1}{4} \left[ (\ln^2 b + 4\zeta_2) + (\ln^2 c + 4\zeta_2) \right] \,.
\label{PhitildeyuB}
\end{align}
In the form~\eqref{PhitildeyuA} we can see how the
condition~\eqref{branchcutoddcondition} holds:
the functions $\Li_2(1-w)$ and $\ln v\ln w$ vanish, while $\Li_2(1-u)$
and $\Li_2(1-v)$ both approach $\zeta_2$, forcing the last
term to be $+2\zeta_2$.
In the second form~\eqref{PhitildeyuB}, we have rewritten the result
in the basis of \eqn{eq:wt2Steinmann}.
In section~\ref{sec:intconsts}, we will see that the $\zeta_2$ factors
can be absorbed into the $\ln^2 a_i$ functions as indicated.

This construction then continues, iteratively in the weight. It turns out, however, that not all zeta values are required to appear in hexagon functions to fix bad branch cuts in this way. We now turn to cosmic Galois theory, which will provide the appropriate tools for understanding the implications of this observation. 

%%%%%%%%%%%%%%%%%%%%%%%%%%%%%%%%%%%%%%%%%%%%%
\section{Cosmic Galois Theory}
\label{sec:cgg}
Feynman integrals correspond to integrals of rational functions over rational contours (that is, domains specified by rational inequalities). As such, they should be described by a Galois theory of periods. While the existence of such a theory remains strictly conjectural~\cite{2008arXiv0805.2568A,2008arXiv0805.2569A}, this issue can be sidestepped by studying the motivic avatars of Feynman integrals, which in the polylogarithmic case realize all known functional relations as shuffle and stuffle relations~\cite{Brown1102.1312,Brown:2015fyf}. In particular, motivic polylogarithms come endowed with a coaction that enforces the shuffle and stuffle relations algebraically and allows one to algorithmically (via fibration bases~\cite{FBThesis,Anastasiou:2013srw,Panzer:2014caa}) expose all functional equations. These properties have already proven useful for studying Feynman integrals and amplitudes in diverse contexts, ranging from $\phi^4$ theory~\cite{Panzer:2016snt}, QED~\cite{Schnetz:2017bko}, and QCD~\cite{Anastasiou:2013srw} to maximally supersymmetric gauge theory~\cite{Goncharov:2010jf,Golden:2013xva} and string theory~\cite{Schlotterer:2012ny}. They have also played a central role in the amplitude bootstrap program. However, in this context only some of the power of the coaction has been utilized---namely, the part that has a natural physical interpretation in terms of branch cuts and derivatives. In this section, we expand our use of the coaction to take into account coaction restrictions on the transcendental constants that appear in the amplitude. These in turn prove to be an essential ingredient in pushing the computation of the planar six-point amplitude in $\mathcal{N}=4$ super Yang-Mills theory to six and seven loops for the NMHV and MHV helicity configurations, respectively, which we have carried out in a companion paper \cite{Caron-Huot:2019vjl}. While these more general coaction restrictions don't have a clear physical interpretation, they may point to some graph-theoretic property respected by all Feynman diagrams contributing to these amplitudes.

\subsection{The coaction on multiple polylogarithms}
\label{sec:coactionMPLsubsection}

Multiple polylogarithms, considered abstractly as functions that map from a kinematic domain to the complex numbers, are extremely complicated multi-valued objects. For special values of the kinematics, they evaluate to interesting numerical constants. It has proven famously hard for mathematicians to show that even the simplest constants in this space---the odd Riemann zeta values---are transcendental. The only odd zeta value proven to be irrational is $\zeta_3$~\cite{Apery:1979}. (Although it is also known that ``many'' of the odd zeta values are irrational; for example, for any $\varepsilon>0$, at least $2^{(1-\varepsilon)\ln s/\ln\ln s}$ of the odd zeta values between 3 and $s$ are irrational~\cite{FSZ}.) Nothing is proven about whether they are actually transcendental, i.e.~not algebraic numbers.

This situation is greatly ameliorated by considering instead the motivic
versions of multiple polylogarithms, as all identities between these motivic
objects can be trivialized\footnote{Note that the coaction only trivializes
these identities up to algebraic identities between symbol letters, which can be
arbitrarily complex. This will not concern us here, since we are only
considering polylogarithms with the hexagon symbol alphabet, as defined in
eq.~\eqref{eq:hex_letters}.} using the coaction~\cite{Gonch2,2011arXiv1101.4497D}, as further refined
in~\cite{Brown:2011ik,Duhr:2011zq,Duhr:2012fh}. The coaction is easiest to express in the notation
\begin{equation} \label{eq:I_def}
I(a_0;a_1,\dots, a_n; a_{n+1}) = \int_{a_0}^{a_{n+1}} \frac{dt}{t-a_n} I(a_0;a_1,\dots, a_{n-1}; t) ,
\end{equation}
of which the (possibly more familiar) notation
\begin{equation}
G(a_n, \dots, a_1; a_{n+1}) = I(0;a_1,\dots,a_n; a_{n+1})
\label{Gdef}
\end{equation}
is a special case (note the reversal of arguments). The coaction then corresponds to the operation
\begin{align} \label{def:coaction}
\Delta &I(a_0;a_1,\dots,a_n;a_{n+1}) = \\
& \sum_{0=i_1<\dots<i_{k+1}=n} \!\!\!\!\!\!\!\! I(a_0;a_{i_1},\dots,a_{i_k};a_{n+1}) \otimes \left[\prod_{p=0}^k I(a_{i_p}; a_{i_p+1},\dots,a_{i_{p+1}-1};a_{i_{p+1}}) 
\,\,\text{mod}\,\, i\pi\right], \nonumber
\end{align}
which breaks up polylogarithms into tensor products of functions of lower transcendental weight (where the total weight in each term in the sum is conserved).
The above definition contains trivial terms corresponding to the decomposition
of the polylogarithm into itself. It is therefore useful to define the
\emph{reduced coproduct} $\Delta'$ through 
\be
\Delta(I) = 1\otimes I + I \otimes 1 + \Delta'(I).
\ee
An element $a$ of the Hopf algebra of multiple polylogarithms  with
$\Delta'(a)=0$ is referred to as a \emph{primitive} element.

The coaction can be applied iteratively, until what remains is a tensor product of weight-one functions---namely, logarithms. In this way, all identities between polylogarithms can first be reduced to identities between logarithms, and then built back up to identities between higher-weight polylogarithms systematically~\cite{Duhr:2011zq,Duhr:2012fh}.\footnote{In order for this procedure to be well-defined one must use shuffle regularization~\cite{Gonch3,Gonch2} to handle functions in the coaction which would na\"ively diverge. We omit the details of this procedure here.}

Strictly speaking, the left and right factors in the tensor product of~\eqref{def:coaction} exist within different spaces. The left factor maps back to the original space of (motivic) polylogarithms, while functions appearing in the right factor are de Rham periods. These de Rham periods are actually functions on a group---namely, the cosmic Galois group~\cite{2015arXiv151206410B}---and are thus dual to its generators. Correspondingly, while the cosmic Galois group acts on the space of motivic periods (here, our polylogarithms), these dual objects coact as seen in the operation~\eqref{def:coaction}. In particular, these dual objects have no knowledge of the integration contour of the original polylogarithm and as such are invariant under deformations of said contour, even when the contour deformation crosses a branch point of the original function. The back entries of the coaction therefore need to be invariant under analytic continuation of the original function, which corresponds to deforming the contour of integration around the branch points of the integrand to change its homotopy class. Since all monodromies of the multiple polylogarithms are proportional to powers of $(i\pi)$, the space of de Rham periods can be simply realized for the coaction on multiple polylogarithms by working modulo $(i\pi)$ in the back entry of the coaction. We can therefore almost entirely ignore the distinction between the two spaces and write the coaction in the final form~\eqref{def:coaction}. In practice, we can furthermore neglect the distinction between polylogarithms and their motivic avatars, since every identity resulting from shuffle and stuffle relations constitutes a valid identity between (non-motivic) polylogarithms; what remains conjectural is merely that there exist no other identities between these functions---a fact that in practice we can safely ignore.

\subsection{The coaction principle}
\label{sec:coactionprinciplesubsection}

The hexagon function bootstrap program~\cite{Dixon:2011pw,Dixon:2013eka,Dixon:2011nj,Dixon:2014voa,Dixon:2014xca,Dixon:2014iba,Dixon:2015iva,Caron-Huot:2016owq,Caron-Huot:2019vjl} takes advantage of the algebraic structure of the coaction~\eqref{def:coaction} to construct the six-point amplitude directly from its analytic and kinematic properties. It starts from the assumption (supported both by explicit computation at low loops~\cite{Cachazo:2008hp,DelDuca:2009au,DelDuca:2010zg,Goncharov:2010jf} and an all-orders analysis of the Landau equations~\cite{Prlina:2018ukf}), that the polylogarithmic part of these amplitudes can be expressed in terms of multiple polylogarithms with symbol letters drawn from the set~\eqref{eq:hex_letters}, or equivalently~\eqref{eq:new_hex_letters}. As described in section~\ref{sec:constructing_the_space}, this space of functions (in particular, the span of such functions that have physical branch cuts and obey the extended Steinmann relations) can be built directly at the level of their coproduct, supplemented with (integration) boundary data. This construction is recursive in the weight, implying that the only functions that appear in the first entry of the coaction of higher-weight functions are those that have already appeared at lower weight. This can be phrased formally as a coaction principle~\cite{Brown:2015fyf,Panzer:2016snt,Schnetz:2017bko}:
\be
\boxed{
\Delta {\cal H}^{\text{hex}} \subset {\cal H}^{\text{hex}} \otimes {\cal K}^\pi 
\,.}
\label{eq:coaction_principle}
\ee
Namely, the coaction maps a generic function in the Steinmann hexagon function space back to the same space tensored with the space of de Rham periods discussed above. The functions appearing in ${\cal K}^\pi$ are more general than ${\cal H}^{\text{hex}}$; for instance, their first symbol entries can be any of the nine letters of the alphabet~\eqref{eq:hex_letters}, implying that they can have additional logarithmic branch points when $1-u_i$ and $y_i$ vanish. 

At symbol level, the fact that the Steinmann hexagon function space satisfies a coaction principle is true by construction. Therefore, once we have accepted the conjecture that the six-point amplitude can be expressed in the basis constructed in section~\ref{sec:constructing_the_space}, it directly follows that the symbol of the amplitude also satisfies this coaction principle. When the construction based on the $\Delta_{n,1}$ coaction ansatz~\eqref{n1_ansatz} is lifted to function level, as described in section~\ref{sec:symbols_to_functions}, then the coaction principle also must be satisfied for all components of the form $\Delta_{n-k,1,\ldots,1}$, corresponding to an arbitrary number of iterated derivatives.  The novel import of~\eqn{eq:coaction_principle} resides in the fact that transcendental constants such as Riemann zeta values also exhibit structure under the coaction map~\cite{Brown:2011ik}, even though they are in the kernel of the projection $\Delta_{n-k,1,\ldots,1}$. We now explain why such constants are required to appear in the hexagon function space, and investigate what it means for these constants to respect (or not respect) the coaction principle~\eqref{eq:coaction_principle}.

\subsection{Integration constants and branch cut conditions}
\label{sec:intconsts}

The branch cut conditions~\eqref{branchcutevencondition} and~\eqref{branchcutoddcondition} only require the addition of specific zeta values to the coproducts of Steinmann hexagon symbols to upgrade them to functions. As shown in section~\ref{sec:symbols_to_functions}, nonzero values are only forced by the conditions~\eqref{branchcutoddcondition} on the $y_i$ coproducts of parity-odd functions.
For instance, at weight two we see from \eqn{PhitildeyuB} that a contribution proportional to $\zeta_2$ must be added to the $y_i$ coproduct entries of the first parity-odd function in the hexagon function space, $\tilde \Phi_6$. However, because $\tilde \Phi_6$ is fully symmetric under all permutations of the six-particle cross ratios, $\zeta_2$ is only required to appear in a single linear combination of weight-two functions and its images under the dihedral group. From examining \eqn{PhitildeyuB} alone, we might consider adding it to either $\Li_2(1-1/u_i)$ or $\ln^2 a_i$.  However, the $1-u_i$ coproduct of $\Li_3(1-1/u_i)$ is $\Li_2(1-1/u_i)$, and so if we added $\zeta_2$ to $\Li_2(1-1/u_i)$ we would spoil its vanishing at $u_i=1$, which is required by~\eqn{branchcutevencondition}.  Therefore we must add $\zeta_2$ to $\ln^2 a_i$. Dihedral symmetry and the condition~\eqref{branchcutoddcondition} fix the normalization to be as shown in \eqn{PhitildeyuB}.  That is, $\zeta_2$ always appears in the specific linear combinations
\be
\ln^2 a_i + 4 \zeta_2 \,,
\qquad i=1,2,3.
\label{eq:zeta2_combination}
\ee
Thus we are not actually forced to include $\zeta_2$ as an independent weight-two function---rather, we just shift the relevant orbit of weight-two functions to include this contribution, as given in~\eqn{eq:zeta2_combination}.
In summary, there are only six functions in $\Hhex$ at weight 2,
\be
\Hhex_2\ =\
\biggl\{
\Li_2\left(1-\frac{1}{u_i}\right),\ \ln^2 a_i + 4 \zeta_2 \biggr\} \,,
\qquad i=1,2,3,
\label{eq:wt2Hhex}
\ee
not the seven we might na\"ively have expected.

Now let us consider the branch-cut conditions for weight-four functions.
We find that the conditions~\eqref{branchcutevencondition} on
the even functions are so strong that they force {\it all} the even weight
three functions to vanish at $(1,1,1)$, and so, rather surprisingly,
$\Hhex_3(1,1,1)$ is empty!  (The
constraints~\eqref{branchcutoddcondition} applied to the two
parity-odd weight-four functions are consistent with this fact, of course.)
Because all higher-weight functions are constructed on top of the
weight-four basis,
the coaction principle~\eqref{eq:coaction_principle} implies that $\zeta_3$
does not appear in the first entry of the coaction on any hexagon function.

On the other hand, the promotion of the weight-five basis from symbols to
functions does require the addition of $\zeta_4$ contributions.
In fact, so many linearly independent combinations of weight-four functions
must be shifted by $\zeta_4$ contributions that $\zeta_4$ must be included
as an independent function in the weight-four space. That is, it is not
possible to just shift the existing weight-four functions by a multiple
of $\zeta_4$: fixing the branch cuts in some of the weight-five functions
in this way makes it impossible to fix the branch cuts in other
functions.\footnote{We might entertain the alternate possibility that such
functions should just be removed from the space. However, we know from
Table~\ref{tab:MHVNMHVdim} that all weight-five functions are required
to describe the derivatives of the five-loop amplitude.}
This impossibility is entirely associated with the three weight-four
even functions that contain parity-odd letters in their symbols, which
are associated with the double pentagon integral $\Omega^{(2)}(u,v,w)$
and its two cyclic images.  That is, the branch-cut
conditions~\eqref{branchcutevencondition} for the even weight-five functions
force all the other weight-four functions, the ones with no parity-odd
letters, to vanish at $(1,1,1)$.

At first sight, the fact that $\zeta_4$ is an independent constant might
seem slightly puzzling, considering that $\zeta_4 = \tfrac{2}{5}\zeta_2^2$
and one might thus expect the addition of a free $\zeta_4$ to spoil terms in the coaction involving $\zeta_2$. However, it is important to remember that
the second entry of the coaction is modulo $(i\pi)$ and thus
$\Delta_{2,2} (\zeta_4) = 0$, so that this apparent contradiction is resolved.
In general, all even Riemann zeta values $\zeta_{2k}$ are primitive,
or indecomposable, under the coaction, so their appearance can never
be forbidden by the coaction principle.

\begin{table}
\begin{center}
\begin{tabular}{  r  c  c  c  } 
    \hline\hline
{\footnotesize Weight} & {\footnotesize \ \ Multiple Zeta Values \ \ } & {\footnotesize Appear in $\Hzeta(1,1,1)$} & {\footnotesize Independent Constants in $\Hzeta$} \\ 
\hline
0 & 1 & 1 & 1 \\ 
\hline
1 & $-$ & $-$ & $-$ \\ 
\hline
2 & $\zeta_2$ & $\zeta_2$ & $-$ \\ 
\hline
3 & $\zeta_3$ & $-$ & $-$ \\ 
\hline
4 & $\zeta_4$ & $\zeta_4$ & $\zeta_4$ \\ 
\hline
5 & $\zeta_5$, $\zeta_2 \zeta_3$ & $5 \zeta_5 - 2 \zeta_2 \zeta_3$ & $-$ \\ 
\hline
6 & $(\zeta_3)^2$, $\zeta_6$ & $\zeta_6$ & $\zeta_6$ \\ 
\hline
7 & $\zeta_7$, $\zeta_2 \zeta_5$, $\zeta_4 \zeta_3$ 
& $7 \zeta_7 - \zeta_2 \zeta_5 - 3\zeta_4 \zeta_3$,
$\zeta_7 - 4 \zeta_4 \zeta_3$ & $\zeta_7 - 4 \zeta_4 \zeta_3$ \\ 
\hline
8 & $\zeta_{5,3}$, $\zeta_3 \zeta_5$, $\zeta_2 (\zeta_3)^2$, $\zeta_8$
& $\zeta_{5,3} + 5 \zeta_3 \zeta_5 - \zeta_2 (\zeta_3)^2$, $\zeta_8$
& $\zeta_8$ \\ 
\hline\hline
\end{tabular} 
\caption{Through weight 8, we display first the complete set of MZVs, followed
by the linear combinations that appear in the intermediate function space
$\Hzeta \supset \Hhex$
when the functions are evaluated at $(1,1,1)$, followed
by the independent constants that are required in $\Hzeta$.}
\label{table:H_zeta_space_constants}
\end{center} 
\end{table}

The branch cut conditions can be solved in an analogous way at each higher weight; in practice we carried out this construction through weight eight. We refer to the space of hexagon functions constructed in this way (where only the zeta values required to solve the branch cut conditions are introduced) as $\Hzeta$. Our final, minimal space $\Hhex$ will be slightly smaller than $\Hzeta$, because not all functions with non-vanishing symbols appear in the amplitudes' coproducts, starting at weight eight. 

The zeta values that appear in $\Hzeta$ are given through weight eight in Table~\ref{table:H_zeta_space_constants}. In this table, we distinguish between zeta values that appear in the span of all functions in $\Hzeta$ evaluated at the point $u=v=w=1$, and those that are required to appear in this function space as independent constant functions. We see from the table that the space of weight-five constants is similar to weight-three---the branch-cut conditions at one higher weight can be satisfied by shifting the existing (symbol-level) basis of functions.  Note that only one of the two possible linear combinations of $\zeta_5$ and $\zeta_2 \zeta_3$ appears. Weight six is also similar to weight four, insofar as the branch cut conditions one weight higher cannot be solved just by shifting the existing weight-six basis. However, there is now a two-dimensional space of constants we can consider adding to our basis. Since we want to add the smallest number of free zetas to the space, we first try to solve these branch cut conditions after adding just a single linear combination of $\zeta_6$ and $(\zeta_3)^2$ to the space, as well as allowing further shifts to be absorbed into individual basis functions. This gives rise to a nonlinear system of equations that can only be solved if the independent constant is chosen to be $\zeta_6$. A similar analysis yields the results at weight seven and eight in Table~\ref{table:H_zeta_space_constants}. 

While the six-particle amplitudes are known to be expressible in this basis at the level of their symbol, there is no guarantee they will exist within the span of this basis as functions. In fact, the BDS-like-normalized amplitudes do not. However, the MHV and NMHV amplitudes in this normalization are misaligned with $\Hzeta$ by the same exact amount. This is seen first at three loops, where the BDS-like-normalized MHV and NMHV amplitudes evaluate to
\be
\EE^{{\rm old}\, (3)}(1,1,1) = \frac{413}{3} \, \zeta_6 + 8 (\zeta_3)^2 \,,
\qquad
E^{{{\rm old}\, (3)}}(1,1,1) = - \frac{940}{3} \zeta_6 + 8 (\zeta_3)^2 \,.
\label{old3loops111}
\ee
These numbers are not in the span of $\Hzeta(1,1,1)$ due to the appearance of $(\zeta_3)^2$. However, we have the freedom to normalize the amplitudes differently, for instance shifting them by $-8 (\zeta_3)^2$ at three loops. This amounts to multiplying the BDS-like ansatz by a constant factor $\rho(g^2)$, which allows us to adjust the amplitudes' normalization by a constant at each loop order. Through seven loops, this factor can be chosen to be \cite{Caron-Huot:2019vjl} 
\bea
\rho(g^2) &=& 1 + 8 (\zeta_3)^2 \, g^6  - 160 \zeta_3 \zeta_5 \, g^8
+ \Bigl[ 1680 \zeta_3 \zeta_7 + 912 (\zeta_5)^2 - 32 \zeta_4 (\zeta_3)^2 \Bigr]
\, g^{10}
\nonumber\\
&&\null\hskip0.0cm
- \Bigl[ 18816 \zeta_3 \zeta_9 + 20832 \zeta_5 \zeta_7
  - 448 \zeta_4 \zeta_3 \zeta_5 - 400 \zeta_6 (\zeta_3)^2 \Bigr] \, g^{12}
\nonumber\\
&&\null\hskip0.0cm
+ \Bigl[ 221760 \zeta_3 \zeta_{11} + 247296 \zeta_5 \zeta_9 + 126240 (\zeta_7)^2
   - 3360 \zeta_4 \zeta_3 \zeta_7 - 1824 \zeta_4 (\zeta_5)^2
	\nonumber\\
&&\null\hskip0.7cm
 - 5440 \zeta_6 \zeta_3 \zeta_5 - 4480 \zeta_8 (\zeta_3)^2 \Bigr] \, g^{14}
\ +\ {\cal O}(g^{16}).
\label{rho}
\eea
We emphasize that this ``cosmic normalization'' only works because the
parity-even parts of the MHV and NMHV amplitudes, evaluated at $u=v=w=1$,
are misaligned by exactly the same factor at each loop order.
The choice of the factor $\rho$ is then unique, given the conditions
described in our companion paper~\cite{Caron-Huot:2019vjl}.
 
The fact that the six-particle amplitude can be shifted in the above way through six loops motivates an all-loop conjecture:
\begin{quotation}
\noindent {\bf Branch Cut (Over-)Completeness:} The space of hexagon functions $\Hhex$ needed to describe $\EE$, $E$ and $\Et$ is contained within the minimal space required to upgrade extended Steinmann hexagon symbols to functions, namely $\Hzeta$.
\end{quotation}
This conjecture requires that the difference $\EE^{(L)}(1,1,1)-E^{(L)}(1,1,1)$, computed using only the value of $\rho$ truncated at one lower loop order, is within $\Hzeta(1,1,1)$ to all loop orders $L$. We have no proof of this assertion. Perhaps it can be argued for from the perspective of the graph-theoretic properties of the Feynman diagrams contributing to these amplitudes (cf.~the `small graphs principle' for $\phi^4$ theory~\cite{Brown:2015fyf}).

%%%%%%%%%%%%%%%%%%%%%%%%%%%%%%%%%%%%%%%%

\subsection{Restrictions from cosmic Galois theory}
\label{sec:MZV_coaction_principle}

While the conjecture of the last section may seem modest, it puts strong, all-loop-order constraints on the transcendental constants that can appear in the six-point amplitude and its derivatives. The constraints follow from the coaction on multiple zeta values, which breaks down these constants into simpler primitives, just as the symbol breaks down full polylogarithms into logarithmic primitives.  In section~\ref{sec:hexagon_limits}, we will verify that the coaction principle also holds for more general spaces of transcendental constants, such as alternating sums and multiple polylogarithms evaluated at higher roots of unity, by evaluating the functions in $\Hhex$ at other points besides $(1,1,1)$.

Multiple zeta values are a generalization of the Riemann zeta values to include multiple (nested) infinite sums. A finite multiple zeta value can be associated with every string of positive integers $\vec w$ by the definition 
\be
\zeta_{\vec{w}} = \zeta_{w_1, \dots, w_d} \equiv \sum_{k_1 > \cdots > k_d > 0} \frac{1}{k_1^{w_1} \cdot \cdot \cdot k_d^{w_d}} \, ,\label{eq:MZV_def}
\ee
whenever $w_1>1$.  The depth is $d$ and the weight is $n=\sum_{i=1}^d w_i$. These constants satisfy many shuffle and stuffle relations, and the dimension $d^{\rm MZV}_n$ of the vector space they form over $\mathbb Q$ at weight $n$ is given by the generating function
\be
d^{\rm MZV}(t) \equiv \sum_{n=0}^\infty d^{\rm MZV}_n t^n = \frac{1}{1 - t^2 - t^3}
= 1 + t^2 + t^3 + t^4 + 2 t^5 + 2 t^6 + \ldots \,,
\label{eq:MZVindep}
\ee
at least motivically~\cite{Zagier:1994,Broadhurst:1996kc,Brown:2011ik}.

The multiple zeta values also exist in one-to-one correspondence with HPLs with indices $\{0,1\}$ evaluated at unity, namely (up to sign conventions) the restriction of~\eqn{Gdef} to indices taking the value 0 or 1, and evaluated at $a_{n+1}=1$. (The $w_i$ in \eqn{eq:MZV_def} correspond to $w_i-1$ `0's followed by a `1' in the $G$ function notation, or a `1' followed by $w_i-1$ `0's in the $I$ notation.)  As a result, MZVs inherit the coaction structure of polylogarithms~\cite{Brown:2011ik}. For instance, we can take the coaction of the multiple zeta value $\zeta_{5,3} = I(0;1,0,0,1,0,0,0,0;1)$ using~\eqn{def:coaction}. It is found that
\begin{align} \label{eq:coaction_z53}
\Delta' \zeta_{5,3} &= - 5 \ I(0;1,0,0;1) \otimes I(0;1,0,0,0,0;1) \nonumber \\
&= - 5 \ \zeta_3 \otimes \zeta_5 \,,
\end{align}
after shuffle regularization. Since $\zeta_3$ is absent from the weight-three basis in $\Hzeta(1,1,1)$, we immediately conclude that $\zeta_{5,3}$ cannot appear by itself in $\Hzeta(1,1,1)$. And indeed, by reference to Table~\ref{table:H_zeta_space_constants}, we see that $\zeta_{5,3}$ appears only in the linear combination $\zeta_{5,3} +5 \zeta_5 \zeta_3 - \zeta_2 (\zeta_3)^2$. As can be checked via~\eqn{def:coaction}, $\zeta_5$ and $\zeta_3$ are primitives under the coaction (i.e.~they don't decompose into simpler objects), and since the coproduct of the product is the product of coproducts, we simply have
\begin{equation} \label{eq:coaction_z5z3}
\Delta' (\zeta_5 \zeta_3) = \zeta_5 \otimes \zeta_3 + \zeta_3 \otimes \zeta_5 \,.
\end{equation}
The $\zeta_3 \otimes \zeta_5$ term of the coaction thus cancels in the combination $\zeta_{5,3} +5 \zeta_5 \zeta_3- \zeta_2 (\zeta_3)^2 $, as needed.  Indeed,
\begin{equation} \label{eq:coaction_wt8comb}
\Delta_{5,3} \left(\zeta_{5,3} +5 \zeta_5 \zeta_3- \zeta_2 (\zeta_3)^2 \right) 
= (5 \zeta_5 - 2 \zeta_2 \zeta_3 ) \otimes \zeta_3 \,, 
\end{equation}
is also consistent with the linear combination that appears at weight five in  
$\Hzeta(1,1,1)$.

This type of reasoning gives rise to an increasingly large number of constraints
as one moves up in weight. In practice, these constraints are easiest to impose
at the point $u=v=w=1$, as we have done above, although the coaction
principle~\eqref{eq:coaction_principle} holds for generic values of $u$, $v$,
and $w$. To apply the constraints most efficiently,
it is useful to translate the MZVs into an `$f$-alphabet' in
which each odd Riemann zeta value $\zeta_{2 k+1}$ is mapped to the letter
$f_{2k +1}$~\cite{Brown:2011ik}. The letters $f_{2k+1}$ form a free algebra over
the rationals ${\mathbb Q} \langle f_{2 k+1} \rangle$ that,
when supplemented by powers of $\pi^2$, is isomorphic to the
vector space over the rationals formed by the multiple zeta values. 
In other words, products of $f$'s in different orders are independent objects
(words), while even Riemann zeta values can be commuted at will across
the strings of $f$'s.  We will adopt the shorthand notation for products,
$f_{2k+1,2l+1,2m+1} \equiv f_{2k+1} f_{2l+1} f_{2m+1}$. Also, we will adopt
the ordering convention in refs.~\cite{Panzer:2016snt,HyperlogProcedures},
which unfortunately is reversed from our tensor product notation
for the coaction.

The coaction on multiple polylogarithms simply becomes deconcatenation in
the $f$-alphabet. This means that the $f$-alphabet representation of any
multiple zeta value can be read directly off of its coaction, up to the
contribution coming from generators of the same weight as that of the original
constant. For instance, it can be seen from eqs.~\eqref{eq:coaction_z53}
and~\eqref{eq:coaction_z5z3} that
$\zeta_{5,3} \rightarrow -5 f_5 f_3 \equiv - 5 f_{5,3}$
(due to the reversed ordering for the $f$ notation)
and $\zeta_5 \zeta_3 \rightarrow f_{3,5} + f_{5,3}$, up to primitives of weight
8. In the latter case, we see that multiplication is represented in the $f$-alphabet by the shuffle product---any product of multiple zeta values $\zeta_{\vec{w}_1} \zeta_{\vec{w}_2}$ is mapped to the shuffle product of the $f$-alphabet representations of $\zeta_{\vec{w}_1}$ and $\zeta_{\vec{w}_2}$. 

While there are no primitives of the form $f_{2k+1}$ at even weights, an additional letter should be added to our $f$-alphabet to account for the appearance of even zeta values, $\zeta_{2k}$. These constants are semi-simple under the coaction, meaning that they are mapped to zero in the de Rham factor of the coproduct~\cite{Brown:2011ik,Duhr:2012fh,2015arXiv151206410B}. In equation form, we have
\begin{equation}
\Delta \zeta_{2k} = \zeta_{2k} \otimes 1 \, .
\end{equation}
Because even zeta values cannot appear in the de Rham factor of the coaction,
their position in words formed out of the $f$-alphabet doesn't encode any
information; thus we may use a convention to write $\zeta_{2k}$ in front of
all $f$'s. Also, we will use a single even Riemann zeta value $\zeta_{2k}$
instead of $k$ powers of $\zeta_2$ or $\pi^2$, as it
tends to simplify the rational numbers that appear.

The $f$-alphabet representations of the MZVs have been tabulated to high weight~\cite{Schlotterer:2012ny,HyperlogProcedures}. (Note that the first reference defines MZVs with indices reversed from our convention, although the $f$ ordering is the same as ours.) The translation of single odd zeta values (and their products) follows directly from the definition
\begin{align}
\zeta_{2k+1} &\rightarrow f_{2k+1} \, ,
\end{align}
and the translation of multiplication to the shuffle product, for example
\be
(\zeta_3)^2 \zeta_5\ \to
\ f_3 \shuffle f_3 \shuffle f_5\ = 
\ 2 f_{3,3,5} + 2 f_{3,5,3} + 2 f_{5,3,3} \,.
\label{eq:shuffleexample}
\ee
The decomposition of multiple zeta values is computed via the coaction~\eqref{def:coaction}, which has a single ambiguity due to the appearance of a new $f_{2k+1}$ ($\zeta_{2k}$) letter at weight $2k+1$ ($2k$), which belongs to its kernel. This ambiguity can be fixed numerically~\cite{Brown:2011ik}. 

Using the $f$-alphabet, it is easy to determine the space of allowed constants at $u=v=w=1$, given which constants have appeared at all lower weights. Since the coaction acts as deconcatenation on words in this alphabet, constraints following from the coaction principle~\eqref{eq:coaction_principle} can be derived by isolating all terms with a given sequence of odd indices on the left. This corresponds to taking a sequence of `derivations' $\partial_{2k+1}$, each of which returns the left factor of the coaction~\eqref{def:coaction} whenever a specific odd zeta value appears in the right (de Rham) factor, and zero otherwise. Since our coaction and $f$-alphabet conventions have reversed order with respect to each other, this means the derivations $\partial_{2k+1}$ act on the left as
\begin{align}
\partial_{2k+1} \left(f_{i_1,i_2,\dots,i_r} \right) = \begin{cases} f_{i_2,\dots,i_{r}} & \text{ if $i_1 = 2k+1$, }\\ 0 & \text{ otherwise. } \end{cases}
\label{eq:oddzetaderivation}
\end{align}
No such derivations exist for the even zeta values, which don't appear in the de Rham factor of the coaction. Correspondingly, the coaction principle does not forbid terms such as $\zeta_4 \, f_{3}$ from appearing in $\Hzeta(1,1,1)$, because $\zeta_4$ is in $\Hzeta(1,1,1)$ at weight four, and there is no coaction term in which $\zeta_3$ appears alone in the first entry,
i.e.~$\Delta'(\zeta_4 \zeta_3) = \zeta_4 \otimes \zeta_3$.

The operation~\eqref{eq:oddzetaderivation} is at the heart of how we apply
cosmic Galois theory in this paper: in addition to taking derivatives
with respect to dynamical variables, it allows us to formally take
derivatives with respect to odd zeta values.  
\Eqn{eq:oddzetaderivation} can be loosely thought of as an infinitesimal
version of the coaction~\eqref{def:coaction}, or as its specialization
to MZV points.
As far as we understand, $\partial_{2k+1}$
is interpreted in the mathematics literature
as dual to an infinitesimal generator of the cosmic Galois
group~\cite{2015arXiv151206410B}.
For our purposes, the group structure amounts to saying
that it suffices to study \eqn{eq:oddzetaderivation} together with
the constraints from usual partial derivatives discussed in
section~\ref{sec:stein}.  That is, we expect that inspecting the
action of $\partial_{2k+1}$ at the point $(1,1,1)$ will exhaust all
additional constraints from the coaction principle.
As a check, the properties of the coaction at other kinematic points
and along various lines will be analyzed explicitly in
section~\ref{sec:hexagon_limits}.

%%%%%%%%%%%%%%%%%%%%%%%%%%%%%%%%%%%%%%%%%%%%
\begin{table}
\begin{center}
\begin{tabular}{ | r | c | c | } 
\hline
{\footnotesize Weight} & {\footnotesize \ \ Multiple Zeta Values \ \ } & {\footnotesize Appear in $\Hzeta(1,1,1)$} \\ 
\hline
0 & \tikzmark{l0}1\tikzmark{r0} & 1 \\ 
\hline
1 & $-$ & $-$ \\ 
\hline
2 & \tikzmark{l2}$\zeta_2$\tikzmark{r2} & \tikzmark{L2}$\zeta_2$\tikzmark{R2} \\ 
\hline
3 & \tikzmark{l3}$f_3$\tikzmark{r3} & \tikzmark{L3}$-$\tikzmark{R3} \\ 
\hline
4 & \tikzmark{l4}$\zeta_4$\tikzmark{r4} & \tikzmark{L4}$\zeta_4$\tikzmark{R4} \\ 
\hline
5 & \tikzmark{l5}$f_5$, $\zeta_2 f_3$\tikzmark{r5} & \tikzmark{L5}$5 f_5 {-} 2 \zeta_2 f_3$\tikzmark{R5} \\ 
\hline
6 & \tikzmark{l6}$f_{3,3}$, $\zeta_6$\tikzmark{r6} & $\zeta_6$ \\ 
\hline
7 & \tikzmark{l7}$f_7$, \tikzmark{m7}$\zeta_2 f_5$, $\zeta_4 f_3$\tikzmark{r7} & \ \  \tikzmark{L7} \ $7 f_7 {-} \zeta_2 f_5 {-} 3 \zeta_4 f_3$, $f_7 {-} 4\zeta_4 f_3$  \tikzmark{R7} \quad \\ 
\hline
8 & \ \ \tikzmark{l8}$f_{5,3}$, \tikzmark{c8}$f_{3,5}$, $\zeta_2 f_{3,3}$\tikzmark{r8} \!\!, $\zeta_8$ \ \ &  \tikzmark{L8}$5 f_{3,5} -{} 2 \zeta_2 f_{3,3}$\tikzmark{R8} \!\!, $\zeta_8$ \\ 
\hline
\end{tabular}
\begin{tikzpicture}[overlay, remember picture, shorten >=.5pt, shorten <=.5pt]
    \draw [thick,->] ([yshift=4pt,xshift=0pt]{pic cs:l8}) [draw=blue,bend left] to[out=60,in=130] node[pos=0.7]{{\color{blue} \ \ \ \ \ \ $\partial_5$}} ([yshift=3pt,xshift=-1pt]{pic cs:l3}) ;
    \draw [thick,->] ([yshift=3pt,xshift=0pt]{pic cs:c8}) [draw=green,bend left] to[out=70,in=110] node[pos=0.62]{{\color{green} $\partial_3$ \, \ \ }} ([yshift=3pt,xshift=-1pt]{pic cs:l5}) ;
    \draw [thick,->] ([yshift=4pt,xshift=0pt]{pic cs:r8}) [draw=green,bend right] to[out=-70,in=-110] node[pos=0.584]{{\color{green} \ \ \ \ \ $\partial_3$}} ([yshift=3pt,xshift=0pt]{pic cs:r5}) ;
    \draw [thick,-] ([yshift=3pt,xshift=-1pt]{pic cs:L8}) [draw=blue,bend left] to[out=40,in=140] ([yshift=15pt,xshift=00pt]{pic cs:L7}) ;
    \draw [thick,->] ([yshift=13.99pt,xshift=-.04pt]{pic cs:L7}) [draw=blue,bend left] to[out=41,in=150] node[pos=0.32]{{\color{blue} $\partial_5$ \ \ \, }} ([yshift=3pt,xshift=-1pt]{pic cs:L3}) ;
    \draw [thick,-] ([yshift=5pt,xshift=-1pt]{pic cs:R8}) [draw=green,bend right] to[out=10,in=-106] ([yshift=8pt,xshift=0pt]{pic cs:R7}) ;
    \draw [thick,->] ([yshift=7pt,xshift=0.16pt]{pic cs:R7}) [draw=green,bend right] to[out=-45,in=-140] node[pos=0.26]{{\color{green} $\partial_3$ \ \ \ }} ([yshift=3pt,xshift=0pt]{pic cs:R5}) ;
  \end{tikzpicture}
\end{center}
\caption{The left columns in Table~\ref{table:H_zeta_space_constants},
rewritten in the $f$-alphabet.  The arrows illustrate the action of the derivations $\partial_3$ and $\partial_5$.} \label{table:MZV_coaction_restrictions}
\end{table}

The constraints implied by the coaction principle can be formulated as a system of linear constraints on the general space of weight-$w$ multiple zeta values, by taking all possible derivations and requiring the resulting words to lie within the span of the relevant space at lower weight. This is illustrated in Table~\ref{table:MZV_coaction_restrictions}, where the action of $\partial_3$ and $\partial_5$ on the weight-eight MZVs is shown. Since only $f_{5,3}$ is mapped to $f_3$ by $\partial_5$, and $f_3$ isn't in the span of the (cosmically normalized) amplitudes, $f_{5,3}$ cannot appear at weight eight. Similarly, only the combination $5f_{3,5} - 2 \zeta_2 f_{3,3}$ maps to the allowed combination of weight-five constants under $\partial_3$.  Note that we don't need to consider taking multiple derivations, because the lower-weight spaces already respect the coaction principle, by construction.

The space $\Hzeta$ that we constructed through weight eight obeys all
the restrictions of the coaction principle at $(u,v,w)=(1,1,1)$.
Imposing the coaction principle simplifies the branch cut
conditions~\eqref{branchcutevencondition}
and~\eqref{branchcutoddcondition}, to an increasing degree at higher weights,
because it limits which constants can appear at $(1,1,1)$.
For instance, it immediately follows from these restrictions that $(\zeta_3)^2$
could not have appeared in $\Hzeta(1,1,1)$, a fact that we arrived at by a
more complicated means in the last section.

On the other hand, it becomes increasingly cumbersome to fix all the zeta
valued constants at $(1,1,1)$ from the ``bottom up'' as we did in the last
section.  Also, the space of functions $\Hzeta$ may still be larger than
$\Hhex$, which we defined to be the minimal space containing the
cosmically normalized amplitudes and all of their derivatives
($\{n-k,1,\ldots,1\}$ coproducts). We will return to this issue in the
next section.

%%%%%%%%%%%%%%%%%%%%%%%%%%%%%%%%%%%%%%%%%%%%%%%%%%

\section{The Saturation of \texorpdfstring{$\Hhex$}{Hhex}}
\label{sec:saturation}

\subsection{Saturation of full functions}
\label{sec:saturationfull}

Having computed the NMHV amplitude through six loops and
the MHV amplitude through seven loops, we can construct a large
number of weight-$n$ functions in $\Hhex$ by taking all $\{n,1,1,\ldots,1\}$
coproducts. In principle, there are $9^{2L-n}$ possibilities, i.e~we can
choose a different symbol letter for each of the $2L-n$  weight-one coproduct
entries. In practice, a much smaller number of functions are needed,
due to integrability, the extended Steinmann relations,
final-entry conditions (for small values of $2L-n$), and so on.
The numbers of linearly independent weight-$n$ functions generated in this
way is shown in Table~\ref{tab:MHVNMHVdim}, where each successive row
gives the number using both MHV and NMHV amplitudes at $L$ loops, except
for the last line which combines the information from all amplitudes
together, including seven-loop MHV. For a given loop order, reading from
right to left, the numbers first increase and then decrease. The increase
is because there are nine letters, so each function could have
several linearly independent functions among its first coproducts.
The decrease is because eventually all the functions have to fit into
a fixed space, $\Hhex$, whose dimension decreases as the weight decreases.
At a fixed weight $n$, as $L$ increases, the dimension shown in the table
increases until it \emph{saturates}.  At this point, $\Hhex_n$ is spanned
by the iterated coproducts of the $L$-loop amplitude,
for all higher loop orders.

%%%%%%%%%%%%%%%%%%%%%%%%%%%%%%%%%%%%%%%%%%%%%%%%%%
\renewcommand{\arraystretch}{1.25}
\begin{table}[!t]
\centering
\begin{tabular}[t]{l c c c c c c c c c c c c c c c}
\hline\hline
weight $n$
& 0 & 1 & 2 & 3 & 4 &  5 &  6 &  7 &  8 &  9 & 10 & 11 & 12 & 13 & 14
\\\hline\hline
$L=1$
& \green{1} & \green{3} & 4 &  &  &  &  &  &  &  &  &  &  &  & 
\\\hline
$L=2$
& \green{1} & \green{3} & \green{6} & 10 & 6 &  &  &  &  &  &  &  &  &  & 
\\\hline
$L=3$
& \green{1} & \green{3} & \green{6} & \green{13} & 24 & 15 & 6 &  &
&  &  &  &  &  & 
\\\hline
$L=4$
& \green{1} & \green{3} & \green{6} & \green{13} & \green{27} & 53
& 50 & 24 & 6 &  &  &  &  &  & 
\\\hline
$L=5$
& \green{1} & \green{3} & \green{6} & \green{13} & \green{27} & \green{54}
& 102 & 118 & 70 & 24 & 6 &  &  &  & 
\\\hline
$L=6$
& \green{1} & \green{3} & \green{6} & \green{13} & \green{27} & \green{54}
& \green{105} & 199 & 269 & 181 & 78 & 24 & 6 &  & 
\\\hline
$L=7+$
& \green{1} & \green{3} & \green{6} & \green{13} & \green{27} & \green{54}
& \green{105} & \green{200} & 338 & 331 & 210 & 85 & 27 & 6 & 1
\\\hline\hline
\end{tabular}
\caption{The number of independent $\{n,1,1,\ldots,1\}$ coproducts
of the MHV and NMHV amplitudes through $L=6$ loops. A green number
denotes saturation. The final line gives the number
using {\it all} known loop orders together, including 7 loop MHV.}
\label{tab:MHVNMHVdim}
\end{table}

In Table~\ref{tab:MHVNMHVdim},
we use a green color to denote numbers where saturation has been achieved.
If the next loop order is available, we suppose that saturation has
been achieved if the number does not grow with the addition of that
additional information, i.e.~if the next number below is the same. 
We can also ask if the green (saturated) number agrees with the
number constructed from the ``bottom-up'' approach,
i.e.~with the dimension of $\Hzeta$.
These numbers always agree, until one hits the `200' at weight 7
and $L=7+$.
Indeed, combining the constants in
Table~\ref{table:H_zeta_space_constants} with the symbols in
Table~\ref{tab:SymbolDim} would have produced 201 weight-7 functions.
However, we find that the constant $\zeta_7-4\zeta_4\zeta_3$ displayed
in the `Independent Constants' column in
Table~\ref{table:H_zeta_space_constants} is not in the span of
the 200 weight-7 parity-even amplitude coproducts in
Table~\ref{tab:MHVNMHVdim}.   
This independent constant was needed in $\Hzeta$ in order to prevent
the branch-cut constraints from removing a particular weight 8 parity-odd
function, $O_8$, which is allowed by the symbol-level constraints.
However, in Table~\ref{tab:MHVNMHVodddim} we can see from the
repeated `59' that the weight-8 parity-odd space already appears to saturate
at 6 loops; that is, the seven-loop MHV amplitude did not require
any more such functions---and $O_8$ is not in the span
of these 59 functions.  We conclude that $\Hhex$ starts to
be smaller than $\Hzeta$ beginning with an independent constant at weight 7,
and going on to actual \emph{dropout functions} starting at weight 8.
A dropout function is any function whose symbol is allowed, but the function is forbidden by the branch-cut constraints, once we have restricted the independent constants to those in~\eqn{eq:indepzetas}.

%%%%%%%%%%%%%%%%%%%%%%%%%%%%%%%%%%%%%%%%%%%%%%%%%%
\renewcommand{\arraystretch}{1.25}
\begin{table}[!t]
\centering
\begin{tabular}[t]{l c c c c c c c c c c c c c c c}
\hline\hline
weight $n$
& 0 & 1 & 2 & 3 & 4 &  5 &  6 &  7 &  8 &  9 & 10 & 11 & 12 & 13 & 14
\\\hline\hline
$L=1$
& \green{0} & \green{0} & \green{0} &  &  &  &  &  &  &  &  &  &  &  & 
\\\hline
$L=2$
& \green{0} & \green{0} & \green{0} & \green{1} & \green{2}
&  &  &  &  &  &  &  &   &  &
\\\hline
$L=3$
& \green{0} & \green{0} & \green{0} & \green{1} & \green{2}
& \green{6} & 2 &  &  &  &  &  &  &  &
\\\hline
$L=4$
& \green{0} & \green{0} & \green{0} & \green{1} & \green{2}
& \green{6} & \green{13} & 12 & 2 &  &  &  &  &  &
\\\hline
$L=5$
& \green{0} & \green{0} & \green{0} & \green{1} & \green{2}
& \green{6} & \green{13} & \green{30} & 30 & 12 & 2 &  &  &  &
\\\hline
$L=6$
& \green{0} & \green{0} & \green{0} & \green{1} & \green{2}
& \green{6} & \green{13} & \green{30} & \green{59} & 82 & 36 & 12 & 2  &  &
\\\hline
$L=7+$
& \green{0} & \green{0} & \green{0} & \green{1} & \green{2}
& \green{6} & \green{13} & \green{30} & \green{59} & 110 & 98 & 43 & 11
& 3 & 0
\\\hline\hline
\end{tabular}
\caption{Same as Table~\ref{tab:MHVNMHVdim}, but just the parity odd
$\{n,1,1,\ldots,1\}$ coproducts of the MHV and NMHV amplitudes.
Note that saturation of the odd functions now begins two loops earlier.}
\label{tab:MHVNMHVodddim}
\end{table}

In Table~\ref{tab:MHVNMHVKfns} we show the number of $\{n,1,\ldots,1\}$
coproducts of the $L$ loop amplitudes which have no parity-odd $y_i$ letters
in their symbols, which we call `$K$'.  (The remaining $y_i$-containing
functions, we call `non-$K$').  Rather interestingly, at high loop order $L$
one has to take a large number of iterated coproducts of an amplitude,
$L-2$ to be precise, before one encounters a $K$ function.

%%%%%%%%%%%%%%%%%%%%%%%%%%%%%%%%%%%%%%%%%%%%%%%%%%
\renewcommand{\arraystretch}{1.25}
\begin{table}[!t]
\centering
\begin{tabular}[t]{l c c c c c c c c c c c c c c c}
\hline\hline
weight $n$
& 0 & 1 & 2 & 3 & 4 &  5 &  6 &  7 &  8 &  9 & 10 & 11 & 12 & 13 & 14
\\\hline\hline
$L=1$
& \green{1} & \green{3} & 4 &  &  &  &  &  &  &  &  &  &  &  & 
\\\hline
$L=2$
& \green{1} & \green{3} & \green{6} & 9 & 1 &  &  &  &  &  &  &  &  &  & 
\\\hline
$L=3$
& \green{1} & \green{3} & \green{6} & \green{12} & 19 & 4 & 0 &  &
&  &  &  &  &  & 
\\\hline
$L=4$
& \green{1} & \green{3} & \green{6} & \green{12} & \green{22} & 38
& 15 & 0 & 0 &  &  &  &  &  & 
\\\hline
$L=5$
& \green{1} & \green{3} & \green{6} & \green{12} & \green{22} & \green{39}
& \green{67} & 36 & 0 & 0 & 0 &  &  &  & 
\\\hline
$L=6$
& \green{1} & \green{3} & \green{6} & \green{12} & \green{22} & \green{39}
& \green{67} & 113 & 94 & 0 & 0 & 0 & 0 &  & 
\\\hline
$L=7+$
& \green{1} & \green{3} & \green{6} & \green{12} & \green{22} & \green{39}
& \green{67} & \green{114} & 156 & 32 & 0 & 0 & 0 & 0 & 0
\\\hline\hline
\end{tabular}
\caption{Same as Table~\ref{tab:MHVNMHVdim}, but just for the parity even
$K$ functions that do not contain $y_i$ in their symbols.
Note that for loop order $L>2$, the first $L-2$ coproducts of the amplitudes
do not include any $K$ functions.}
\label{tab:MHVNMHVKfns}
\end{table}

Because the parity-even part of the function space is only saturated
through weight 7, we have to extrapolate somewhat to say
that the space of independent constants is really
$\zeta_4,\,\zeta_6,\,\zeta_8,\ldots$.
In fact, $\zeta_8$ by itself is not in the span of the 338 weight 8
functions shown in Table~\ref{tab:MHVNMHVdim}.
(Of these functions, 279 are parity-even, whereas 313 would be needed
to span the full expected weight 8 parity-even space. On
the other hand, the set of 279 even functions does include all of the
123 more complicated, $y_i$-containing `non-$K$' functions
shown in Table~\ref{tab:Hdim}.)

%%%%%%%%%%%%%%%%%%%%%%%%%%%%%%
\subsection{Saturation at \texorpdfstring{$(1,1,1)$}{(1,1,1)}}
\label{sec:saturation111}

What is easier to identify to higher weights
is the correct space of zeta values in $\Hhex$ at $(1,1,1)$
because there is no issue of mixing with all the other functions,
as there is in determining the independent constants.
In Table~\ref{tab:MHVNMHV111dim} we show that the weight-8 space is
saturated by four loops. (We can only get 2 values at weight $2L$,
one from $\EE^{(L)}(1,1,1)$ and one from $E^{(L)}(1,1,1)$;
this is enough at weight 8, but not at weight 10.)
Odd weights are harder to saturate because the final-entry conditions
on the MHV and NMHV amplitudes, together with the branch-cut condition,
imply that all the weight $2L-1$ first coproducts of the amplitudes
vanish at $(1,1,1)$.  (For example, $\EE^{1-u_i}(1,1,1)=0$ by the branch-cut
condition~\eqref{branchcutevencondition}, but 
$\EE^{u_i}(1,1,1)=-\EE^{1-u_i}(1,1,1)=0$
by the final-entry condition, and $\EE^{y_i}(1,1,1)=0$ by parity.)
Weight 9 is saturated by 7 loops, although it is a bit marginal because
we don't have any 8-loop data.  Weight 10 is also saturated at 7 loops.
This case is more secure, because only three linear combinations of
weight 10 zeta values are allowed by the coaction principle.

In summary, the space $\Hhex(1,1,1)$ is spanned by the following
elements through weight 12, in the $f$ alphabet of
ref.~\cite{HyperlogProcedures}, from weights 0 through 12:
\bea
&&1\label{finalHhex111}\\
&& - \nonumber\\
&& \zeta_2 \nonumber\\
&& - \nonumber\\
&& \zeta_4 \nonumber\\
&& 5 f_5 - 2\zeta_2 f_3 \nonumber\\
&& \zeta_6 \nonumber\\
&& 7 f_7 - \zeta_2 f_5 - 3\zeta_4 f_3 \nonumber\\
&& \zeta_8 \,,\ \ 5 f_{3,5} - 2\zeta_2 f_{3,3} \nonumber\\
&& 7 f_9 - 6\zeta_4f_5 \,,\ \ 5 f_9 - 3\zeta_6 f_3,\ \ \zeta_2 f_7-\zeta_6 f_3
\nonumber\\
&& \zeta_{10} \,,\ \ 7 f_{3,7}-\zeta_2 f_{3,5}-3\zeta_4 f_{3,3}  \,,\ \
5 f_{5,5}-2\zeta_2f_{5,3}
\nonumber\\ 
&& 33 f_{11} - 20 \zeta_8 f_3\,,
\ \zeta_2 f_9 - \zeta_8 f_3\,,
\ 3 \zeta_4 f_7 - 2 \zeta_8 f_3\,,
\ 3 \zeta_6 f_5 - 2 \zeta_8 f_3\,,
\ 5 f_{3,3,5} - 2 \zeta_2 f_{3,3,3} + \frac{5611}{132} \zeta_8 f_3
\nonumber\\
&&\zeta_{12} \,,
\ 7 f_{3,9}-6 \zeta_4 f_{3,5}\,,
\ 5 f_{3,9}-3 \zeta_6 f_{3,3}\,,
\ \zeta_2 f_{3,7}-\zeta_6 f_{3,3}\,,
\ 7 f_{5,7}-\zeta_2 f_{5,5}-3 \zeta_4 f_{5,3}\,,
\ 5 f_{7,5}-2 \zeta_2 f_{7,3} \,.
\nonumber
\eea
In appendix~\ref{appendix:fbasis}, we provide the conversion between the
$f$-alphabet and MZVs through weight 11. In the ancillary file
{\tt ftoMZV.txt} we do the same through weight 14.

%%%%%%%%%%%%%%%%%%%%%%%%%%%%%%%%%%%%%%%%%%%%%%%%%%
\renewcommand{\arraystretch}{1.25}
\begin{table}[!t]
\centering
\begin{tabular}[t]{l c c c c c c c c c c c c c c c}
\hline\hline
weight $n$
& 0 & 1 & 2 & 3 & 4 &  5 &  6 &  7 &  8 &  9 & 10 & 11 & 12 & 13 & 14
\\\hline\hline
$L=1$
& \green{1} & \green{0} & \green{1} &  &  &  &  &  &  &  &  &  &  &  & 
\\\hline
$L=2$
& \green{1} & \green{0} & \green{1} & \green{0} & \green{1}
&  &  &  &  &  &  &  &   &  &
\\\hline
$L=3$
& \green{1} & \green{0} & \green{1} & \green{0} & \green{1}
& 0 & \green{1} &  &  &  &  &  &  &  &
\\\hline
$L=4$
& \green{1} & \green{0} & \green{1} & \green{0} & \green{1}
& \green{1} & \green{1} & 0 & \green{2} &  &  &  &  &  &
\\\hline
$L=5$
& \green{1} & \green{0} & \green{1} & \green{0} & \green{1}
& \green{1} & \green{1} & \green{1} & \green{2} & 0 & 2 &  &  &  &
\\\hline
$L=6$
& \green{1} & \green{0} & \green{1} & \green{0} & \green{1}
& \green{1} & \green{1} & \green{1} & \green{2} & 1 & 2 & 0 & 2  &  &
\\\hline
$L=7+$
& \green{1} & \green{0} & \green{1} & \green{0} & \green{1}
& \green{1} & \green{1} & \green{1} & \green{2} & \green{3} & \green{3} & 1 & 2
& 0 & 1
\\\hline\hline
\end{tabular}
\caption{Same as Table~\ref{tab:MHVNMHVdim}, but just the space
of values of $\{n,1,1,\ldots,1\}$ coproducts of the MHV and NMHV amplitudes
at $(1,1,1)$.  Saturation of $\Hhex(1,1,1)$ is achieved through weight 10.}
\label{tab:MHVNMHV111dim}
\end{table}
%%%%%%%%%%%%%%%%%%%%%%%%%

At weight 11, we make use of a subspace of the hexagon functions that
can be defined to all weights, which is related to, but is
larger than, the $\Omega$ space associated with double pentaladder
integrals~\cite{Caron-Huot:2018dsv}.  This subspace saturates $\Hhex(1,1,1)$
through weight 10, and we assume it does so at weight 11.
This assumption removes one of the weight 11 zeta values allowed
by the coaction principle.
The values at weight 11 are also consistent with an analysis of
the branch-cut constraints for the general function space that takes
into account the triple coproducts of $\EE^{(7)}$.  And they are consistent
with the computed $\EE^{(7)}(1,1,1)$ and the nontrivial existence of a
suitable seven-loop $\rho$ to make it compatible with the coaction principle.

In appendix~\ref{appendix:fbasis}, we provide the values of the MHV
and NMHV amplitudes at $u=v=w=1$, $\EE^{(L)}(1,1,1)$ and $E^{(L)}(1,1,1)$,
through seven and six loops respectively, in terms of the $f$-basis
given in \eqn{finalHhex111} and rational number coefficients.
Most of the coefficients are actually integers.

From \eqn{finalHhex111} one can count how many combinations of zeta values
disappear {\it without} being forced to by the coaction principle. Without
such disappearances, the coaction principle would be trivially satisfied.
The only such disappearances are at odd weights $3,5,7,9,11,\ldots$,
and the number missing are $1,1,2,1,1,\ldots$.  We assume that there
are no such disappearances at weight 12, since there were none at smaller
even weights.  We have no ``amplitudes data'' at weight 13,
and only 1 data point at weight 14, namely $\EE^{(7)}(1,1,1)$. 
The coaction principle,
given \eqn{finalHhex111}, allows 9 independent combinations at weight 13,
and 12 combinations at weight 14. In comparison, the total number of MZVs at
these weights is $d^{\rm MZV}_{13} = 16$ and $d^{\rm MZV}_{14} = 21$,
or almost twice the dimension.

Returning to Table~\ref{tab:MHVNMHVdim}, one can see another kind
of saturation taking place:  the number of weight $2L-1$ entries,
or single coproducts of the MHV and NMHV amplitudes together, saturates at 24,
of which 12 are parity-even and 12 are odd
(using also Table~\ref{tab:MHVNMHVodddim}).
(The last line of these tables should be disregarded in this analysis,
since it does not include the unknown 7-loop NMHV amplitude.)
On the other hand, the set of weight $2L-2$ double coproducts has not
yet clearly reached a maximum at 6 loops, at 78.
If we look at the same tables for just the MHV amplitude,
Tables~\ref{tab:MHVdim} and \ref{tab:MHVodddim}, we see that
the MHV double coproducts have saturated at 21, of which 12 are
parity-even and 9 are odd.  It is not yet clear if the MHV triple
coproducts have saturated.  This kind of saturation provides
very useful information; the saturation of the MHV double
coproducts at 21 next-to-final-entries was assumed in constructing
the initial ansatz for $\EE^{(7)}$ in ref.~\cite{Caron-Huot:2019vjl}.

%%%%%%%%%%%%%%%%%%%%%%%%%%%%%%%%%%%%%%%%%%%%%%%%%%
\renewcommand{\arraystretch}{1.25}
\begin{table}[!t]
\centering
\begin{tabular}[t]{l c c c c c c c c c c c c c c c}
\hline\hline
weight $n$
& 0 & 1 & 2 & 3 & 4 &  5 &  6 &  7 &  8 &  9 & 10 & 11 & 12 & 13 & 14
\\\hline\hline
$L=1$
& \green{1} & \green{3} & 1 &  &  &  &  &  &  &  &  &  &   &  & 
\\\hline
$L=2$
& \green{1} & \green{3} & \green{6} & 4 & 1 &  &  &  &  &  &  &  &   &  & 
\\\hline
$L=3$
& \green{1} & \green{3} & \green{6} & \green{13} & 14 & 6 & 1 &  &  &  &  &  & 
&  & \\\hline
$L=4$
& \green{1} & \green{3} & \green{6} & \green{13} & \green{27} & 35
& 20 & 6 & 1 &  &  &  &  &  & 
\\\hline
$L=5$
& \green{1} & \green{3} & \green{6} & \green{13} & \green{27} & \green{54}
& 78 & 51 & 21 & 6 & 1 &  &  &  & 
\\\hline
$L=6$
& \green{1} & \green{3} & \green{6} & \green{13} & \green{27} & \green{54}
& \green{105} & 170 & 128 & 58 & 21 & 6 & 1 &  & 
\\\hline
$L=7$
& \green{1} & \green{3} & \green{6} & \green{13} & \green{27} & \green{54}
& \green{105} & \green{200} & 338 & 300 & 159 & 62 & 21 & 6 & 1
\\\hline\hline
\end{tabular}
\caption{The number of independent $\{n,1,1,\ldots,1\}$ coproducts
of the MHV amplitudes through $L=7$ loops. A green color indicates
saturation.}
\label{tab:MHVdim}
\end{table}

%%%%%%%%%%%%%%%%%%%%%%%%%%%%%%%%%%%%%%%%%%%%%%%%%%
\renewcommand{\arraystretch}{1.25}
\begin{table}[!t]
\centering
\begin{tabular}[t]{l c c c c c c c c c c c c c c c}
\hline\hline
weight $n$
& 0 & 1 & 2 & 3 & 4 &  5 &  6 &  7 &  8 &  9 & 10 & 11 & 12 & 13 & 14
\\\hline\hline
$L=1$
& \green{0} & \green{0} & \green{0} &  &  &  &  &  &  &  &  &  &  &  &
\\\hline
$L=2$
& \green{0} & \green{0} & \green{0} & \green{1} & 0 &  &  &  &  &  &  &  & 
&  &\\\hline
$L=3$
& \green{0} & \green{0} & \green{0} & \green{1} & \green{2}
& 3 & 0 &  &  &  &  &  &  &  &
\\\hline
$L=4$
& \green{0} & \green{0} & \green{0} & \green{1} & \green{2}
& \green{6} & 8 & 3 & 0 &  &  &  &  &  &
\\\hline
$L=5$
& \green{0} & \green{0} & \green{0} & \green{1} & \green{2}
& \green{6} & \green{13} & 21 & 9 & 3 & 0 &  &  &  &
\\\hline
$L=6$
& \green{0} & \green{0} & \green{0} & \green{1} & \green{2}
& \green{6} & \green{13} & \green{30} & 50 & 27 & 9 & 3 & 0 &  &
\\\hline
$L=7$
& \green{0} & \green{0} & \green{0} & \green{1} & \green{2}
& \green{6} & \green{13} & \green{30} & \green{59} & 110 & 75 & 31 & 9 & 3 & 0
\\\hline\hline
\end{tabular}
\caption{Same as Table~\ref{tab:MHVdim}, but just the parity odd
$\{n,1,1,\ldots,1\}$ coproducts of the MHV amplitude.
Note that saturation of the odd functions begins one loop earlier.}
\label{tab:MHVodddim}
\end{table}

In Table~\ref{tab:dropoutdim} we show the number of dropout functions,
which are not in $\Hhex$ even though their symbols satisfy
all symbol-level constraints.  As mentioned earlier, the
first such dropout is a unique (dihedrally symmetric)
weight-8 parity-odd function.  At weight 9, there is a unique parity-even
dropout.  At weight 10, there are two dropouts, now parity odd,
and also two at weight 11 parity even.  The situation at weight 12,
and especially beyond, is less clear.

In Table~\ref{tab:Hdim} we show the dimension of $\Hhex$, graded by parity.
We split the parity-even functions into the $K$ functions with no
parity-odd letters in their symbols and the remaining $y_i$-containing
functions (non-$K$).

%%%%%%%%%%%%%%%%%%%%%%%%%%%%%%%%%%%%%%%%%%%%
\renewcommand{\arraystretch}{1.25}
\begin{table}[!t]
\centering
\begin{tabular}[t]{l c c c c c c c c c c c c c c}
\hline\hline
weight $n$
& 0 & 1 & 2 & 3 & 4 &  5 &  6 &  7 &  8 &  9 & 10 & 11 & 12 & 13 \\
\hline\hline
P even dropouts
& 0 & 0 & 0 & 0 & 0 & 0 & 0 & 0 & 0 & 1 & 0 & 2 & 0? & 3??
\\\hline
P odd dropouts
& 0 & 0 & 0 & 0 & 0 & 0 & 0 & 0 & 1 & 0 & 2 & 0 & 2? & 0??
\\\hline\hline
\end{tabular}
\caption{The number of dropouts:
functions that do {\it not} appear in $\Hhex$ even though they
satisfy all the constraints at symbol level.
The numbers at weights 12 and 13 are slightly uncertain.}
\label{tab:dropoutdim}
\end{table}

%%%%%%%%%%%%%%%%%%%%%%%%%%%%%%%%%%%%%%%%%%%%%%%
\renewcommand{\arraystretch}{1.25}
\begin{table}[!t]
\centering
\begin{tabular}[t]{l c c c c c c c c c c c c c c}
\hline\hline
weight $n$
& 0 & 1 & 2 & 3 & 4 &  5 &  6 &  7 &  8 &  9 & 10 & 11 & 12 & 13 \\
\hline\hline
total
& 1 & 3 & 6 & 13 & 27 & 54 & 105 & 200 & 372 & 679 & 1214 & 2136 & 3693? & 6292?
\\\hline\hline
P even, $K$
& 1 & 3 & 6 & 12 & 22 & 39 & 67 & 114 & 190 & 315 & 517 &  846 & 1378 & 2241
\\\hline
P even, non $K$
& 0 & 0 & 0 &  0 &  3 &  9 & 25 &  56 & 123 & 244 & 474 &  872 & 1573 & 2740?
\\\hline
P odd
& 0 & 0 & 0 &  1 &  2 &  6 & 13 &  30 &  59 & 120 & 223 &  418 &  742? & 1311?
\\\hline\hline
\end{tabular}
\caption{The dimension of the extended Steinmann hexagon function space $\Hhex$, graded by parity and by $K$ {\it vs.} non-$K$ in the P-even case.  Beyond weight 7, the coproducts of known amplitudes do not saturate all of the functions, and so the numbers may be further reduced eventually. At weights 12 and 13, the numbers may be off by one or two.}
\label{tab:Hdim}
\end{table}
%%%%%%%%%%%%%%%%%%%%%%%%%

%%%%%%%%%%%%%%%%%%%%%
\subsection{\texorpdfstring{$K$}{K} functions and asymptotic growth}
\label{sec:Ksubsection}

The $K$ functions can be constructed systematically. (A similar set
of $K$ functions was constructed in ref.~\cite{Caron-Huot:2016owq},
but that set was too large; it included many functions
that did not satisfy the extended Steinmann relations.)
The basis for constructing the $K$ functions
is a set of HPLs of the form $H_{\vec{w}}(x)$,
where $x=1-1/u$ and $w_i\in \{0,1\}$.  The extended Steinmann condition forbids
two adjacent `$u$'s in the symbol, which means there cannot be two adjacent `$1$'s in the list of $w_i$. (This restriction is equivalent
to the $A_1$ cluster algebra adjacency restriction, and
so the counting of functions will be the same~\cite{DFGPrivate}.)
In the compressed notation (where
$k-1$ `0's followed by a `1' is represented by `$k$'), a `1' can only appear
at the beginning of the string, and at weight $n$ the string is a
partition of $n$.  So the first few functions are
\begin{align}
&H_1(x),  \nonumber\\
&H_2(x),  \nonumber\\
&H_3(x),\ H_{1,2}(x),  \nonumber\\
&H_4(x),\ H_{1,3}(x),\ H_{2,2}(x) \label{Ku0}  \\
&H_5(x),\ H_{1,4}(x),\ H_{2,3}(x),\ H_{3,2}(x),\ H_{1,2,2}(x), \nonumber\\
& \quad \vdots \nonumber
\end{align}
Notice that if the last element in the string for an HPL 
at weight $n$ is a `2', then it corresponds
to appending a `2' to one of the functions at weight $n-2$;
otherwise it corresponds to adding `1' to the last entry of one
of the functions at weight $n-1$.  In other words, the number of
such functions is given by the sum of the two previous numbers,
i.e.~it is enumerated by the Fibonacci sequence.

The full set of $K$ functions based on $u$ also has dependence
on $v/w$.  At weight $n$, one can construct a suitable function
for every function in \eqn{Ku0} with weight less than or equal to $n$
by multiplying by powers of $\ln(v/w)$ and adding some correction terms.
For example, suppressing the argument $x$ of the HPLs, the
first few are
\begin{align}
&\hbox{weight 1:} \quad H_1, \quad \ln(v/w), \label{Ku}\\
&\hbox{weight 2:} \quad H_2, \quad H_1 \ln(v/w),
\quad \tfrac{1}{2} \ln^2(v/w) + H_{1,1} , \nonumber\\
&\hbox{weight 3:} \quad H_3, \quad H_{1,2}, \quad H_2 \ln(v/w),
\quad \tfrac{1}{2} H_1 \ln^2(v/w) + H_{1,1,1},
\quad \tfrac{1}{6} \ln^3(v/w) + H_{1,1} \ln(v/w). \nonumber
\end{align}
Since the sum of the first $n$ terms in a Fibonacci sequence is
also a Fibonacci sequence, we again get a Fibonacci sequence
for the dimensions of this space.
The sequence of dimensions in \eqn{Ku0} is generated by $1+t/(1-t-t^2)$,
while the one in \eqn{Ku} is generated by $(1+t)/(1-t-t^2)$.

To get the complete set of $K$ functions, we need to consider also
cyclic permutations of the functions in \eqn{Ku}, i.e.~functions whose
HPL arguments are $1-1/v$ or $1-1/w$.  At each weight,
the cyclic permutations include a double-count of three pure-log functions
that have to be removed, so altogether we get a generating function of
\be
1 + 3 \biggl[ \frac{1+t}{1-t-t^2} - \frac{1}{1-t} \biggr]
= 1 + \frac{3\,t}{(1-t)(1-t-t^2)} \,.
\label{almostkdim}
\ee
Finally, at weight $n$
the independent constants $\zeta_4$, $\zeta_6$, $\zeta_8$, etc.,
can multiply $K$ functions of lower weight $n-4$, $n-6$, $n-8$, etc.
We can take them into account by multiplying the generating
function~\eqref{almostkdim} by the generating function counting
this sequence.  That is, the generating function for the sequence of
dimensions $k_n$ of all possible $K$ functions is
\be
k(t) = \sum_{n=0}^\infty k_n t^n
= \biggl[ 1 + \frac{3\,t}{(1-t)(1-t-t^2)} \biggr]
 (1+t^4+t^6+t^8+t^{10}+\ldots).
\label{kdim}
\ee
Series expanding $k(t)$ gives the dimensions in the line `P even, $K$'
in Table~\ref{tab:Hdim}.

The asymptotic growth rate of any Fibonacci sequence involves
the golden ratio $\phi = (1+\sqrt{5})/2$ = 1.618$\ldots$,
i.e. $k_n/k_{n-1} \sim \phi$ as $n\to\infty$.
This growth rate can be computed from the generating function $k(t)$
by finding the singularity on the positive $t$ axis closest to the origin,
which comes from the factor $1-t-t^2$ and is located at $t=1/\phi$,
and taking its inverse.
What about the growth rate of the dimensions $h_n$ of $\Hhexn$,
the weight $n$ part of $\Hhex$?
We don't have a closed formula generating $h_n$, but the last several
ratios $h_n/h_{n-1}$ from Table~\ref{tab:Hdim} are
1.8600, 1.8253, 1.7879, 1.7595, 1.7289, 1.7037.
It is tempting to think that this sequence might be approaching
the golden ratio asymptotically.

%%%%%%%%%%%%%%%%%%%%%%%%%%%%%%%%%%%%%%%%%%%%%%%%%%%%%

\section{The coaction principle at work on special lines and points}
\label{sec:hexagon_limits}

As described earlier, the coaction principle is built into the construction of the space of hexagon
functions $\Hhex$ at the level of the $\Delta_{n-1,1}$ coaction. Ideally, we would also like to explore its validity for general coaction components $\Delta_{n-m,m}$, as well as for arbitrary values of the cross ratios $u$, $v$, and $w$ in the bulk.
For all weights $n\leq8$, we have verified the coaction principle in the bulk
for arbitrary $m$, using the generalized polylogarithmic representations
of hexagon functions that can be computed, for example, with the
package \textsc{PolyLogTools}~\cite{Duhr:2019tlz}.
However, beyond weight eight, explicit representations for the elements
of $\Hhex$ in terms of generalized polylogarithms become so large that the
construction of their coproducts in the bulk becomes infeasible. 

As an alternative, we can check the coaction principle on lower-dimensional
surfaces within the three-dimensional bulk.
The focus on lower-dimensional surfaces is not a conceptual restriction for the
study of the coaction principle; the
coassociativity of the Hopf algebra of multiple polylogarithms,
cf.~ref.~\cite{Duhr:2012fh}, promotes the built-in coaction principle for the
components $\Delta_{n-1,1}$, to a coaction principle for all components
$\Delta_{n-m,m}$ for which the weight $m$ component in the second entry
of the coaction has a non-vanishing $\Delta_{1,\dots,1}$ component.
Hence the non-trivial checks of the coaction principle
arise from components of $\Delta$ for which the second entry vanishes when
acting again with $\Delta_{\bullet,1}$ --- for example, a transcendental constant
such as a MZV.  We are therefore particularly interested 
in studying the coproduct structure of the hexagon function space in the
presence of constants in the second entry. Studying the hexagon function space
in kinematic limits, such as lower-dimensional surfaces, allows
such constants to survive and provides a particularly rich laboratory for
our studies.

\begin{figure}
\begin{center}
\includegraphics[width=5.5in]{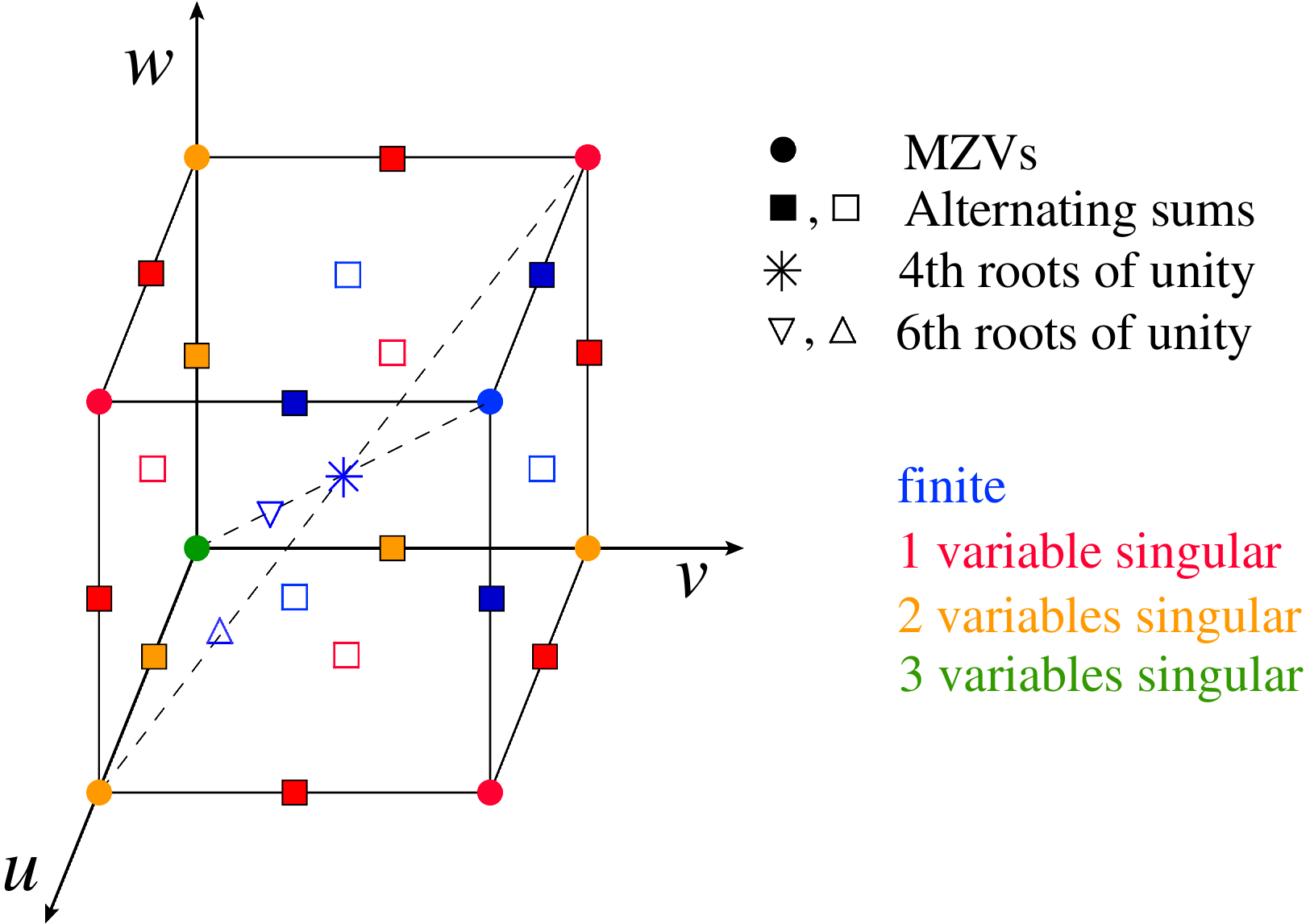}
\end{center}
\caption{Points associated with the unit cube in $(u,v,w)$ where the functions in $\Hhex$ evaluate to interesting transcendental numbers associated with polylogarithms with indices that are square, fourth and sixth roots of unity, as indicated by the shape of the symbol. The color of the symbol indicates how many of the three cross ratios are singular (equal to zero) at that point.}
\label{fig:cubespecial}
\end{figure}

In this section, we will first discuss the spaces of functions obtained
when we collapse
$\Hhex$ onto various one-dimensional lines, where the functions become
either harmonic polylogarithms (HPLs)~\cite{Remiddi:1999ew} or their
generalizations, cyclotomic polylogarithms (CPLs)~\cite{Ablinger:2011te}.
\Tab{tab:lines} shows several examples of such lines, as well as
special points along the line where the functions evaluate to MZVs,
alternating sums (ASums), or cyclotomic polylogarithms whose
weights include 4$^{\rm th}$
or 6$^{\rm th}$ roots of unity, evaluated at 1 (4$^{\rm th}$ Roots
or 6$^{\rm th}$ Roots, for short).
Interestingly, the latter two spaces of numbers
are also found~\cite{Schnetz:2017bko} in the analytic formula
for the four-loop electron anomalous
magnetic moment~\cite{Laporta:2017okg}.
Some of the special points are plotted in Figure~\ref{fig:cubespecial}.

%%%%%%%%%%%%%%%%%%%%%%%%%%%%%%%%%%%%%%%%%%%%%%%%%%%%%%%%%
\renewcommand{\arraystretch}{1.25}
\begin{table}[!t]
\centering
\begin{tabular}[t]{l l l c}
\hline\hline
$(u,v,w)$ for line
& symbol letters & special points & more special points \\\hline\hline
$(u,u,1)$ & $u,1-u$ & $u=0,1,\infty \, \Rightarrow$ MZVs
                  & $u=\frac{1}{2},2 \, \Rightarrow$ ASums \\\hline
$(u,1,1)$ & $u,1-u$ & $u=0,1,\infty \, \Rightarrow$ MZVs
                  & $u=\frac{1}{2},2 \, \Rightarrow$ ASums \\\hline
$(u,0,1)$ & $u,1-u$ & $u=0,1,\infty \, \Rightarrow$ MZVs
                  & $u=\frac{1}{2},2 \, \Rightarrow$ ASums \\\hline
$(u,0,0)$ & $u,1-u$ & $u=0,1,\infty \, \Rightarrow$ MZVs
                  & $u=\frac{1}{2},2 \, \Rightarrow$ ASums \\\hline\hline
$\Bigl(\frac{y}{1+y},0,\frac{y}{1+y}\Bigr)$ & $y,1-y,1+y$ & $y=1,-1 \, \Rightarrow$ ASums
                  & $-$ \\\hline
$\Bigl(\frac{(1+y)^2}{4y},\frac{1}{2},\frac{1}{2}\Bigr)$ & $y,1-y,1+y$ & $y=1,-1 \, \Rightarrow$ ASums
                  & $y=i \, \Rightarrow$ $4^{\rm th}$ Roots  \\\hline\hline
$\Bigl(\frac{y}{(1+y)^2},\frac{y}{(1+y)^2},\frac{y}{(1+y)^2}\Bigr)$ & $y,1+y,y-\omega,y-\bar\omega$ & $y=1 \, \Rightarrow$ $6^{\rm th}$ Roots  & $-$ \\\hline
$\Bigl(\frac{1+y+y^2}{(1+y)^2},\frac{1+y+y^2}{(1+y)^2},\frac{y}{(1+y)^2}\Bigr)$ & $y,1+y,y-\omega,y-\bar\omega$ & $y=1 \, \Rightarrow$ $6^{\rm th}$ Roots  & $-$ \\\hline\hline
\end{tabular}
\caption{Examples of special lines through the space of cross ratios where the function space collapses to cyclotomic polylogarithms, and special points
where the functions evaluate to MZVs or generalizations thereof.
Here $\omega=\exp(2\pi i/3)$, $\bar\omega=\exp(-2\pi i/3)$.}
\label{tab:lines}
\end{table}
%%%%%%%%%%%%%%%%%%%%%%%%%%%%%%%%%%%%%%%%%%%%%%%

%%%%%%%%%%%%%%%%%%%%%%%%%%%%%%%%%%%
%%%%%%%%%%%%%%%%%%%%%%%%%%%%%%%%%%%%%%%%%%%%%%%%%%%%%%

%%%%%%%%%%%%%%%%%%%%%%%%%%%%%%%%%
\subsection{Lines with symbol alphabet \texorpdfstring{$\{u,1-u\}$}{\{u,1-u\}}}

The first four lines of Table~\ref{tab:lines} are all similar in that
there are only two symbol letters, $u$ and $1-u$.  The functions must
be HPLs $H_{\vec{w}}$ with weight vectors $\vec{w}$ for which all the components
$w_i \in \{0,1\}$.  For argument of the HPLs, we use the variable $x=1-1/u$.
Because there are no cuts at $u=1$, the last weight vector index is always 1.
The dimensionality of $\Hhex$, restricted to each of the lines, is different
in each case, as shown in Table~\ref{tab:MZVlines}.
The line `maximal dim.' in the table refers to only imposing
the branch-cut constraint on the HPLs, and allowing for all possible
MZVs to be present as independent constants.
The generating function for HPLs with no branch cuts at $u=1$ is
\be
d^{\rm H}(t) = \frac{1-t}{1-2t} = 1 + t + 2 t^2 + 4 t^3 + 8 t^4 + \ldots,
\label{HPLbranchcutgenfn}
\ee
while the generating function for the MZVs was given in \eqn{eq:MZVindep}.
The generating function for the maximal set of functions with
symbol letters $u$, $1-u$ and no branch cuts at $u=1$ is just the product
\be
d^{\rm H}(t) d^{\rm MZV}(t)
= 1 + t + 3 t^2 + 6 t^3 + 12 t^4 + 25 t^5 + 50 t^6 + \ldots,
\label{maxgenfn}
\ee
as shown in the first row of Table~\ref{tab:MZVlines}.

%%%%%%%%%%%%%%%%%%%%%%%%%%%%%%%%%%%%%%%%%%%%%%%%%%%%%%%%%
\renewcommand{\arraystretch}{1.25}
\begin{table}[!t]
\centering
\begin{tabular}[t]{l c c c c c c c c c c c c}
\hline\hline
weight & 1 & 2 & 3 & 4 & 5 & 6 & 7 & 8 & 9 & 10 \\\hline\hline
maximal dim.   & 1 & 3 & 6 & 12 & 25 & 50 & 101 & 203 & 407 & 816 \\\hline
$(u,u,1)$ dim. & 1 & 3 & 4 &  8 & 15 & 26 &  48 &  84 & 150 & 256 \\\hline
$(u,1,1)$ dim. & 1 & 2 & 3 &  6 & 10 & 18 &  30 &  52 &  90 & 152 \\\hline
$(u,0,1)$ dim. & 1 & 3 & 5 & 10 & 19 & 36 &  68 & 129 & 240 & 443 \\\hline 
$(u,0,0)$ dim. & 1 & 3 & 6 & 11 & 21 & 38 &  68 & 120 & 207 & 352 \\\hline\hline
\end{tabular}
\caption{Dimensions of $\Hhex$ when restricted to four lines with
symbol letters $u$, $1-u$, for weights up to 10.
The maximal dimension corresponds to allowing all MZVs and all HPLs with
no branch cuts at $u=1$. For the last two lines, only the finite parts
of the singular limits onto the lines are used.}
\label{tab:MZVlines}
\end{table}
%%%%%%%%%%%%%%%%%%%%%%%%%%%%%%%%%%%%%%%%%%%%%%%

On all four lines shown in Table~\ref{tab:MZVlines}, the number of functions
that hexagon functions $\Hhex$ approach in the limit is considerably
less than for the maximal set.  The last two lines, $(u,0,1)$ and $(u,0,0)$,
are short-hand for $(u,v,1)$ with $v\to0$ and $(u,v,w)$ with $v,w\to0$.
On these lines, the limiting behavior of functions in $\Hhex$
also includes powers of the singular logarithm $\ln v $ (and in the second
case, also $\ln w$), multiplied by lower-weight functions of the same type.
For simplicity, the table just counts the dimension of the finite terms,
i.e.~we ignore the terms with positive powers of $\ln v$ or $\ln w$.

Some of the four sequences of dimensions are strictly smaller than others;
however, none of the four function spaces is contained in the others.
To illustrate this, we provide bases for the various function spaces
through weight 4:
\bea
&&(u,u,1): \nonumber\\
&& H_1 \nonumber\\
&& H_2,\ H_{1,1},\ \zeta_2 \label{uu1basis} \\
&& H_3,\ H_{2,1},\ H_{1,2},\ H_{1,1,1} + \tfrac{1}{2} \zeta_2 H_1 \nonumber \\
&& H_4,\ H_{3,1},\ H_{2,2},\ H_{2,1,1} + \tfrac{1}{2} \zeta_2 H_2,\
         H_{1,3},\ H_{1,1,1,1} + \tfrac{1}{2} \zeta_2 H_{1,1},\ 
         H_{1,2,1} + H_{1,1,2},\ \zeta_4 \nonumber
\eea
\bea
&&(u,1,1): \nonumber\\
&& H_1 \nonumber\\
&& H_2,\ H_{1,1} + 2 \zeta_2 \label{u11basis} \\
&& H_3,\ H_{1,2},\ H_{1,1,1} + 2 \zeta_2 H_1 \nonumber \\
&& H_4,\ H_{2,2},\ H_{1,3},\ H_{1,1,1,1} + 2 \zeta_2 H_{1,1},\ 
         H_{1,1,2} + H_{2,1,1} + 2 \zeta_2 H_2,\ \zeta_4 \nonumber
\eea
\bea
&&(u,0,1): \nonumber\\
&& H_1 \nonumber\\
&& H_2,\ H_{1,1},\ \zeta_2 \label{u01basis} \\
&& H_3,\ H_{1,2},\ H_{1,1,1},\ \zeta_2 H_1,\ \zeta_3 \nonumber \\
&& H_4,\ H_{2,2},\ H_{2,1,1},\ \zeta_2 H_2,\ H_{1,3},\ H_{1,1,2},\
         H_{1,1,1,1},\ \zeta_2 H_{1,1},\ \zeta_3 H_1,\ \zeta_4 \nonumber
\eea
\bea
&&(u,0,0): \nonumber\\
&& H_1 \nonumber\\
&& H_2,\ H_{1,1},\ \zeta_2 \label{u00basis} \\
&& H_3,\ H_{2,1},\ H_{1,2},\ H_{1,1,1},\ \zeta_2 H_1,\ \zeta_3 \nonumber \\
&& H_4,\, H_{3,1},\, H_{2,2},\, H_{2,1,1} + H_{1,2,1},\, H_{1,3},\,
         H_{1,2,1} + H_{1,1,2},\, H_{1,1,2} + \zeta_2 H_2,\, 
         H_{1,1,1,1},\, \zeta_2 H_{1,1},\, \zeta_3 H_1,\, \zeta_4 .
\nonumber
\eea
On the line $(u,1,1)$, there is a dropout at weight 2,
in that $H_{1,1}$ and $\zeta_2$ do not appear separately,
but only in the combination $H_{1,1}+2\zeta_2$.  This reflects
a similar combination of $\zeta_2$ with logarithms in the bulk.
Also, the function $H_{2,1}$ does not appear at weight 3.
At weight 4, two HPLs have to be combined into a sum.
Similar dropouts happen on the other lines.
Even though there are always fewer functions on the line $(u,1,1)$
than on the line $(u,u,1)$, the former space is not a subset of the latter,
starting at weight 3, because the coefficients `$r$' in 
$H_{1,1,1} + r \zeta_2 H_1$ are different in the two cases.
Similarly, the $(u,u,1)$ functions are not contained in the $(u,0,0)$
functions, beginning at weight 4.

One can see the coaction principle at work by examining
the lists of functions.  For example, on the line $(u,1,1)$,
once the function $H_{2,1}$ does not appear at weight 3,
then the function $H_{3,1}$ cannot appear at weight 4,
because
\be
\Delta_{3,1} H_{3,1}\ =\ H_{2,1} \otimes \ln x
\label{DeltaH31}
\ee
Similarly, the combination $H_{1,1,1} + 2 \zeta_2 H_1$ at weight 3
is dictated by the combination $H_{1,1} + 2 \zeta_2$ at weight 3.

These examples just illustrate the $\{n-1,1\}$ component
of the coaction. However, using the iterated integral representations
of the HPLs, we can verify that the coaction principle holds on
these lines for $\{n-m,m\}$ for generic $m$ for sufficiently large $n$.
The restriction to large enough $n$ ensures that the dimension of the space in
the first entry of the coproduct is at least as large as the dimension of the
space in the second entry.

In the process, we find that the space ${\cal K}^\pi$
represented by the second, de Rham term in the
coaction~\eqref{eq:coaction_principle} on these lines seems to be
totally unrestricted.  That is, all HPLs with
weight-vector components $\{0,1\}$ appear and all MZVs appear, except for
powers of $\pi^2$ which never appear in the second entry of $\Delta$
by construction.  The generating function for this space is
\be
d^{{\rm dR}_{\{0,1\}}}(t) = \frac{1-t^2}{1-t^2-t^3}\frac{1}{1-2t}
=  1 + 2 t + 4 t^2 + 9 t^3 + 18 t^4 + 37 t^5 + 75 t^6 + \ldots.
\label{ulinedeRhamdimension}
\ee
In Table~\ref{tab:deRhamSpaceuuone} we give the dimensions deduced
for this space by performing the coaction on elements of $\Hhex_n$,
where $n$ is the overall weight.  Again a green color denotes
saturation, i.e.~reaching the dimensions predicted by
\eqn{ulinedeRhamdimension}.  Even going to overall weight 10,
we can only saturate through de Rham weight 4.  The problem is that the
number of de Rham entries is growing much faster with weight
than the number of first entries, but one cannot see more independent
elements of ${\cal K}^\pi$ than there are first-entry functions
with which to pair them.

\begin{table}[!t]
\centering
\begin{tabular}[t]{r || c c c c c c c c c c c}
\ & \multicolumn{11}{c}{\qquad $\longleftarrow$ de Rham weight $\longrightarrow$ \quad\ } \\
\hline\hline
overall weight & 1 & 2 & 3  & 4 & 5 & 6 & 7 & 8 & 9 \\\hline\hline
     2 & 1       &          &         &   &   &   &   &   & \\\hline
     3 &\green{2}& 1        &         &   &   &   &   &   & \\\hline
     4 &\green{2}& 3        & 1       &   &   &   &   &   & \\\hline
     5 &\green{2}&\green{4} & 3       & 1 &   &   &   &   & \\\hline
     6 &\green{2}&\green{4} & 4       & 3 & 1 &   &   &   & \\\hline
     7 &\green{2}&\green{4} & 8       & 4 & 3 & 1 &   &   & \\\hline
     8 &\green{2}&\green{4} &\green{9}& 8 & 4 & 3 & 1 &   & \\\hline
     9 &\green{2}&\green{4} &\green{9}&15 & 8 & 4 & 3 & 1 & \\\hline
    10 &\green{2}&\green{4} &\green{9}&\green{18} &15 & 8 & 4 & 3 & 1
  \\\hline\hline
\end{tabular}
\caption{Dimensions of the space $\mathcal{K}^{\pi}$ of de Rham entries of
    the coaction of hexagon functions restricted to the line $(u,u,1)$.
    The green entries mark
    spaces that are saturated, in the sense that all possible functions in
    $\mathcal{K}^{\pi}$ at the given weight, cf.~\eqn{ulinedeRhamdimension},
    appear as independent de Rham entries of the coaction.}
\label{tab:deRhamSpaceuuone}
\end{table}

There are three values of $u$ for which the values of functions in $\Hhex$
on the four lines approach MZVs: $u=0,1,\infty$.  At $u=0$ and $\infty$,
there can be associated singular factors of $\ln u$.
At all of these points {\it except} for the base point $(u,v,w)=(1,1,1)$, the values of the functions span the complete set of MZVs through weight 10,
achieving the dimension given by \eqn{eq:MZVindep}.  In other words,
the only MZV point that we have found where there are
dropout MZV values --- and a nontrivial coaction principle ---
is $(1,1,1)$.  (However, as we will discuss further
in section~\ref{sec:conclusions},
there are indications based on the flux tube
expansion~\cite{Basso:2013vsa,Basso:2013aha,Basso:2014hfa}
that there {\it should} be dropouts at the point 
$(u,v,w)=(1,0,0)$ and its cyclic images.)
In contrast, we will find multiple points
exhibiting nontrivial coaction features in the alternating sum
and cyclotomic cases.

%%%%%%%%%%%%%%%%%%%%%%%%%%%%%%%%%%%%%%%%%%%%%%

\subsection{Lines with symbol alphabet \texorpdfstring{$\{y,1-y,1+y\}$}{\{y,1-y\}}}

The next class of lines in Table \ref{tab:lines} are the two lines with symbol
letters $y, 1-y$, and $1+y$. Functions built from this alphabet must be HPLs
$H_{\vec{w}}(y)$ with weight vectors $\vec{w}$
drawn from the set $\{0,1,-1\}$, and argument $y$.
The first-entry condition ensures that there is only a single weight one
function in each case: $\ln\big(\frac{y}{1+y}\big)=H_0-H_{-1}$ in the case of the first line,
and $\ln\big(\frac{(1+y)^2}{4y}\big)=2H_{-1}-H_0-2\ln{2}$ in the case of the
second line.
Consequently, at weight $n$, there are at most $3^{n-1}$ different functions
that can be built from this symbol alphabet.

Specializing the space of HPLs with weight vectors drawn from $\{0,1,-1\}$ to unit
argument results in the space of alternating sums. Alternating sums can be defined as
harmonic sums evaluated at infinity,
\be
S_{k_1,\dots,k_d}=\sum_{1\le{}n_d\le{}n_{d-1}\le\dots\le{}n_1\le\infty}\frac{\sign(k_1)^{n_1}}{n_1^{k_1}}\dots\frac{\sign(k_d)^{n_d}}{n_d^{k_d}}\,.
\ee
For positive indices, this definition reduces to (linear combinations of) the ordinary MZVs.
One choice of basis for alternating sums at the first few weights is shown in
Table~\ref{tab:asums}.  The $f$-alphabet representation is also
provided~\cite{HyperlogProcedures}, using the same notation as
in the MZV case except for the superscript `2' on the $f$ to indicate
the alternating sum case.

%%%%%%%%%%%%%%%%%%%%%%%%%%%%%%%%%%%%%%%%%%
\begin{table}[t!]
    \begin{center}
\begin{tabular}[t]{c c c}
    weight & basis elements & conversion to $f^2$-alphabet \\\hline\hline
    1 & $\ln2$   &  $-f^2_1$\\\hline
    2 & $\zeta_2$ & $\zeta_2$\\\hline
    3 & $\zeta_3$  & $-\frac{4}{3}f^2_3$\\\hline
    4 & $\zeta_4$  & $\zeta_4$
       \\
      & $\Li_4(\tfrac{1}{2})$
      & $\tfrac{15}{16}\zeta_4+\tfrac{1}{2}\zeta_2f^2_{1,1}-\tfrac{7}{6}f^2_{1,3}-f^2_{1,1,1,1}$\\\hline
    5 & $\zeta_5$ & $-\tfrac{16}{15}f^2_5$\\
      & $\Li_5(\tfrac{1}{2})$
      & $-\tfrac{11}{20}f^2_5+\tfrac{15}{16}\zeta_4f^2_1+\tfrac{1}{2}\zeta_2f^2_{1,1,1}-\tfrac{7}{6}f^2_{1,1,3}-f^2_{1,1,1,1,1}$\\\hline
    6 & $\zeta_6$  & $\zeta_6$\\
      & $\Li_6(\tfrac{1}{2})$
      & $\tfrac{53}{64}\zeta_6+\tfrac{15}{16}\zeta_4f^2_{1,1}+\tfrac{1}{2}\zeta_2f^2_{1,1,1,1}-\tfrac{11}{20}f^2_{1,5}-\tfrac{7}{6}f^2_{1,1,1,3}-f^2_{1,1,1,1,1,1}$\\
      & $\textrm{S}_{-5,-1}$& $ - \tfrac{23}{16}\zeta_6+\tfrac{4}{3}f^2_{3,3}+\tfrac{31}{15}f^2_{1,5}$ \\\hline
\hline
\end{tabular}
\caption{Indecomposable basis elements for alternating sums at the first few
    weights.}
\label{tab:asums}
\end{center}
\end{table}
%%%%%%%%%%%%%%%%%%%%%%%%%%%%%%%%%%%%%%%%%%

The number of all basis elements (including products of lower weight constants)
for alternating sums at a given weight $n$ is counted by the
Fibonacci number $F_{n+1}$~\cite{Zagier:1994,Broadhurst:1996kc},
and the generating function for these dimensions is
\be
d^{\textrm{alt}}(t)
= \frac{1}{1-t-t^2}
= 1 + t + 2t^2 + 3t^3 + 5t^4 + 8t^5 + 13t^6 + \ldots\,.
\ee
The generating function for HPLs with indices drawn from $\{0,1,-1\}$ and no
branch cuts except at $u=0,\infty$ is
\be
d^{H(\pm 1)}(t) = \frac{1-2t}{1-3t} = 1 + t + 3t^2 + 9t^3 + 27t^4 + 81t^5 + 243t^6
+ \dots\,.
\ee
The generating function for the maximal set of functions with symbol letters
$y,1-y,1+y$ and no branch cuts except at $u=0,\infty$ is then just the product,
\be
d^{H(\pm 1)}(t) d^{\textrm{alt}}(t)
= 1 + 2t + 6t^2 + 17t^3 + 50t^4 + 148t^5 + 441t^6 + \ldots\,.
\ee

As was the case for the lines in Table~\ref{tab:MZVlines},
the hexagon functions actually span a much smaller set of functions
when constrained to these particular lines. In Table~\ref{tab:altdim}
we tabulate the dimensions of the spaces obtained from the
hexagon functions. This table shows that the dimensions of the spaces obtained
from restricting the hexagon functions to lines with a three letter alphabet are all 
significantly smaller than the maximal dimension possible for this alphabet.
(It could not really be otherwise, since the number of independent
functions cannot be greater than the total number of hexagon functions,
which grows by a factor of about 1.7 for each additional weight,
while the three-letter space grows by a factor of 3 for each additional weight.)
Thus, much of the rich structure of the space of hexagon
functions survives when limiting to either line.
Restricting to the line
$\Big(\tfrac{(1+y)^2}{4y},\tfrac{1}{2},\tfrac{1}{2}\Big)$, the basis for the
space of functions can be expressed most conveniently in terms of HPLs with indices $\pm 1$ and
$0$, and argument $y$. For the first few weights we have then,
\bea
&&\Big\{H_0-2H_{-1},\quad \ln2\Big\} \,, \nonumber\\
\nonumber\\
&&\Big\{H_{-1}^2-H_{-1,0}+2H_{1,-1}-H_{1,0}-2\ln2(H_{-1}+H_1),\quad
(H_0-2H_{-1})^2,\nonumber\\
&&\phantom{\{}\ln2(H_0-2H_{-1})+\ln^22,\quad \zeta_2\Big\} \,, \nonumber\\
\nonumber\\
&&\Big\{2H_{0,-1,-1}-H_{0,-1,0}-H_{0,0,-1}+4H_{1,-1,-1}-2H_{1,-1,0}+4H_{1,1,-1}-2H_{1,1,0}\nonumber\\
&&\quad+\tfrac{1}{12}H_0^3+\zeta_2(H_{-1}-H_1)-2\ln2(H_1^2+2H_{1,-1}),\nonumber\\
&&\phantom{\{}2H_{-1,-1,0}+2H_{-1,0,1}-H_{-1,0,0}+2H_{0,-1,-1}-H_{0,-1,0}-H_{0,0,-1}-\tfrac{2}{3}H_{-1}^3+\tfrac{1}{12}H_0^3,\nonumber\\
&&\phantom{\{}2H_{-1,-1,0}+4H_{-1,1,-1}-2H_{-1,1,0}-2H_{0,-1,-1}+H_{0,-1,0}-2H_{0,1,-1}+H_{0,1,0}\nonumber\\
&&\quad+\tfrac{2}{3}H_{-1}^3+2\ln2(H_{0,-1}+H_{0,1}-2H_{-1,1}-H_{-1}^2),\nonumber\\
&&\phantom{\{}H_{0,0,-1}-2H_{0,1,-1}+H_{0,1,0}+\tfrac{1}{12}H_0^3+2\zeta_2H_{-1}+2\ln2(H_{0,-1}+H_{0,1}),\nonumber\\
&&\phantom{\{}\ln2(H_0-2H_{-1})^2+2\ln^22(H_0-2H_{-1})+\tfrac{4}{3}\ln^32,\nonumber\\
&&\phantom{\{}\zeta_2(H_0-2H_{-1}),\quad \zeta_2\ln2,\quad \zeta_3\Big\}\,.
\eea
\begin{table}[t!]
    \begin{center}
        \begin{tabular}[t]{l c c c c c c c c c c}
            \hline\hline
   weight    & 1 & 2 &  3 &  4 &   5 &   6 &    7 &    8 &     9 & 10
\\\hline\hline
maximal dim. & 2 & 6 & 17 & 50 & 148 & 441 & 1318 & 3946 & 11825 & 35454
\\\hline
            $\left(\tfrac{y}{1+y},0,\tfrac{y}{1+y}\right)$
             & 1 & 3 & 6  & 11 &  24 &  45 &  88 &   163 &   301 & 539\\\hline
            $\left(\tfrac{(1+y)^2}{4y},\tfrac{1}{2},\tfrac{1}{2}\right)$
             & 2 & 4 & 8  & 15 &  28 &  52 &  96 &   174 &   319 & 567\\
            \hline\hline
        \end{tabular}
        \caption{Dimensions of $\mathcal{H}^{\textrm{hex}}$ when restricted to
            the two lines with with symbol letters $y,1-y,1+y$. The maximal
            dimension corresponds to allowing all alternating sums and all HPLs
            with no branch cuts except at $u=0,\infty$.
            We only count the finite parts
            of the functions on the first, singular line.}
        \label{tab:altdim}
    \end{center}
\end{table}

Using explicit representations of the hexagon functions on the line, we can
study the structure of the coaction on the hexagon functions. We once again
verify that the coaction principle holds on these lines as well for $\{n-m,m\}$
components of the coaction for generic $m$ and sufficiently large $n$.
In the process of verifying the coaction principle, we can study the space of
functions appearing in the second term of the coaction. In Table
\ref{tab:deRhamSpaceHpl} we tabulate the dimensions of the space of
functions observed in the back (de Rham) entry. Once again we observe that
the number of functions that can
appear in the back entry is considerably larger at a given weight than the space
of functions on the line at the same weight. Again the explanation is
that the back-entry functions are not required to fulfill a first-entry
condition. Because the space of back-entry functions is larger than the
space of hexagon functions and grows faster with increasing weight,
the functions appearing in the back entry of the coaction saturate very slowly,
as can be seen in Table \ref{tab:deRhamSpaceHpl}.

A basis for the saturated space of
back entries at weights one and two can be written as,
\bea
&&\{H_0,\ H_1,\ H_{-1},\ \ln2\} \,,\nonumber\\
&&\{H_0^2,\ H_{-1}^2,\ H_{-1,1},\ H_0H_1,\ H_0H_{-1},\ H_{0,1},\ H_{0,-1},\ H_0\ln2,
\nonumber\\
&&\phantom{\{}~~~~H_1^2-2H_{1,-1},\ H_{-1}\ln2-\tfrac{1}{2}\ln^22,
\ H_{-1}H_1-H_1\ln2-\tfrac{1}{2}\ln^22 \} \,. \label{eq:uhhwt1and2}
\eea
From the explicit representation of the back-entry functions,
it is clear that the back-entry space is not completely unrestricted
but still seems to retain some residual constraints from the full space:
two of the 13 potential weight two functions
(from the generating function $(1-t^2)/(1-t-t^2)/(1-3t) = 1 + 4t + 13t^2 + \ldots$)
are missing from \eqn{eq:uhhwt1and2}.
This behavior is contrary to what was observed on the simpler lines with
symbol alphabet $\{u,1-u\}$.

\begin{table}[!t]
\centering
\begin{tabular}[t]{r | c c c c c c c c c}
\ & \multicolumn{8}{c}{\ $\longleftarrow$ de Rham weight $\longrightarrow$ \ } \\
\hline\hline
overall weight & 1       & 2        & 3       & 4        & 5 & 6 & 7  \\\hline\hline
     2 & 2       &          &         &          &   &   &   &   \\\hline
     3 &\green{4}& 2        &         &          &   &   &   &   \\\hline
     4 &\green{4}& 4        & 2       &          &   &   &   &   \\\hline
     5 &\green{4}& 8        & 4       & 2        &   &   &   &   \\\hline
     6 &\green{4}&\green{11} & 8       & 4       & 2 &   &   &   \\\hline
     7 &\green{4}&\green{11} & 15       & 8      & 4 & 2 &   &   \\\hline
     8 &\green{4}&\green{11} & 28       & 15     & 8 & 4 & 2 &   \\\hline
\end{tabular}
\caption{Dimensions of the space $\mathcal{K}^{\pi}$ of de Rham entries of the
    hexagon functions restricted to the line
    $(\tfrac{(1+y)^2}{4y},\tfrac{1}{2},\tfrac{1}{2})$. The colored entries mark
    spaces that are saturated, in the sense that no more functions should
    appear in $\mathcal{K}^{\pi}$ at the given de Rham weight,
    even when the overall weight is increased further.}
\label{tab:deRhamSpaceHpl}
\end{table}

%%%%%%%%%%%%%%%%%%%%%%%%%%%%%%%
\subsection{Lines with symbol alphabet \texorpdfstring{$\{y,1+y,y-\omega,y-\bar\omega\}$}{\{y,1+y,y-omega,y-omegabar\}}}

The final pair of lines in Table \ref{tab:lines} have the
four-letter symbol alphabet $\{y,1+y,y-\omega,y-\bar{\omega}\}$. Here $\omega
= \exp(2\pi i/3)$ is a sixth root of unity arising as a zero of the cyclotomic
polynomial $1+y+y^2$.  (It is also a cube root of unity, of course,
but since $1+y$ also appears as a letter,
it is better to consider it a sixth root, along with $-1$.)
The functions built from this alphabet are cyclotomic polylogarithms that
can be expressed as
$G$ functions with indices drawn from the set $\{0,-1,\omega,\bar{\omega}\}$
with argument $y$. 
The first entry condition allows only branch cuts starting at
$u=0$, which means that there is only a single weight one function in the case
of the first line ($\ln u$), and two functions in the case of the second line
($\ln u$ and $\ln w = \ln(1-u)$).
The generating functions for cyclotomic polylogarithms with these first
entry conditions are,
\bea
d^{{\textrm{C}_1}}(t) &=& \frac{1-3t}{1-4t}= 1+  t+4 t^2+16 t^3+ 64 t^4+\dots,\\
d^{{\textrm{C}_2}}(t) &=& \frac{1-2t}{1-4t}= 1+2 t+8 t^2+32 t^3+128 t^4+\dots.
\eea
These formulas are significant
overcounts, though, because $(y-\omega)$ and $(y-\bar\omega)$
do not appear independently in the derivatives of functions in $\Hhex$;
only the product $(y-\omega)(y-\bar\omega) = 1+y+y^2$ appears.

At the base point of integration for the construction of these lines, $(0,0,0)$,
respectively $(1,1,0)$, the hexagon functions degenerate to MZVs. The possible
appearance of these boundary values needs to be taken into account when counting
the maximal number of independent functions that can appear on these lines.
The generating function for the MZVs is given in \eqn{eq:MZVindep}. If we assume
that all MZVs can appear independently, we can obtain a generating function for
the maximum number of functions that can appear on the second
four-letter line as the product,
\be
d^{\textrm{C}_2}(t)d^{\textrm{MZV}}(t) = 1+2 t+9 t^2+35 t^3+139 t^4+556 t^5+2222 t^6+\dots,
\ee
as also shown in the first row of Table~\ref{tab:cycLines}.

%%%%%%%%%%%%%%%%%%%%%%%%%%%%%%%%%%%%%%%%%%%%%%%%%%%%%%%%%
\renewcommand{\arraystretch}{1.25}
\begin{table}[!t]
\centering
\begin{tabular}[t]{l c c c c c c c c c c}
\hline\hline
weight         & 1 & 2 & 3  & 4   & 5   & 6    & 7    & 8 \\\hline\hline
maximal dim. ($2^{\rm nd}$ line)
               & 2 & 9 & 35 & 139 & 556 & 2222 & 8887 & 35546 \\\hline
    $\Bigl(\frac{y}{(1+y)^2},\frac{y}{(1+y)^2},\frac{y}{(1+y)^2}\Bigr)\,$ dim. 
               & 1 & 2 & 4 &  7 & 13 & 25 &  43 &  77\\\hline
 $\Bigl(\frac{1+y+y^2}{(1+y)^2},\frac{1+y+y^2}{(1+y)^2},\frac{y}{(1+y)^2}\Bigr)\,$
    dim.       & 2 & 4 & 8 & 16 & 31 & 59 & 110 &  ?  \\\hline\hline
\end{tabular}
\caption{Dimensions of $\Hhex$ when restricted to two lines with
    symbol letters $y,1+y,y-\omega,1-\bar{\omega}$, for weights up to 8.
The maximal dimension corresponds to allowing all MZVs and all cyclotomic
HPLs with no branch cuts on $(u,u,1-u)$ other than at $u=0,1$.}
\label{tab:cycLines}
\end{table}
%%%%%%%%%%%%%%%%%%%%%%%%%%%%%%%%%%%%%%%%%%%%%%%

In addition to the theoretical maximal dimension, we also show the actual
dimensions of the lines in the hexagon space in Table~\ref{tab:cycLines}.
Once again the dimension of the space of hexagon functions grows considerably
more slowly than the theoretical maximum. As in the case of the lines discussed
previously, this is due to the structure of the full space of hexagon functions that
survives when restricting to the lines.
To illustrate we show a possible basis choice for the line $(\tfrac{y}{(1+y)^2},
\tfrac{y}{(1+y)^2},\tfrac{y}{(1+y)^2})$ at low weight in terms of $G$ functions
with implicit argument $y$:
\bea
&&\{G_0-2G_{-1}\} \,, \nonumber\\
&&\{-2 G_{\bar{\omega },-1}+G_{\bar{\omega },0}-2 G_{\omega ,-1}+G_{\omega ,0}+2
G_{0,-1}-G_{0,0}-\zeta _2,\nonumber\\
&&\phantom{\{}~~4 G_{-1,-1}-2 G_{-1,0}-2 G_{0,-1}+G_{0,0}+2 \zeta_2\} \,.
\eea
We can observe that at weight two, $\zeta_2$ does not appear as an independent
function, but rather only in specific combinations.

%%%%%%%%%%%%%%%%%%%%%%%%%%%%

\subsection{Alternating sum points}

Finally we specialize from lines to points.
Figure~\ref{fig:cubespecial} shows a host of points where
the hexagon functions reduce to numbers associated
with cyclotomic polylogarithms~\cite{Ablinger:2011te}
with unit argument and indices that are various roots of unity.
These points can be classified by how many cross ratios
are vanishing, leading to logarithmic singularities,
as well as by which roots of unity are involved.

In this subsection we consider points where the hexagon functions
reduce to alternating sums.  There are at least two different ways to
generate alternating sums from the lines displayed in \Tab{tab:lines}.
One way is to set $u=1/2$ or $u=2$ on
one of the lines with symbol alphabet $\{u,1-u\}$.  The other is
to set $y=1$ on a line with symbol alphabet $\{y,1-y,1+y\}$.
Four examples of the first type are the points $(\tfrac{1}{2},1,1)$,
$(2,1,1)$, $(\tfrac{1}{2},\tfrac{1}{2},1)$, and $(2,2,1)$.
These four points are all nonsingular, as no cross ratio vanishes.
Through weight 10, the spaces of alternating
sum values at these points exhibit no missing values whatsoever;
the dimension is generated precisely by the Fibonacci
sequence, i.e.~by $d^{\rm alt}(t)$.

There is also a singular point, $(\tfrac{1}{2},0,1)$, which has
very similar behavior:  ignoring coefficients
of the $\ln v$ singular factors,
the finite parts again exhibit no missing values through weight 10.

At the doubly singular point $(\tfrac{1}{2},0,0)$, the situation
looks identical at first, through weight 8. (Again we focus on the
finite parts and ignore the coefficients of positive powers of
$\ln v$ and $\ln w$.) However, at weight 9 the first missing value occurs.
Instead of having the six independent values,
\be
f^2_{3,1,3,1,1},\ f^2_{3,1,1,3,1},\
f^2_{1,3,3,1,1},\ f^2_{1,3,1,1,3},\
f^2_{1,1,3,3,1},\ f^2_{1,1,3,1,3},
\label{eq:h00wt9before}
\ee
only five of the six appear, in the following linear combinations:
\be
f^2_{3,1,3,1,1} + f^2_{3,1,1,3,1},\ 
f^2_{1,3,3,1,1} + f^2_{1,3,1,1,3},\ 
f^2_{1,1,3,3,1} + f^2_{1,1,3,1,3},\ 
f^2_{3,1,3,1,1} + f^2_{1,3,3,1,1},\ 
f^2_{1,3,1,1,3} + f^2_{1,1,3,1,3}.
\label{eq:h00wt9after}
\ee
Table~\ref{tab:Altpoints} displays the dimension that
$\Hhex$ reduces to at $(\tfrac{1}{2},0,0)$, as well as
the number of values that are absent on this line, beyond those
predicted by the coaction principle.
At weight 10 there are two new missing values, which
like \eqn{eq:h00wt9after} involve taking linear combinations of
words with two $f^2_3$ letters, and the remaining (four)
letters are $f^2_1$.

%%%%%%%%%%%%%%%%%%%%%%%%%%%%%%%%%%%%%%%%%%%%%%%%%%%%%%%%%
\renewcommand{\arraystretch}{1.25}
\begin{table}[!t]
\centering
\begin{tabular}[t]{l c c c c c c c c c c c c}
\hline\hline
weight & 1 & 2 & 3 & 4 & 5 & 6 & 7 & 8 & 9 & 10 \\\hline\hline
maximal dim.   & 1 & 2 & 3 & 5 & 8 & 13 & 21 & 34 & 55 & 89 \\\hline
$(\tfrac{1}{2},0,0)$ dim.
               & 1 & 2 & 3 & 5 & 8 & 13 & 21 & 34 & 54 & 86 \\
%$(\tfrac{1}{2},0,0)$ drop
new missing    & 0 & 0 & 0 & 0 & 0 &  0 &  0 &  0 &  1 &  2 \\\hline
$(\tfrac{1}{2},0,\tfrac{1}{2})$ dim.
               & 1 & 2 & 3 & 5 & 8 & 12 & 19 & 29 & 44 & 67 \\
%$(\tfrac{1}{2},0,\tfrac{1}{2})$ drop
new missing    & 0 & 0 & 0 & 0 & 0 &  1 &  1 &  3 &  5 &  9 \\\hline
$(\infty,0,\infty)$ dim.
               & 0 & 1 & 2 & 2 & 4 &  7 & 11 & 18 & 29 & 47 \\
%$(\infty,0,\infty)$ drop
new missing    & 1 & 0 & 0 & 1 & 0 &  0 &  0 &  0 &  0 &  0 \\\hline\hline
\end{tabular}
\caption{Dimensions of $\Hhex$ when restricted to various alternating-sum
points, for weights up to 10.
The maximal dimension corresponds to all alternating sums and is given by
the Fibonacci sequence. It is attained through weight 10
by the points $(\tfrac{1}{2},1,1)$, $(2,1,1)$, $(\tfrac{1}{2},\tfrac{1}{2},1)$,
$(2,2,1)$ and $(\tfrac{1}{2},0,1)$.
The `new missing' lines refer to the number of values that are absent at
a given weight that are {\it not} predicted to be absent
by the coaction principle.}
\label{tab:Altpoints}
\end{table}
%%%%%%%%%%%%%%%%%%%%%%%%%%%%%%%%%%%%%%%%%%%%%%%

Next we turn to two alternating-sum points on the line
$(u,0,u)=(\tfrac{y}{1+y},0,\tfrac{y}{1+y})$, again focusing
on the finite values, ignoring any values multiplied by $\ln v$ factors.
The first point has $u=\tfrac{1}{2}$ ($y=1$).
As shown in Table~\ref{tab:Altpoints}, the first missing value at
$(\tfrac{1}{2},0,\tfrac{1}{2})$ is at weight 6.
It corresponds to replacing $f^2_{1,3,1,1}$ and $f^2_{1,1,3,1}$
with the single linear combination
\be
f^2_{1,3,1,1} + f^2_{1,1,3,1} \,.
\label{eq:h0hwt6after}
\ee
At weight 7, $f^2_{1,3,1,1,1}$, $f^2_{1,1,3,1,1}$ and $f^2_{1,1,1,3,1}$
are similarly replaced by their sum,
\be
f^2_{1,3,1,1,1} + f^2_{1,1,3,1,1} + f^2_{1,1,1,3,1} \,.
\label{eq:h0hwt7after}
\ee
One of the two removed combinations ($f^2_{1,1,3,1,1} + f^2_{1,1,1,3,1}$)
is predicted by the coaction principle, given \eqn{eq:h0hwt6after}, while
the other is new.  At weight 8, the three new dropouts are associated
with
\bea
&f^2_{1,3,1,1,1,1} + f^2_{1,1,3,1,1,1} + f^2_{1,1,1,3,1,1} + f^2_{1,1,1,1,3,1} \,,
\nonumber\\
&f^2_{1,5,1,1} + f^2_{1,1,5,1} \,, \nonumber\\
&\zeta_2 \, (f^2_{1,3,1,1} + f^2_{1,1,3,1}),
\label{eq:h0hwt8after}
\eea
and so on.  The missing values at the point $(\tfrac{1}{2},0,\tfrac{1}{2})$
have a very characteristic pattern, but its significance is not clear to us.

The final alternating-sum point we have examined is from setting
$y=-1$ ($u\to\infty$), which we denote by $(\infty,0,\infty)$.
We also ignore singular factors of $\ln u$ (or $\ln(1+y)$) in this limit.
Here the first dropout is at weight one: $f^2_1 = -\ln 2$ is missing.
Through the coaction principle, this one low-weight missing value
causes a huge reduction in the dimension of $\Hhex(\infty,0,\infty)$.
There is also a missing value at weight 4, in that $f^2_{1,3}$
and $\zeta_2 \, f^2_{1,1}$ get replaced by the linear combination
\be
7 f^2_{1,3} - 9 \zeta_2 f^2_{1,1} \,.
\label{eq:i0iwt4after}
\ee
Remarkably, that is the last new missing value at this point through
weight 10.  The contrast between the behavior at this point
and the previous ones in Table~\ref{tab:Altpoints} is striking,
and we have no explanation for it.

%%%%%%%%%%%%%%%%%%%%%%%%%%%%%%%%%%%%%%%

\subsection{\texorpdfstring{4$^{\rm th}$}{4th} root of unity point}

Next we examine the point $(\tfrac{1}{2},\tfrac{1}{2},\tfrac{1}{2})$ at
the center of the cube in Figure~\ref{fig:cubespecial}.
As indicated in Table~\ref{tab:lines},
this point can be reached by setting $y=i$ on the line
$(u,\tfrac{1}{2},\tfrac{1}{2})$ for $u=(1+y)^2/(4y)$.
However, a better parametrization for the line $(u,\tfrac{1}{2},\tfrac{1}{2})$
for $u<1$ is to let $u=1/(r^2+1)=1/[(r+i)(r-i)]$ ($y=(r+i)/(r-i)$).
The alphabet is $\{r,r+i,r-i\}$.  As $r$ goes
from 0 to 1, $u$ goes from 1 (an alternating-sum point)
down to $\tfrac{1}{2}$. This parametrization puts the complex values
into the indices rather than the argument of the $G$ functions.

In contrast to most of the other points we have considered, this
point is {\it not} on the parity-odd vanishing surface $\Delta(u,v,w)=0$.
The parity odd functions are pure imaginary at this point, while
the parity even functions are real.

%%%%%%%%%%%%%%%%%%%%%%%%%%%%%%%%%%%%%%%%%%%%%%%%%%%%%%%%%
\renewcommand{\arraystretch}{1.25}
\begin{table}[!t]
\centering
\begin{tabular}[t]{l c c c c c c c c c c c c}
\hline\hline
weight & 1 & 2 & 3 & 4 &  5 &  6 &  7 &  8 &  9 &  10 \\\hline\hline
``maximal'' dim.
       & 1 & 3 & 5 & 11 & 21 & 43 & 85 & 171 & 341 & 683 \\\hline
$(\tfrac{1}{2},\tfrac{1}{2},\tfrac{1}{2})$ dim.
       & 1 & 2 & 4 &  5 & 11 & 17 & 32 &  53 &  99 & 167 \\
new ``missing''
       & 0 & 1 & 0 &  2 &  2 &  8 &  9 &  21 &  27 &  59 \\
\hline\hline
\end{tabular}
\caption{Dimensions of $\Hhex$ when restricted to the 4$^{\rm th}$ root
of unity point $(\tfrac{1}{2},\tfrac{1}{2},\tfrac{1}{2})$,
for weights up to 10. The ``maximal'' dimension is defined in the text.}
\label{tab:hhhpoint}
\end{table}
%%%%%%%%%%%%%%%%%%%%%%%%%%%%%%%%%%%%%%%%%%%%%%%

The dimensions of $\Hhex$ at this point are shown in
Table~\ref{tab:hhhpoint}.  The $f$-alphabet for 4$^{\rm th}$ roots
of unity has a separate letter at each weight, $f^4_1$, $f^4_2$, $f^4_3$, $f^4_4$,
etc.   The generating function for these words is
$1/(1-t-t^2-t^3-t^4-\cdots) = (1-t)/(1-2t)$.
There are also both odd and even powers of $i\pi$.
The types of constant values coming from the parity
even and parity odd sectors are quite different.
If we define the words of even weight, ($f^4_2$, $f^4_4$, etc.) and $i\pi$,
to have odd parity, and the words of odd weight
($f^4_1$, $f^4_3$, $f^4_5$, etc.) to have even parity,
then that parity always agrees with the parity of the function from which
the constant originated.

There is a subspace of the even parity values that involve only the
words of odd weight and Riemann zeta values $\zeta_{2k}$.  There are no
missing values in this subspace; all new missing values are associated with
the odd subspace.  We also find that the odd powers of $\pi$ are not
independent, but are coupled to other odd $f$-alphabet words the first
time they appear.  With this property in mind, we define a ``maximal'' dimension
which only counts powers of $\pi^2$ along with the $f$-alphabet. 
The generating function is then:
\be
\frac{1}{1-t^2}\,\frac{1-t}{1-2t} 
= 1 + t + 3 t^2 + 5 t^3 + 11 t^4 + 21 t^5 + 43 t^6 + \ldots,
\label{eq:4throotmax}
\ee
as shown in Table~\ref{tab:hhhpoint}.
With respect to this definition of ``maximal'', the first missing value
is at weight 2, where $f^4_2$ does not appear (because there are no
parity-odd weight 2 functions).  At weight 3, the function $\tilde \Phi_6$
evaluates to something proportional to
\be
f^4_{2,1} - \frac{i\pi^3}{48} \,.
\label{eq:wt3oddhhh}
\ee
At weight 4, the two parity odd functions both vanish on the entire line
$(u,u,u)$, and so they also vanish at the point
$(\tfrac{1}{2},\tfrac{1}{2},\tfrac{1}{2})$.
Associated with this, the potential odd values
$f^4_{1,2,1}-(i\pi^3/48)f^4_1$ and $\zeta_2 \, f^4_2$ are missing,
as shown in the table.  There are just two weight 5 odd values,
\be
3f^4_{4,1} + 8 f^4_{2,3} - \frac{79}{5376} i\pi^5 \,, \quad
f^4_{2,1,1,1} + 2\zeta_2 f^4_{2,1} - \frac{59}{2880} i\pi^5 \,,
\label{eq:wt5oddhhh}
\ee
while two are missing.  As the table shows, there is an increasing number
of new missing values at higher weight, and the actual values in
$\Hhex(\tfrac{1}{2},\tfrac{1}{2},\tfrac{1}{2})$ are quite restricted.
The coaction principle is obeyed at this point as far as we
have been able to check it, through weight 10.

%%%%%%%%%%%%%%%%%%%%%%%%%%%%

\subsection{\texorpdfstring{6$^{\rm th}$}{6th} root of unity points}

Finally we examine two points where $\Hhex$ reduces to
6$^{\rm th}$ root of unity values,
$(\tfrac{1}{4},\tfrac{1}{4},\tfrac{1}{4})$
and $(\tfrac{3}{4},\tfrac{3}{4},\tfrac{1}{4})$.
Both points are located on the parity-odd vanishing surface $\Delta(u,v,w)=0$,
so we only have to evaluate the parity-even functions here.
There are two weight 1 letters in the 6$^{\rm th}$ root of unity $f$-alphabet,
$f^6_{\pm1}$
and one letter for each higher integer weight, $f^6_2$, $f^6_3$, $f^6_4$, etc.
However, we find that only the odd weight letters appear at these two
points, $f^6_{\pm1}$, $f^6_3$, $f^6_5$, etc.   The absence of the even
weight letters may be related to being on the $\Delta=0$ surface.
The generating function for the odd weight letters and the
Riemann zeta values $\zeta_{2k}$ is
\bea
\frac{1}{1-t^2} \, \frac{1}{1-2t-t^3-t^5-t^7-\cdots}
&=& \frac{1}{1-2t-t^2+t^3}  \label{eq:6throotgenfn}\\
&=& 1 + 2 t + 5 t^2 + 11 t^3 + 25 t^4 + \ldots.  \nonumber
\eea

%%%%%%%%%%%%%%%%%%%%%%%%%%%%%%%%%%%%%%%%%%%%%%%%%%%%%%%%%
\renewcommand{\arraystretch}{1.25}
\begin{table}[!t]
\centering
\begin{tabular}[t]{l c c c c c c c c c c c c}
\hline\hline
weight & 1 & 2 &  3 &  4 &  5 &   6 &   7 &   8 \\\hline\hline
maximal dim.
       & 2 & 5 & 11 & 25 & 56 & 126 & 283 & 636 \\\hline
$(\tfrac{1}{4},\tfrac{1}{4},\tfrac{1}{4})$ dim.
       & 1 & 2 &  3 &  7 & 11 &  22 &  36 &  66 \\
new missing
       & 1 & 1 &  2 &  1 &  6 &   4 &  18 &  21 \\\hline\hline
$(\tfrac{3}{4},\tfrac{3}{4},\tfrac{1}{4})$ dim.
       & 2 & 4 &  7 & 15 & 27 &  52 &  93 & 170 \\
new missing
       & 0 & 1 &  2 &  2 &  8 &  12 &  31 &  53 \\\hline\hline
\end{tabular}
\caption{Dimensions of $\Hhex$ when restricted to the 6$^{\rm th}$ root
of unity points $(\tfrac{1}{4},\tfrac{1}{4},\tfrac{1}{4})$ and
$(\tfrac{3}{4},\tfrac{3}{4},\tfrac{1}{4})$, for weights up to 8.}
\label{tab:6throotpoints}
\end{table}
%%%%%%%%%%%%%%%%%%%%%%%%%%%%%%%%%%%%%%%%%%%%%%%

Table~\ref{tab:6throotpoints} shows that there are many other missing
values for both of the 6$^{\rm th}$ root of unity points.
A basis for the first three weights of
$\Hhex(\tfrac{1}{4},\tfrac{1}{4},\tfrac{1}{4})$
is given by
\bea
&& \{ f^6_{-1} \} \,,  \label{eq:qqqwt1}\\
&& \Bigl\{ f^6_{-1,-1} + 2\zeta_2 \,, \quad f^6_{1,-1} + \frac{2}{3} \zeta_2 \Bigr\}
\,, \label{eq:qqqwt2}\\
&& \Bigl\{ 3 f^6_{3} - f^6_{1,1,-1} - \frac{2}{3} \zeta_2 f^6_{1} \,, \quad
5 f^6_{3} - 8 f^6_{-1,1,-1} - \frac{16}{3} \zeta_2 f^6_{-1} \,, \quad
f^6_{-1,-1,-1} + 2 \zeta_2 f^6_{-1} \Bigr\} \,.
\label{eq:qqqwt3}
\eea
The corresponding basis at $(\tfrac{3}{4},\tfrac{3}{4},\tfrac{1}{4})$
is given by
\bea
&& \{ f^6_{1} \,, \quad f^6_{-1} \} \,, \label{eq:ttqwt1}\\
&& \Bigl\{ f^6_{1,1} + \zeta_2 \,, \quad
f^6_{1,-1} + \frac{2}{3} \zeta_2 \,, \quad
f^6_{-1,1} + \frac{4}{3} \zeta_2 \,, \quad
f^6_{-1,-1} + 2 \zeta_2 \Bigr\} \,, \label{eq:ttqwt2}\\
&& \Bigl\{ f^6_{1,1,1} + \zeta_2 f^6_{1} \,, \quad
f^6_{-1,-1,-1} + 2 \zeta_2 f^6_{-1} \,, \quad
\frac{3}{4} f^6_{3} + f^6_{1,-1,1} + f^6_{-1,1,1} + \zeta_2 f^6_{-1}
   + \frac{4}{3} \zeta_2 f^6_{1} \,, \nonumber\\
&&\hskip0.1cm
5 f^6_{1,-1,-1} + 10 \zeta_2 f^6_{1} - 31 f^6_{-1,1,-1} - 14 \zeta_2 f^6_{-1}
   + 5 f^6_{-1,-1,1} \,, \quad
5 f^6_{-1,1,1} - 23 \zeta_2 f^6_{-1} - 42 f^6_{-1,1,-1} \,, \nonumber\\
&&\hskip0.1cm
- 5 f^6_{3} + 8 f^6_{-1,1,-1} + \frac{16}{3} \zeta_2 f^6_{-1} \,, \quad
- 6 f^6_{3} + 2 f^6_{1,1,-1} + \frac{4}{3} \zeta_2 f^6_{1} \Bigr\} \,.
\label{eq:ttqwt3}
\eea
The coaction principle is obeyed at these two points as far as we
have been able to check it, through weight 8.

%%%%%%%%%%%%%%%%%%%%%%%%%%%%%%%%%%%%%%%%%%%%%%%%%%%%%
\section{Conclusions}
\label{sec:conclusions}

In this work we have presented a minimal space of functions relevant to six-particle scattering in planar ${\cal N}=4$ super-Yang-Mills theory, at least through six loops in the NMHV sector and seven loops in the MHV sector. This space of functions obeys two novel constraints, the extended Steinmann relations and a cosmic Galois coaction principle---in particular, employing the derivations $\partial_{2k+1}$ in \eqn{eq:oddzetaderivation} acting at the point $(1,1,1)$---which together severely restrict the number of functions that can appear. We have also described how to construct this space of functions order by order in transcendental weight, and have carried out this procedure through weight eleven, with partial results for weight twelve.

The extended Steinmann relations, described in section \ref{sec:stein}, generalize the Steinmann relations to a property that holds on all Riemann sheets. Namely, they correspond to applying the Steinmann relations after carrying out any sequence of analytic continuations, thereby constraining not just the first two discontinuities of the amplitude, but any consecutive pair of discontinuities. The resulting space also exhibits constraints on longer sequences of discontinuities, as described in appendix~\ref{appendix:longrange}. The extended Steinmann relations exhibit a striking resemblance to the recently-discovered phenomenon of cluster adjacency~\cite{Drummond:2017ssj}. While these constraints are equivalent at six points (and the latter implies the former at all $n$~\cite{Golden:2019kks}), the relation between these constraints is still not fully understood. Moreover, while something resembling the extended Steinmann relations ought to hold for a wider class of quantum field theories, we have left this investigation for future work.

We have also described the presence of a coaction principle~\cite{Brown:2015fyf,Panzer:2016snt,Schnetz:2017bko} that is obeyed by the space of functions entering the six-particle amplitude. This property requires the introduction of a new normalization constant $\rho$, which suggests that the coaction principle selects a preferred scheme for subtracting infrared divergences. It would be interesting to identify this scheme in terms of known (or new) physical quantities, and investigate its interplay with the observed positivity of the amplitude~\cite{Arkani-Hamed:2014dca,Dixon:2016apl}. There also remains the question of whether a truly ``bottom-up'' definition of the space of constants present in these amplitudes exists. In particular, it would be interesting to find an explanation for why we have only found it necessary to include even powers of $\pi$ as independent constant functions in $\Hhex$.

We know that $\Hhex$ cannot be any smaller through weight 7, nor for the 
parity-odd part at weight 8, because the coproducts of the amplitudes
we have computed span these parts of $\Hhex$.  However, starting with
the parity even functions at weight 8, there is still the possibility
that a more minimal space should be defined.  Indeed, we have fairly strong
evidence that this possibility will be realized, based on the behavior
of the functions at the point $(u,v,w)=(1,0,0)$ and its cyclic images
$(0,1,0)$ and $(0,0,1)$.  These three points represent combined soft
and collinear limits of the amplitude, which are predicted to all loop
orders by the flux tube or pentagon operator product
expansion~\cite{Basso:2013vsa,Basso:2013aha,Basso:2014hfa}.
This expansion never contains any MZVs with depth greater than one;
only depth-one Riemann zeta values $\zeta_n$ arise.  Since the operator
product expansion can be expressed as a series expansion in all three variables
around $(1,0,0)$, this same conclusion applies to arbitrary derivatives
of the amplitudes evaluated at $(1,0,0)$, i.e.~to arbitrary coproducts:
only Riemann zeta values should ever appear. On the other hand,
the first irreducible depth 2 MZV, $\zeta_{5,3}$, appears in the values
of many of the 313 weight 8 functions at $(1,0,0)$ --- but it
does not appear in the limits of the 279 functions that are actually
coproducts of presently known amplitudes!
We conclude that at least three linear combinations of the 313
functions will have to be removed from $\Hhex$,
one each to kill the $\zeta_{5,3}$ in the
$(1,0,0)$, $(0,1,0)$ and $(0,0,1)$ limits.  Also, because the 279 amplitude
coproducts span all 123 of the non-$K$ functions, the functions
to be removed should be the simpler $K$ functions.

We have looked at a variety of other MZV points to see whether $\zeta_{5,3}$
disappears from the amplitude coproducts.  The only other point we
have found with this property is the origin, $(u,v,w)=(0,0,0)$.
This point is far from the OPE limit, so it is not as clear that
depth 2 MZVs cannot appear here.  Some of the 313 functions in
the basis do have $\zeta_{5,3}$ in their limits at the origin,
though none of the 279 amplitude coproducts do.
However, after we eliminate $\zeta_{5,3}$ from the $(1,0,0)$, $(0,1,0)$
and $(0,0,1)$ limits of the weight 8 functions, by removing the three linear
combinations mentioned above, we find that the remaining 310 functions
at the origin are free of $\zeta_{5,3}$.
A similar phenomenon occurs at weight 9, where
$\zeta_{5,3}$ can be seen accompanying $\ln u_i$ in the limits,
and for the non-Riemann zeta values $\zeta_{7,3}$ and $\zeta_2 \zeta_{5,3}$
appearing in the same limits at weight 10.

It is clear there is still more to learn about the bottom-up construction
of the space $\Hhex$ (defined in the introduction as the minimal space
containing all amplitude coproducts).
What are the proper constraints to impose, beyond the coaction principle
exploited in this paper?  Will we need to compute new seven-
and eight- loop amplitudes to determine precisely which functions
should drop out, at weight 8 and beyond?
Is it obvious that the smaller space will still satisfy a coaction principle?

While we have primarily investigated the coaction principle at kinematic points and on codimension-two surfaces, it is expected to hold in general kinematics. It would be interesting to find out whether higher-point amplitudes also obey a coaction principle for the same choice of $\rho$. For a sufficiently large number of particles, these amplitudes will no longer be polylogarithmic~\cite{CaronHuot:2012ab,ArkaniHamed:2012nw,Nandan:2013ip,Bourjaily:2015jna,Bourjaily:2017wjl,Bourjaily:2017bsb,Bourjaily:2018ycu,Bourjaily:2018yfy}; however, this presents no \emph{a priori} obstacle to the existence of a coaction principle, as a coaction can also be constructed on the more complicated periods that are expected to arise~\cite{2015arXiv151206410B}. This has already been done explicitly for the case of elliptic polylogarithms~\cite{Broedel:2017kkb,Broedel:2018iwv}. While non-supersymmetric amplitudes generically involve more complicated rational prefactors and will not enjoy uniform transcendental weight, there is also no \emph{a priori} obstacle to finding coaction principles in more general quantum field theories, as has already been done in string theory, $\phi^4$ theory, and QED~\cite{Schlotterer:2012ny,Brown:2015fyf,Panzer:2016snt,Schnetz:2017bko}.

%%%%%%%%%%%%%%%%%%%%%%%%%%%%%%%%%%%%%%%%5

\vskip0.5cm
\noindent {\large\bf Acknowledgments}
\vskip0.3cm

\noindent
We thank Francis Brown, Claude Duhr, Erik Panzer and Oliver Schnetz for many
stimulating conversations, and Claude Duhr for sharing his notes on ref.~\cite{2015arXiv151206410B}.
We are grateful to
James Drummond and {\"O}mer G{\"u}rdo{\u{g}}an for sharing their
results on spaces obeying cluster adjacency.
We thank Francis Brown for comments on the manuscript.
This research was supported in part by the National Science Foundation
under Grant No.\ NSF PHY17-48958, by the US Department of Energy under
contract DE--AC02--76SF00515, the Munich Institute for Astro- and Particle Physics (MIAPP) of the DFG cluster of excellence ``Origin and Structure of the Universe'',
the Danish National Research Foundation (DNRF91), a grant from the Villum Fonden, a Starting Grant \mbox{(No.\ 757978)} from the European Research Council, a grant from the Simons Foundation (341344, LA), the European Union's Horizon 2020 research and innovation program under grant agreement \mbox{No.\ 793151}, a Carlsberg Postdoctoral Fellowship (CF18-0641), and a Humboldt Research Award (LD).
SCH’s work was supported in part by the National Science and Engineering
Council of Canada, the Canada Research Chair program, and the
Fonds de Recherche du Qu\'ebec -- Nature et Technologies.
LD thanks the Perimeter Institute, LPTENS, the Institut de Physique Th\'eorique 
Philippe Meyer, the Higgs Centre at U.~Edinburgh, the Simons Foundation,
the Hausdorff Institute for Mathematics, Humboldt University Berlin,
and the University of Freiburg for hospitality.
LD, MvH and AM thank the Pauli Center of ETH Z\"urich and the University
of Z\"urich for hospitality.
LD, MvH, AM, and GP thank the Kavli Institute for Theoretical Physics
for hospitality. We are all grateful to the Galileo Galilei Institute
for hospitality.

%%%%%%%%%%%%%%%%%%%%%%%%%%%%%%%%%%%%%%%%%%%%%%%%%%%%%%%%%%%%%
\appendix

\section{Values of the Amplitudes at \texorpdfstring{$(1,1,1)$}{(1,1,1)} in the \texorpdfstring{$f$-basis}{f-basis}} \label{appendix:fbasis}

The values of the MHV amplitudes $\EE^{(L)}(1,1,1)$ for $L=1$ to 7
in the $f$-basis are:
\bea
\EE^{(1)}(1,1,1) &=& 0 \,,
\label{EXMHVfg1_111}\\
\EE^{(2)}(1,1,1) &=& - 10 \, \zeta_4 \,,
\label{EXMHVfg2_111}\\
\EE^{(3)}(1,1,1) &=& \frac{413}{3} \, \zeta_6 \,,
\label{EXMHVfg3_111}\\
\EE^{(4)}(1,1,1) &=&  - \frac{5477}{3} \, \zeta_8
+ 24 \, \Bigl[ 5 f_{3,5} - 2 \zeta_2 f_{3,3} \Bigr] \,,
\label{EXMHVfg4_111}\\
\EE^{(5)}(1,1,1) &=& \frac{379957}{15} \, \zeta_{10}
- 384 \, \Bigl[ 7 f_{3,7} - \zeta_2 f_{3,5} - 3 \zeta_4 f_{3,3} \Bigr]
- 312 \, \Bigl[ 5  f_{5,5} - 2  \zeta_2  f_{5,3} \Bigr] \,,
\label{EXMHVfg5_111}\\
\EE^{(6)}(1,1,1) &=& - \frac{2273108143}{6219} \zeta_{12}
+ 2264 \, \Bigl[ 7f_{3,9}-6\zeta_4 f_{3,5} \Bigr]
+ 6536 \,\Bigl[ 5 f_{3,9}-3 \zeta_6 f_{3,3} \Bigr]\nonumber\\
&&\null\hskip0.0cm
- 3072 \, \Bigl[ \zeta_2 f_{3,7} - \zeta_6 f_{3,3} \Bigr]
+ 5328 \, \Bigl[ 7 f_{5,7} - \zeta_2 f_{5,5} - 3 \zeta_4 f_{5,3} \Bigr]
\nonumber\\
&&\null\hskip0.0cm
+ 4224 \, \Bigl[ 5 f_{7,5} - 2 \zeta_2f_{7,3} \Bigr]
\,, \label{EXMHVfg6_111}
\eea
\bea
\EE^{(7)}(1,1,1) &=& \frac{2519177639}{1260} \zeta_{14}
- 63968 \Bigl[ 5 f_{9,5} - 2 \zeta_2 f_{9,3} \Bigr]
- 77952 \Bigl[ 7 f_{7,7} - \zeta_2 f_{7,5} - 3 \zeta_4 f_{7,3} \Bigr]
\nonumber\\
&&\null\hskip0.0cm
- 34976 \Bigl[ 7 f_{5,9} - 6 \zeta_4 f_{5,5} \Bigr]
- 95552 \Bigl[ 5 f_{5,9} - 3 \zeta_6 f_{5,3} \Bigr]
+ 44640 \Bigl[ \zeta_2 f_{5,7} - \zeta_6 f_{5,3} \Bigr]
\nonumber\\
&&\null\hskip0.0cm
- \frac{413920}{11} \Bigl[ 33 f_{3,11} - 20 \zeta_8 f_{3,3} \Bigr]
+ 28000 \Bigl[ \zeta_2 f_{3,9} - \zeta_8 f_{3,3} \Bigr]
\nonumber\\
&&\null\hskip0.0cm
+ 62720 \Bigl[ 3 \zeta_4 f_{3,7} - 2 \zeta_8 f_{3,3} \Bigr]
+ \frac{218696}{3} \Bigl[ 3 \zeta_6 f_{3,5} - 2 \zeta_8 f_{3,3} \Bigr]
\nonumber\\
&&\null\hskip0.0cm
- 4992 \Bigl[ 5 f_{3,3,3,5} - 2 \zeta_2 f_{3,3,3,3}
            + \frac{5611}{132} \zeta_8 f_{3,3} \Bigr]
\,.
\label{EXMHVfg7_111}
\eea

The values of the NMHV amplitudes $E^{(L)}(1,1,1)$ for $L=1$ to 6
in the $f$-basis are:
\bea
E^{(1)}(1,1,1) &=& - 2 \, \zeta_2 \,,
\label{EXfg1_111}\\
E^{(2)}(1,1,1) &=& 26 \, \zeta_4 \,,
\label{EXfg2_111}\\
E^{(3)}(1,1,1) &=& - \frac{940}{3} \, \zeta_6 \,,
\label{EXfg3_111}\\
E^{(4)}(1,1,1) &=&  \frac{36271}{9} \, \zeta_8
- 24 \, \Bigl[ 5 f_{3,5} - 2 \zeta_2 f_{3,3} \Bigr] \,,
\label{EXfg4_111}\\
E^{(5)}(1,1,1) &=& - \frac{1666501}{30} \, \zeta_{10}
+ 528  \, \Bigl[ 7 f_{3,7} - \zeta_2 f_{3,5} - 3 \zeta_4 f_{3,3} \Bigr]
+ 384  \,  \Bigl[ 5 f_{5,5} - 2 \zeta_2 f_{5,3} \Bigr]\,,~~~~
\label{EXfg5_111}\\
E^{(6)}(1,1,1) &=& \frac{5066300219}{6219} \zeta_{12}
- 4664 \, \Bigl[ 7 f_{3,9}-6\zeta_4 f_{3,5} \Bigr]
- 11384 \,\Bigl[ 5 f_{3,9}-3 \zeta_6 f_{3,3} \Bigr]\nonumber\\
&&\null\hskip0.0cm
+ 5664 \, \Bigl[ \zeta_2 f_{3,7} - \zeta_6 f_{3,3} \Bigr]
- 8928 \, \Bigl[ 7 f_{5,7} - \zeta_2 f_{5,5} - 3 \zeta_4 f_{5,3} \Bigr]
\nonumber\\
&&\null\hskip0.0cm
- 6528 \, \Bigl[ 5 f_{7,5} - 2 \zeta_2f_{7,3} \Bigr]
\,. \label{EXfg6_111}
\eea
Notice the abundance of integers among the rational-number coefficients.
The ones that are not integers are typically associated with even Riemann
zeta values. Those coefficients might take a simpler form if the even
zeta values were rewritten in terms of sums of products of other
even Riemann zeta values, but we refrain from doing this, since there
is no unique way to do so.

We also provide the conversion between the $f$-alphabet and MZVs through
weight 11:
\bea
f_{3,3} &=& \frac{1}{2} (\zeta_3)^2 \,, \\
f_{5,3} &=& -\frac{1}{5} \zeta_{5,3} \,, \\
f_{3,3,3} &=& \frac{1}{6} (\zeta_3)^3 \,, \\
f_{3,7} &=& \zeta_3 \zeta_7
       + \frac{1}{14} \Bigl[ 3 (\zeta_5)^2 + \zeta_{7,3} \Bigr] \,, \\
f_{7,3} &=& - \frac{1}{14} \Bigl[ 3 (\zeta_5)^2 + \zeta_{7,3} \Bigr] \,, \\
f_{5,5} &=& \frac{1}{2} (\zeta_5)^2 \,, \\
f_{3,3,5} &=& \frac{1}{2} \Bigl[ \zeta_6 + (\zeta_3)^2 \Bigr] \zeta_5
     + \frac{1}{5} \Bigl[ - \zeta_{5,3,3} + \zeta_3 \zeta_{5,3}
                          - 3 \zeta_4 \zeta_7 \Bigr]
     - 9 \zeta_2 \zeta_9 \,, \\
f_{3,5,3} &=& - \zeta_6 \zeta_5
     + \frac{1}{5} \Bigl[ 2 \zeta_{5,3,3} - \zeta_3 \zeta_{5,3}
                        + 6 \zeta_4 \zeta_7 \Bigr]
     + 18 \zeta_2 \zeta_9 \,, \\
f_{5,3,3} &=& \frac{1}{2} \zeta_6 \zeta_5 
     - \frac{1}{5} \Bigl[ \zeta_{5,3,3} + 3 \zeta_4 \zeta_7 \Bigr]
     - 9 \zeta_2 \zeta_9 \,.
\label{ftozeta}
\eea
The ancillary file {\tt ftoMZV.txt} gives the same results through weight 14.

%%%%%%%%%%%%%%%%%%%%%%%%%%%%%%%%%%%%%%

\section{Longer-Range Symbol Restrictions} \label{sec:long_range_symb_restrictions}
\label{appendix:longrange}

In section~\ref{sec:stein}, it was pointed out that only certain combinations of symbol letters appear between pairs of letters, such as $a$ and $b$, that are restricted from appearing in adjacent entries by the Steinmann relations. We here explore this phenomenon further, and show that all sequences of symbol letters that appear between restricted letters (namely, those disallowed by equation~\eqref{eq:adjacent_letters_constraint}) vanish in the kinematic limit where the discontinuities in $a \sim s_{234}$ and $b \sim s_{345}$ are simultaneously accessible. Conversely, between all other pairs of letters, there exist sequences of symbol letters that do not vanish in this limit.

The discontinuities corresponding to the symbol letters $a \sim s_{234}$ and $b \sim s_{345}$ are accessible in the region where both of these Mandelstam invariants vanish. We can take this limit while keeping all other Mandelstam invariants generic by sending
\be \label{abVanishingLimit}
y_u \rightarrow \frac{1}{y_w} + \delta_u, \quad y_v \rightarrow \frac{1}{y_w} + \delta_v,
\ee 
where both $\delta_u$ and $\delta_v$ are infinitesimal. This implies
\begin{gather} 
a \rightarrow \frac{y_w^3}{(1-y_w)^2} (\delta_v)^2, \quad b \rightarrow \frac{y_w^3}{(1-y_w)^2} (\delta_u)^2, \quad c \rightarrow \frac{(1+y_w)^2}{y_w}, \nonumber \\ 
m_u \rightarrow -1, \qquad m_v \rightarrow -1, \qquad m_w \rightarrow -1, \label{eq:ab_vanishing_all_letters} \\  
y_u y_w \rightarrow 1, \qquad y_v y_w \rightarrow 1, \qquad  y_w \rightarrow y_w \, , \nonumber    
\end{gather}
where the $y_u$ and $y_v$ letters have been put into combinations that are independent of $y_w$ in this limit.
It is again easy to see how the two letters $m_w$ and $y_u y_v y_w$ mentioned in section~\ref{sec:stein} behave differently in this limit---any symbol involving $m_w$ will vanish, while those involving $y_u y_v y_w$ in general will not, since $y_u y_v y_w\rightarrow 1/y_w$. 

Next we investigate the weight-four case, in which two symbol entries appear between the letters $a$ and $b$, by constructing the full space of weight-four symbols without the first entry condition imposed. More specifically, we construct the space of symbols involving only the 40 weight-two combinations given in equations~\eqref{oddintegpairs} and~\eqref{evenintegpairs} in all pairs of adjacent entries, but allow any of the nine hexagon symbol letters to appear in the first (and last) entries. We then identify all terms in this space that have first entry $a$ and last entry $b$, after expressing the middle entries in terms of the symbol alphabet in~\eqref{eq:ab_vanishing_all_letters}. Note that these terms will not in general be integrable by themselves, which is why we construct the space of symbols with general first and last entries despite being interested in terms with specific such entries. In this way, we find fifteen pairs of letters:
\begin{gather}
a \otimes m_w, \quad m_v \otimes m_u, \quad m_v \otimes m_w, \quad m_v \otimes y_u y_w, \quad m_w \otimes b, \nonumber \\
m_w \otimes m_u, \quad m_w \otimes m_w, \quad m_w \otimes y_u y_w, \quad m_w \otimes y_v y_w, \quad m_w \otimes y_w, \\
y_u y_w \otimes m_w, \quad y_v y_w \otimes m_u, \quad y_v y_w \otimes m_w, \quad y_v y_w \otimes y_u y_w, \quad y_w \otimes m_w. \nonumber
\end{gather}
By reference to equation~\eqref{eq:ab_vanishing_all_letters}, it is easy to see that every one of these terms will vanish in the limit~\eqref{abVanishingLimit}.  
\begin{table}
\hspace{-.4cm} \begin{tabular}{ l || c | c | c | c | c || c | c | c | c }  
$w$ & \!$\dots \otimes a$\! & \!$\dots \otimes m_v$\! & \!$\dots \otimes m_w$\! & \!$\dots \otimes y_u$\! & \!$\dots \otimes y_v y_w$\! & \!$\dots \otimes b$\! & \!$\dots \otimes c$\! & \!$\dots \otimes m_u$\! & \!$\dots \otimes y_v/y_w$\! \\  \hline \hline
  2 & 1  & 1 & 1  & 1 & 1 & 0 & 0 & 0 & 0  \\  \hline
  3 & 6  & 6 & 6 & 6 & 6 & 1 & 1 & 3 & 3 \\  \hline
  4 & 36  & 40 & 40 & 41 & 41 & 15 & 19 & 29 & 29 \\  \hline
  5 & 227 & 285 & 283 & 302 & 302 & 142 & 172 & 242 & 242 \\  
\end{tabular} 
\caption{The number of distinct terms that have first entry $a$ and a given last entry in the space of weight-$w$ symbols constructed out of the 40 adjacent entry pairs given in eqs.~\eqref{oddintegpairs} and~\eqref{evenintegpairs}. This number depends on the symbol alphabet used; we express these symbols in terms of the alphabet $\{ a, b, c, m_u, m_v, m_w, y_u y_w, y_v y_w, y_w\}$ everywhere but in the last entry.}
\label{table:long_range_symbol_terms}
\end{table}

\begin{table}
\hspace{-.4cm} \begin{tabular}{ l || c | c | c | c | c || c | c | c | c }  
$w$ & \!$\dots \otimes a$\! & \!$\dots \otimes m_v$\! & \!$\dots \otimes m_w$\! & \!$\dots \otimes y_u$\! & \!$\dots \otimes y_v y_w$\! &  \!$\dots \otimes b$\! & \!$\dots \otimes c$\! & \!$\dots \otimes m_u$\! & \!$\dots \otimes y_v/y_w$\! \\  \hline \hline
  2 & 1 & 1 & 1 & 1 & 1 & 0 & 0 & 0 & 0 \\  \hline
  3 & 2 & 2 & 2 & 2 & 2 & 0 & 0 & 0 & 0  \\  \hline
  4 & 4 & 4 & 4 & 4 & 4 & 0 & 0 & 0 & 0 \\  \hline
  5 & 8 & 8 & 8 & 8 & 8 & 0 & 0 & 0 & 0 \\  
\end{tabular} 
\caption{The number of terms in Table~\ref{table:long_range_symbol_terms} whose middle $w-2$ entries do not vanish in the limit~\eqref{abVanishingLimit}.}
\label{table:long_range_symbol_terms_double_limit}
\end{table}
 
To see that this behavior is non-generic, let us consider terms that have the letter $a$ in their first entry, and arbitrary letters (not just $b$) in their last entry. We present the number of such terms for different final entries in Table~\ref{table:long_range_symbol_terms}, where we have separated final entries that are in $\mathcal{S}_{\text{a}}$ in equation~\eqref{eq:adjacent_letters_constraint} and those that are in its complement. Beyond weight 2 (where we know that only the letters in $\mathcal{S}_{\text{a}}$ appear next to $a$) there generically exist terms with first entry $a$ and every possible last entry. However, in Table~\ref{table:long_range_symbol_terms_double_limit} we also give the number of these terms that remain nonzero in the limit~\eqref{abVanishingLimit}. (We ignore whether or not the last entry vanishes in this limit, focusing only on the properties of the middle $w{-}2$ entries.) We see that all the symbol terms with first entry $a$ vanish in this limit if and only if the last entry is not in $\mathcal{S}_{\text{a}}$. 

We can provide evidence that this phenomenon will persist to all weights by constructing the full space generated by the letters that remain non-constant in~\eqref{eq:ab_vanishing_all_letters}.  At any weight, only three types of functions can be formed out of these letters, namely
\begin{gather} \label{eq:ab_zero_functions}
\ln \left( \frac{y_w^3}{(1-y_w)^2} (\delta_v)^2 \right), \qquad \ln \left( \frac{y_w^3}{(1-y_w)^2} (\delta_u)^2 \right), \qquad H_{\vec{w}} (y_w),
\end{gather}
where $\vec{w}$ can be any sequence of `0's and `${-}1$'s. However, only products of $\ln(y_w) = H_{0} (y_w)$ and $\smash{\ln \big( \frac{y_w^3}{(1-y_w)^2} (\delta_v)^2 \big)}$ ever actually appear between the letter $a$ and the final entries in $\mathcal{S}_{\text{a}}$ in the limit~\eqref{abVanishingLimit}. This follows from the fact that instances of $\ln \big( \frac{y_w^3}{(1-y_w)^2} (\delta_u)^2 \big)$ and $H_{\dots,-1,\dots} (y_w)$ can only arise from symbols involving the letters $b$ and $c$. Since, as seen in Table~\ref{table:long_range_symbol_terms_double_limit}, terms in which either letter appears in the last entry always vanish in the limit~\eqref{abVanishingLimit}, it follows that any term in which they appear in one of the middle entries must also vanish. 

This leaves at most $w-1$ functions that can appear between $a$ and the final entries in $\mathcal{S}_{\text{a}}$ after we take the limit~\eqref{abVanishingLimit}, namely the functions
\begin{equation}
\ln^{w-2-n} \left( \frac{y_w^3}{(1-y_w)^2} (\delta_v)^2 \right) \ln^{n} y_w
\end{equation}
for any $0 \le n \le w{-}2$. Since these functions are all products of logs, they give rise to $2^{w-2}$ distinct symbol terms (namely, any length-$(w{-}2)$ sequence of the letters $y_w$ and $\frac{y_w^3}{(1-y_w)^2} (\delta_v)^2$). We find that this number, $2^{w-2}$, is saturated by all the entries in the left half of Table~\ref{table:long_range_symbol_terms_double_limit}. More importantly, we conjecture that the last four columns of Table~\ref{table:long_range_symbol_terms_double_limit} remain 0 for all $w$. It seems likely that this vanishing mechanism protects the amplitude from having (perhaps sub-leading) unphysical branch cuts, but at this time we don't know how to derive these constraints directly from the Steinmann relations. 

%%%%%%%%%%%%%%%%%%%%%%%%%%%%%%%%%%%%%%%%%%%%%%%%%%%%%%%%%%%%%

%\bibliographystyle{physics}
%
%\bibliography{cgg}

\providecommand{\href}[2]{#2}\begingroup\raggedright\begin{thebibliography}{100}

\bibitem{Brink:1976bc}
L.~Brink, J.~H. Schwarz, and J.~Scherk, ``{Supersymmetric Yang-Mills
  Theories},''
\href{http://dx.doi.org/10.1016/0550-3213(77)90328-5}{{\em Nucl. Phys.} {\bf
  B121} (1977)  77--92}.
%%CITATION = NUPHA,B121,77;%%.

\bibitem{Gliozzi:1976qd}
F.~Gliozzi, J.~Scherk, and D.~I. Olive, ``{Supersymmetry, Supergravity Theories
  and the Dual Spinor Model},''
\href{http://dx.doi.org/10.1016/0550-3213(77)90206-1}{{\em Nucl. Phys.} {\bf
  B122} (1977)  253--290}.
%%CITATION = NUPHA,B122,253;%%.

\bibitem{Mandelstam:1982cb}
S.~Mandelstam, ``{Light Cone Superspace and the Ultraviolet Finiteness of
                        the N=4 Model},''
\href{http://doi.org/10.1016/0550-3213(83)90179-7}{{\em Nucl. Phys.} {\bf
  B213} (1983)  149--168}.
%%CITATION = NUPHA,B213,149;%%

\bibitem{Brink:1982wv}
L.~Brink, O.~Lindgren, and B.~E.~W. Nilsson, ``{The Ultraviolet Finiteness of
  the N=4 Yang-Mills Theory},''
\href{http://dx.doi.org/10.1016/0370-2693(83)91210-8}{{\em Phys. Lett.} {\bf
  B123} (1983)  323--328}.
%%CITATION = PHLTA,B123,323;%%.

\bibitem{Howe:1983sr}
P.~S. Howe, K.~S. Stelle, and P.~K. Townsend, ``{Miraculous Ultraviolet
  Cancellations in Supersymmetry Made Manifest},''
\href{http://dx.doi.org/10.1016/0550-3213(84)90528-5}{{\em Nucl. Phys.} {\bf
  B236} (1984)  125--166}.
%%CITATION = NUPHA,B236,125;%%.

\bibitem{Drummond:2006rz}
J.~M. Drummond, J.~Henn, V.~A. Smirnov, and E.~Sokatchev, ``{Magic identities
  for conformal four-point integrals},''
  \href{http://dx.doi.org/10.1088/1126-6708/2007/01/064}{{\em JHEP} {\bf 01}
  (2007)  064},
\href{http://arxiv.org/abs/hep-th/0607160}{{ arXiv:hep-th/0607160 [hep-th]}}.
%%CITATION = HEP-TH/0607160;%%.

\bibitem{Bern:2006ew}
Z.~Bern, M.~Czakon, L.~J. Dixon, D.~A. Kosower, and V.~A. Smirnov, ``{The
  Four-Loop Planar Amplitude and Cusp Anomalous Dimension in Maximally
  Supersymmetric Yang-Mills Theory},''
  \href{http://dx.doi.org/10.1103/PhysRevD.75.085010}{{\em Phys. Rev.} {\bf
  D75} (2007)  085010},
\href{http://arxiv.org/abs/hep-th/0610248}{{ arXiv:hep-th/0610248 [hep-th]}}.
%%CITATION = HEP-TH/0610248;%%.

\bibitem{Bern:2007ct}
Z.~Bern, J.~Carrasco, H.~Johansson, and D.~Kosower, ``{Maximally supersymmetric
  planar Yang-Mills amplitudes at five loops},''
  \href{http://dx.doi.org/10.1103/PhysRevD.76.125020}{{\em Phys.Rev.} {\bf D76}
  (2007)  125020},
\href{http://arxiv.org/abs/0705.1864}{{ arXiv:0705.1864 [hep-th]}}.
%%CITATION = ARXIV:0705.1864;%%.

\bibitem{Alday:2007hr}
L.~F. Alday and J.~M. Maldacena, ``{Gluon scattering amplitudes at strong
  coupling},'' \href{http://dx.doi.org/10.1088/1126-6708/2007/06/064}{{\em
  JHEP} {\bf 0706} (2007)  064},
\href{http://arxiv.org/abs/0705.0303}{{ arXiv:0705.0303 [hep-th]}}.
%%CITATION = ARXIV:0705.0303;%%.

\bibitem{Drummond:2008vq}
J.~M. Drummond, J.~Henn, G.~P. Korchemsky, and E.~Sokatchev, ``{Dual
  superconformal symmetry of scattering amplitudes in N=4 super-Yang-Mills
  theory},'' \href{http://dx.doi.org/10.1016/j.nuclphysb.2009.11.022}{{\em
  Nucl. Phys.} {\bf B828} (2010)  317--374},
\href{http://arxiv.org/abs/0807.1095}{{ arXiv:0807.1095 [hep-th]}}.
%%CITATION = ARXIV:0807.1095;%%.

\bibitem{Drummond:2007aua}
J.~Drummond, G.~Korchemsky, and E.~Sokatchev, ``{Conformal properties of
  four-gluon planar amplitudes and Wilson loops},''
  \href{http://dx.doi.org/10.1016/j.nuclphysb.2007.11.041}{{\em Nucl.Phys.}
  {\bf B795} (2008)  385--408},
\href{http://arxiv.org/abs/0707.0243}{{ arXiv:0707.0243 [hep-th]}}.
%%CITATION = ARXIV:0707.0243;%%.

\bibitem{Brandhuber:2007yx}
A.~Brandhuber, P.~Heslop, and G.~Travaglini, ``{MHV amplitudes in
  $\mathcal{N}=4$ super Yang-Mills and Wilson loops},''
  \href{http://dx.doi.org/10.1016/j.nuclphysb.2007.11.002}{{\em Nucl.Phys.}
  {\bf B794} (2008)  231--243},
\href{http://arxiv.org/abs/0707.1153}{{ arXiv:0707.1153 [hep-th]}}.
%%CITATION = ARXIV:0707.1153;%%.

\bibitem{Drummond:2007cf}
J.~Drummond, J.~Henn, G.~Korchemsky, and E.~Sokatchev, ``{On planar gluon
  amplitudes/Wilson loops duality},''
  \href{http://dx.doi.org/10.1016/j.nuclphysb.2007.11.007}{{\em Nucl.Phys.}
  {\bf B795} (2008)  52--68},
\href{http://arxiv.org/abs/0709.2368}{{ arXiv:0709.2368 [hep-th]}}.
%%CITATION = ARXIV:0709.2368;%%.

\bibitem{Drummond:2007au}
J.~Drummond, J.~Henn, G.~Korchemsky, and E.~Sokatchev, ``{Conformal Ward
  identities for Wilson loops and a test of the duality with gluon
  amplitudes},'' \href{http://dx.doi.org/10.1016/j.nuclphysb.2009.10.013}{{\em
  Nucl.Phys.} {\bf B826} (2010)  337--364},
\href{http://arxiv.org/abs/0712.1223}{{ arXiv:0712.1223 [hep-th]}}.
%%CITATION = ARXIV:0712.1223;%%.

\bibitem{Alday:2008yw}
L.~F. Alday and R.~Roiban, ``{Scattering Amplitudes, Wilson Loops and the
  String/Gauge Theory Correspondence},''
  \href{http://dx.doi.org/10.1016/j.physrep.2008.08.002}{{\em Phys.Rept.} {\bf
  468} (2008)  153--211},
\href{http://arxiv.org/abs/0807.1889}{{ arXiv:0807.1889 [hep-th]}}.
%%CITATION = ARXIV:0807.1889;%%.

\bibitem{Adamo:2011pv}
T.~Adamo, M.~Bullimore, L.~Mason, and D.~Skinner, ``{Scattering Amplitudes and
  Wilson Loops in Twistor Space},''
  \href{http://dx.doi.org/10.1088/1751-8113/44/45/454008}{{\em J.Phys.} {\bf
  A44} (2011)  454008},
\href{http://arxiv.org/abs/1104.2890}{{ arXiv:1104.2890 [hep-th]}}.
%%CITATION = ARXIV:1104.2890;%%.

\bibitem{Bern:2005iz}
Z.~Bern, L.~J. Dixon, and V.~A. Smirnov, ``{Iteration of planar amplitudes in
  maximally supersymmetric Yang-Mills theory at three loops and beyond},''
  \href{http://dx.doi.org/10.1103/PhysRevD.72.085001}{{\em Phys. Rev.} {\bf
  D72} (2005)  085001},
\href{http://arxiv.org/abs/hep-th/0505205}{{ arXiv:hep-th/0505205 [hep-th]}}.
%%CITATION = HEP-TH/0505205;%%.

\bibitem{Bartels:2008ce}
J.~Bartels, L.~Lipatov, and A.~Sabio~Vera, ``{BFKL Pomeron, Reggeized gluons
  and Bern-Dixon-Smirnov amplitudes},''
  \href{http://dx.doi.org/10.1103/PhysRevD.80.045002}{{\em Phys.Rev.} {\bf D80}
  (2009)  045002},
\href{http://arxiv.org/abs/0802.2065}{{ arXiv:0802.2065 [hep-th]}}.
%%CITATION = ARXIV:0802.2065;%%.

\bibitem{Bern:2008ap}
Z.~Bern, L.~Dixon, D.~Kosower, R.~Roiban, M.~Spradlin, C.~Vergu, and
  A.~Volovich, ``{The Two-Loop Six-Gluon MHV Amplitude in Maximally
  Supersymmetric Yang-Mills Theory},''
  \href{http://dx.doi.org/10.1103/PhysRevD.78.045007}{{\em Phys.Rev.} {\bf D78}
  (2008)  045007},
\href{http://arxiv.org/abs/0803.1465}{{ arXiv:0803.1465 [hep-th]}}.
%%CITATION = ARXIV:0803.1465;%%.

\bibitem{Drummond:2008aq}
J.~Drummond, J.~Henn, G.~Korchemsky, and E.~Sokatchev, ``{Hexagon Wilson loop =
  six-gluon MHV amplitude},''
  \href{http://dx.doi.org/10.1016/j.nuclphysb.2009.02.015}{{\em Nucl.Phys.}
  {\bf B815} (2009)  142--173},
\href{http://arxiv.org/abs/0803.1466}{{ arXiv:0803.1466 [hep-th]}}.
%%CITATION = ARXIV:0803.1466;%%.

\bibitem{Dixon:2011pw}
L.~J. Dixon, J.~M. Drummond, and J.~M. Henn, ``{Bootstrapping the three-loop
  hexagon},'' \href{http://dx.doi.org/10.1007/JHEP11(2011)023}{{\em JHEP} {\bf
  1111} (2011)  023},
\href{http://arxiv.org/abs/1108.4461}{{ arXiv:1108.4461 [hep-th]}}.
%%CITATION = ARXIV:1108.4461;%%.

\bibitem{Chen}
K.-T. Chen, ``Iterated path integrals,'' {\em Bull. Amer. Math. Soc.} {\bf 83}
  (1977) no. 5, 831--879.
  \url{http://projecteuclid.org/euclid.bams/1183539443}.

\bibitem{G91b}
A.~B. Goncharov, ``Geometry of configurations, polylogarithms, and motivic
  cohomology,''
  \href{https://doi.org/10.1006/aima.1995.1045}{{\em Adv.
  Math.} {\bf 114} (1995) no. 2, 197--318}.
  \url{http://www.sciencedirect.com/science/article/pii/S0001870885710456}.

\bibitem{Goncharov:1998kja}
A.~B. Goncharov, ``{Multiple polylogarithms, cyclotomy and modular
  complexes},'' \href{http://dx.doi.org/10.4310/MRL.1998.v5.n4.a7}{{\em Math.
  Res. Lett.} {\bf 5} (1998)  497--516},
\href{http://arxiv.org/abs/1105.2076}{{ arXiv:1105.2076 [math.AG]}}.
%%CITATION = ARXIV:1105.2076;%%.

\bibitem{Remiddi:1999ew}
E.~Remiddi and J.~Vermaseren, ``{Harmonic polylogarithms},''
  \href{http://dx.doi.org/10.1142/S0217751X00000367}{{\em Int.J.Mod.Phys.} {\bf
  A15} (2000)  725--754},
\href{http://arxiv.org/abs/hep-ph/9905237}{{ arXiv:hep-ph/9905237 [hep-ph]}}.
%%CITATION = HEP-PH/9905237;%%.

\bibitem{Borwein:1999js}
J.~M. Borwein, D.~M. Bradley, D.~J. Broadhurst, and P.~Lisonek, ``{Special
  values of multiple polylogarithms},''
  \href{http://dx.doi.org/10.1090/S0002-9947-00-02616-7}{{\em Trans. Am. Math.
  Soc.} {\bf 353} (2001)  907--941},
\href{http://arxiv.org/abs/math/9910045}{{ arXiv:math/9910045 [math-ca]}}.
%%CITATION = MATH/9910045;%%.

\bibitem{Moch:2001zr}
S.~Moch, P.~Uwer, and S.~Weinzierl, ``{Nested sums, expansion of transcendental
  functions and multiscale multiloop integrals},''
  \href{http://dx.doi.org/10.1063/1.1471366}{{\em J.Math.Phys.} {\bf 43} (2002)
   3363--3386},
\href{http://arxiv.org/abs/hep-ph/0110083}{{ arXiv:hep-ph/0110083 [hep-ph]}}.
%%CITATION = HEP-PH/0110083;%%.

\bibitem{Goncharov:2010jf}
A.~B. Goncharov, M.~Spradlin, C.~Vergu, and A.~Volovich, ``{Classical
  Polylogarithms for Amplitudes and Wilson Loops},''
  \href{http://dx.doi.org/10.1103/PhysRevLett.105.151605}{{\em Phys.Rev.Lett.}
  {\bf 105} (2010)  151605},
\href{http://arxiv.org/abs/1006.5703}{{ arXiv:1006.5703 [hep-th]}}.
%%CITATION = ARXIV:1006.5703;%%.

\bibitem{Duhr:2011zq}
C.~Duhr, H.~Gangl, and J.~R. Rhodes, ``{From polygons and symbols to
  polylogarithmic functions},''
  \href{http://dx.doi.org/10.1007/JHEP10(2012)075}{{\em JHEP} {\bf 1210} (2012)
   075},
\href{http://arxiv.org/abs/1110.0458}{{ arXiv:1110.0458 [math-ph]}}.
%%CITATION = ARXIV:1110.0458;%%.

\bibitem{Golden:2013xva}
J.~Golden, A.~B. Goncharov, M.~Spradlin, C.~Vergu, and A.~Volovich, ``{Motivic
  Amplitudes and Cluster Coordinates},''
  \href{http://dx.doi.org/10.1007/JHEP01(2014)091}{{\em JHEP} {\bf 1401} (2014)
   091},
\href{http://arxiv.org/abs/1305.1617}{{ arXiv:1305.1617 [hep-th]}}.
%%CITATION = ARXIV:1305.1617;%%.

\bibitem{Golden:2014xqa}
J.~Golden, M.~F. Paulos, M.~Spradlin, and A.~Volovich, ``{Cluster
  Polylogarithms for Scattering Amplitudes},''
  \href{http://dx.doi.org/10.1088/1751-8113/47/47/474005}{{\em J. Phys.} {\bf
  A47} (2014) no. 47, 474005},
\href{http://arxiv.org/abs/1401.6446}{{ arXiv:1401.6446 [hep-th]}}.
%%CITATION = ARXIV:1401.6446;%%.

\bibitem{Drummond:2014ffa}
J.~M. Drummond, G.~Papathanasiou, and M.~Spradlin, ``{A Symbol of Uniqueness:
  The Cluster Bootstrap for the 3-Loop MHV Heptagon},''
  \href{http://dx.doi.org/10.1007/JHEP03(2015)072}{{\em JHEP} {\bf 03} (2015)
  072},
\href{http://arxiv.org/abs/1412.3763}{{ arXiv:1412.3763 [hep-th]}}.
%%CITATION = ARXIV:1412.3763;%%.

\bibitem{Golden:2014pua}
J.~Golden and M.~Spradlin, ``{A Cluster Bootstrap for Two-Loop MHV
  Amplitudes},'' \href{http://dx.doi.org/10.1007/JHEP02(2015)002}{{\em JHEP}
  {\bf 02} (2015)  002},
\href{http://arxiv.org/abs/1411.3289}{{ arXiv:1411.3289 [hep-th]}}.
%%CITATION = ARXIV:1411.3289;%%.

\bibitem{1021.16017}
S.~Fomin and A.~Zelevinsky, ``{Cluster algebras I: Foundations},''
  \href{http://dx.doi.org/10.1090/S0894-0347-01-00385-X}{{\em J. Am. Math.
  Soc.} {\bf 15} (2002) no. 2, 497--529},
  \href{http://arxiv.org/abs/math/0104151}{{ arXiv:math/0104151 [math.RT]}}.

\bibitem{1054.17024}
S.~Fomin and A.~Zelevinsky, ``{Cluster algebras II: Finite type
  classification},'' \href{http://dx.doi.org/10.1007/s00222-003-0302-y}{{\em
  Invent. Math.} {\bf 154} (2003) no. 1, 63--121},
  \href{http://arxiv.org/abs/math/0208229}{{ arXiv:math/0208229 [math.RA]}}.

\bibitem{GSV}
M.~Gekhtman, M.~Shapiro, and A.~Vainshtein, ``{Cluster algebras and Poisson
  geometry},'' {\em Mosc.\ Math.\ J.} {\bf 3} (2003) no. 3, 899,
  \href{http://arxiv.org/abs/math/0208033}{{ math/0208033}}.

\bibitem{FG03b}
V.~V. Fock and A.~B. Goncharov, ``Cluster ensembles, quantization and the
  dilogarithm,'' {\em Ann. Sci. \'Ec. Norm. Sup\'er. (4)} {\bf 42} (2009) no.
  6, 865--930, \href{http://arxiv.org/abs/math/0311245}{{ arXiv:math/0311245
  [math.AG]}}.

\bibitem{Dixon:2011nj}
L.~J. Dixon, J.~M. Drummond, and J.~M. Henn, ``{Analytic result for the
  two-loop six-point NMHV amplitude in $\mathcal{N}=4$ super Yang-Mills
  theory},'' \href{http://dx.doi.org/10.1007/JHEP01(2012)024}{{\em JHEP} {\bf
  1201} (2012)  024},
\href{http://arxiv.org/abs/1111.1704}{{ arXiv:1111.1704 [hep-th]}}.
%%CITATION = ARXIV:1111.1704;%%.

\bibitem{Dixon:2013eka}
L.~J. Dixon, J.~M. Drummond, M.~von Hippel, and J.~Pennington, ``{Hexagon
  functions and the three-loop remainder function},''
  \href{http://dx.doi.org/10.1007/JHEP12(2013)049}{{\em JHEP} {\bf 1312} (2013)
   049},
\href{http://arxiv.org/abs/1308.2276}{{ arXiv:1308.2276 [hep-th]}}.
%%CITATION = ARXIV:1308.2276;%%.

\bibitem{Dixon:2014voa}
L.~J. Dixon, J.~M. Drummond, C.~Duhr, and J.~Pennington, ``{The four-loop
  remainder function and multi-Regge behavior at NNLLA in planar $\mathcal{N} =
  4$ super-Yang-Mills theory},''
  \href{http://dx.doi.org/10.1007/JHEP06(2014)116}{{\em JHEP} {\bf 1406} (2014)
   116},
\href{http://arxiv.org/abs/1402.3300}{{ arXiv:1402.3300 [hep-th]}}.
%%CITATION = ARXIV:1402.3300;%%.

\bibitem{Dixon:2014xca}
L.~J. Dixon, J.~M. Drummond, C.~Duhr, M.~von Hippel, and J.~Pennington,
  ``{Bootstrapping six-gluon scattering in planar $\mathcal{N}=4$
  super-Yang-Mills theory},'' {\em PoS} {\bf LL2014} (2014)  077,
\href{http://arxiv.org/abs/1407.4724}{{ arXiv:1407.4724 [hep-th]}}.
%%CITATION = ARXIV:1407.4724;%%.

\bibitem{Dixon:2014iba}
L.~J. Dixon and M.~von Hippel, ``{Bootstrapping an NMHV amplitude through three
  loops},'' \href{http://dx.doi.org/10.1007/JHEP10(2014)065}{{\em JHEP} {\bf
  1410} (2014)  65},
\href{http://arxiv.org/abs/1408.1505}{{ arXiv:1408.1505 [hep-th]}}.
%%CITATION = ARXIV:1408.1505;%%.

\bibitem{Dixon:2015iva}
L.~J. Dixon, M.~von Hippel, and A.~J. McLeod, ``{The four-loop six-gluon NMHV
  ratio function},'' \href{http://dx.doi.org/10.1007/JHEP01(2016)053}{{\em
  JHEP} {\bf 01} (2016)  053},
\href{http://arxiv.org/abs/1509.08127}{{ arXiv:1509.08127 [hep-th]}}.
%%CITATION = ARXIV:1509.08127;%%.

\bibitem{Caron-Huot:2016owq}
S.~Caron-Huot, L.~J. Dixon, A.~McLeod, and M.~von Hippel, ``{Bootstrapping a
  Five-Loop Amplitude Using Steinmann Relations},''
  \href{http://dx.doi.org/10.1103/PhysRevLett.117.241601}{{\em Phys. Rev.
  Lett.} {\bf 117} (2016) no. 24, 241601},
\href{http://arxiv.org/abs/1609.00669}{{ arXiv:1609.00669 [hep-th]}}.
%%CITATION = ARXIV:1609.00669;%%.

\bibitem{Caron-Huot:2019vjl}
S.~Caron-Huot, L.~J. Dixon, F.~Dulat, M.~von Hippel, A.~J. McLeod, and
  G.~Papathanasiou, ``{Six-Gluon Amplitudes in Planar ${\cal N}=4$
  Super-Yang-Mills Theory at Six and Seven Loops},''
\href{http://arxiv.org/abs/1903.10890}{{ arXiv:1903.10890 [hep-th]}}.
%%CITATION = ARXIV:1903.10890;%%.

\bibitem{Dixon:2016nkn}
L.~J. Dixon, J.~Drummond, T.~Harrington, A.~J. McLeod, G.~Papathanasiou, and
  M.~Spradlin, ``{Heptagons from the Steinmann Cluster Bootstrap},''
  \href{http://dx.doi.org/10.1007/JHEP02(2017)137}{{\em JHEP} {\bf 02} (2017)
  137},
\href{http://arxiv.org/abs/1612.08976}{{ arXiv:1612.08976 [hep-th]}}.
%%CITATION = ARXIV:1612.08976;%%.

\bibitem{Drummond:2018caf}
J.~Drummond, J.~Foster, {\"O}.~G{\"u}rdo{\u{g}}an, and G.~Papathanasiou,
  ``{Cluster adjacency and the four-loop NMHV heptagon},''
  \href{http://dx.doi.org/10.1007/JHEP03(2019)087}{{\em JHEP} {\bf 03} (2019)
  087},
\href{http://arxiv.org/abs/1812.04640}{{ arXiv:1812.04640 [hep-th]}}.
%%CITATION = ARXIV:1812.04640;%%.

\bibitem{Gonch2}
A.~B. Goncharov,
  \href{http://dx.doi.org/10.1215/S0012-7094-04-12822-2}{``Galois symmetries of
  fundamental groupoids and noncommutative geometry,''{\em Duke Math. J.} {\bf
  128} (06, 2005)  209--284}, \href{http://arxiv.org/abs/math/0208144}{{
  arXiv:math/0208144 [math.AG]}}.

\bibitem{Brown:2011ik}
F.~Brown, ``{On the decomposition of motivic multiple zeta values},'' {\em
  {Adv. Studies in Pure Math.}} {\bf 63} (2012)  31--58,
\href{http://arxiv.org/abs/1102.1310}{{ arXiv:1102.1310 [math.NT]}}.
%%CITATION = ARXIV:1102.1310;%%.

\bibitem{Brown1102.1312}
F.~Brown, ``Mixed {T}ate motives over {$\mathbb{Z}$},''
  \href{http://dx.doi.org/10.4007/annals.2012.175.2.10}{{\em Ann. of Math. (2)}
  {\bf 175} (2012) no. 2, 949--976}, \href{http://arxiv.org/abs/1102.1312}{{
  arXiv:1102.1312 [math.AG]}}.

\bibitem{Duhr:2012fh}
C.~Duhr, ``{Hopf algebras, coproducts and symbols: an application to Higgs
  boson amplitudes},'' \href{http://dx.doi.org/10.1007/JHEP08(2012)043}{{\em
  JHEP} {\bf 1208} (2012)  043},
\href{http://arxiv.org/abs/1203.0454}{{ arXiv:1203.0454 [hep-ph]}}.
%%CITATION = ARXIV:1203.0454;%%.

\bibitem{Chavez:2012kn}
F.~Chavez and C.~Duhr, ``{Three-mass triangle integrals and single-valued
  polylogarithms},'' \href{http://dx.doi.org/10.1007/JHEP11(2012)114}{{\em
  JHEP} {\bf 11} (2012)  114},
\href{http://arxiv.org/abs/1209.2722}{{ arXiv:1209.2722 [hep-ph]}}.
%%CITATION = ARXIV:1209.2722;%%.

\bibitem{Gonch3}
A.~B. Goncharov, ``{Multiple polylogarithms and mixed Tate motives},''
\href{http://arxiv.org/abs/math/0103059}{{ arXiv:math/0103059 [math.AG]}}.
%%CITATION = MATH/0103059;%%.

\bibitem{FBThesis}
F.~C. Brown, ``{Multiple zeta values and periods of moduli spaces
  $\overline{\mathfrak{M}}_{0,n}(\mathbb{R})$},'' {\em Annales Sci.Ecole
  Norm.Sup.} {\bf 42} (2009)  371,
\href{http://arxiv.org/abs/math/0606419}{{ arXiv:math/0606419 [math.AG]}}.
%%CITATION = ARXIV:MATH/0606419;%%.

\bibitem{2015arXiv151206410B}
F.~Brown, ``{Notes on Motivic Periods},''
  \href{http://arxiv.org/abs/1512.06410}{{ arXiv:1512.06410 [math.NT]}}.

\bibitem{Cartier2001}
P.~Cartier, ``{A mad day's work: from Grothendieck to Connes and Kontsevich,
  the evolution of concepts of space and symmetry},''
  \href{http://dx.doi.org/10.1090/S0273-0979-01-00913-2}{{\em Bull. Amer. Math.
  Soc. (N.S.)} {\bf 38} (2001)  389--408}.

\bibitem{2008arXiv0805.2568A}
Y.~Andr\'{e}, ``{Ambiguity theory, old and new},''
  \href{http://arxiv.org/abs/0805.2568}{{ arXiv:0805.2568 [math.GM]}}.

\bibitem{2008arXiv0805.2569A}
Y.~Andr\'{e}, ``{Galois theory, motives and transcendental numbers},''
  \href{http://arxiv.org/abs/0805.2569}{{ arXiv:0805.2569 [math.NT]}}.

\bibitem{Brown:2015fyf}
F.~Brown, ``{Feynman amplitudes, coaction principle, and cosmic Galois
  group},'' \href{http://dx.doi.org/10.4310/CNTP.2017.v11.n3.a1}{{\em Commun.
  Num. Theor. Phys.} {\bf 11} (2017)  453--556},
\href{http://arxiv.org/abs/1512.06409}{{ arXiv:1512.06409 [math-ph]}}.
%%CITATION = ARXIV:1512.06409;%%.

\bibitem{Broadhurst:1996kc}
D.~J. Broadhurst and D.~Kreimer, ``{Association of multiple zeta values with
  positive knots via Feynman diagrams up to 9 loops},''
  \href{http://dx.doi.org/10.1016/S0370-2693(96)01623-1}{{\em Phys. Lett.} {\bf
  B393} (1997)  403--412},
\href{http://arxiv.org/abs/hep-th/9609128}{{ arXiv:hep-th/9609128 [hep-th]}}.
%%CITATION = HEP-TH/9609128;%%.

\bibitem{Schnetz:2013hqa}
O.~Schnetz, ``{Graphical functions and single-valued multiple
  polylogarithms},'' \href{http://dx.doi.org/10.4310/CNTP.2014.v8.n4.a1}{{\em
  Commun. Num. Theor. Phys.} {\bf 08} (2014)  589--675},
\href{http://arxiv.org/abs/1302.6445}{{ arXiv:1302.6445 [math.NT]}}.
%%CITATION = ARXIV:1302.6445;%%.

\bibitem{Panzer:2016snt}
E.~Panzer and O.~Schnetz, ``{The Galois coaction on $\phi^4$ periods},''
  \href{http://dx.doi.org/10.4310/CNTP.2017.v11.n3.a3}{{\em Commun. Num. Theor.
  Phys.} {\bf 11} (2017)  657--705},
\href{http://arxiv.org/abs/1603.04289}{{ arXiv:1603.04289 [hep-th]}}.
%%CITATION = ARXIV:1603.04289;%%.

\bibitem{Laporta:2017okg}
S.~Laporta, ``{High-precision calculation of the 4-loop contribution to the
  electron g-2 in QED},''
  \href{http://dx.doi.org/10.1016/j.physletb.2017.06.056}{{\em Phys. Lett.}
  {\bf B772} (2017)  232--238},
\href{http://arxiv.org/abs/1704.06996}{{ arXiv:1704.06996 [hep-ph]}}.
%%CITATION = ARXIV:1704.06996;%%.

\bibitem{Schnetz:2017bko}
O.~Schnetz, ``{The Galois coaction on the electron anomalous magnetic
  moment},'' \href{http://dx.doi.org/10.4310/CNTP.2018.v12.n2.a4}{{\em Commun.
  Num. Theor. Phys.} {\bf 12} (2018)  335--354},
\href{http://arxiv.org/abs/1711.05118}{{ arXiv:1711.05118 [math-ph]}}.
%%CITATION = ARXIV:1711.05118;%%.

\bibitem{Schlotterer:2012ny}
O.~Schlotterer and S.~Stieberger, ``{Motivic Multiple Zeta Values and
  Superstring Amplitudes},''
  \href{http://dx.doi.org/10.1088/1751-8113/46/47/475401}{{\em J. Phys.} {\bf
  A46} (2013)  475401},
\href{http://arxiv.org/abs/1205.1516}{{ arXiv:1205.1516 [hep-th]}}.
%%CITATION = ARXIV:1205.1516;%%.

\bibitem{Brown:2018omk}
F.~Brown and C.~Dupont, ``{Single-valued integration and superstring amplitudes
  in genus zero},''
\href{http://arxiv.org/abs/1810.07682}{{ arXiv:1810.07682 [math.NT]}}.
%%CITATION = ARXIV:1810.07682;%%.

\bibitem{Gaiotto:2011dt}
D.~Gaiotto, J.~Maldacena, A.~Sever, and P.~Vieira, ``{Pulling the straps of
  polygons},'' \href{http://dx.doi.org/10.1007/JHEP12(2011)011}{{\em JHEP} {\bf
  1112} (2011)  011},
\href{http://arxiv.org/abs/1102.0062}{{ arXiv:1102.0062 [hep-th]}}.
%%CITATION = ARXIV:1102.0062;%%.

\bibitem{Alday:2009dv}
L.~F. Alday, D.~Gaiotto, and J.~Maldacena, ``{Thermodynamic Bubble Ansatz},''
  \href{http://dx.doi.org/10.1007/JHEP09(2011)032}{{\em JHEP} {\bf 09} (2011)
  032},
\href{http://arxiv.org/abs/0911.4708}{{ arXiv:0911.4708 [hep-th]}}.
%%CITATION = ARXIV:0911.4708;%%.

\bibitem{Yang:2010as}
G.~Yang, ``{A simple collinear limit of scattering amplitudes at strong
  coupling},'' \href{http://dx.doi.org/10.1007/JHEP03(2011)087}{{\em JHEP} {\bf
  03} (2011)  087},
\href{http://arxiv.org/abs/1006.3306}{{ arXiv:1006.3306 [hep-th]}}.
%%CITATION = ARXIV:1006.3306;%%.

\bibitem{Steinmann}
O.~Steinmann, ``{\"Uber den Zusammenhang zwischen den Wightmanfunktionen und
  der retardierten Kommutatoren},'' {\em Helv. Physica Acta} {\bf 33} (1960)
  257.

\bibitem{Steinmann2}
O.~Steinmann, ``{Wightman-Funktionen und retardierten Kommutatoren. II},'' {\em
  Helv. Physica Acta} {\bf 33} (1960)  347.

\bibitem{Cahill:1973qp}
K.~E. Cahill and H.~P. Stapp, ``{OPTICAL THEOREMS AND STEINMANN RELATIONS},''
\href{http://dx.doi.org/10.1016/0003-4916(75)90006-8}{{\em Annals Phys.} {\bf
  90} (1975)  438}.
%%CITATION = APNYA,90,438;%%.

\bibitem{Drummond:2017ssj}
J.~Drummond, J.~Foster, and {\"O}.~G{\"u}rdo{\u{g}}an, ``{Cluster Adjacency
  Properties of Scattering Amplitudes in $N=4$ Supersymmetric Yang-Mills
  Theory},'' \href{http://dx.doi.org/10.1103/PhysRevLett.120.161601}{{\em Phys.
  Rev. Lett.} {\bf 120} (2018) no. 16, 161601},
\href{http://arxiv.org/abs/1710.10953}{{ arXiv:1710.10953 [hep-th]}}.
%%CITATION = ARXIV:1710.10953;%%.

\bibitem{Drummond:2018dfd}
J.~Drummond, J.~Foster, and {\"O}.~G{\"u}rdo{\u{g}}an, ``{Cluster adjacency
  beyond MHV},'' \href{http://dx.doi.org/10.1007/JHEP03(2019)086}{{\em JHEP}
  {\bf 03} (2019)  086},
\href{http://arxiv.org/abs/1810.08149}{{ arXiv:1810.08149 [hep-th]}}.
%%CITATION = ARXIV:1810.08149;%%.

\bibitem{Golden:2019kks}
J.~Golden, A.~J. McLeod, M.~Spradlin, and A.~Volovich, ``{The Sklyanin Bracket
  and Cluster Adjacency at All Multiplicity},''
\href{http://arxiv.org/abs/1902.11286}{{ arXiv:1902.11286 [hep-th]}}.
%%CITATION = ARXIV:1902.11286;%%.

\bibitem{HyperlogProcedures}
O.~Schnetz. Computer program {\tt hyperlogprocedures},
  \url{https://www.math.fau.de/person/oliver-schnetz/}.

\bibitem{Hodges:2009hk}
A.~Hodges, ``{Eliminating spurious poles from gauge-theoretic amplitudes},''
  \href{http://dx.doi.org/10.1007/JHEP05(2013)135}{{\em JHEP} {\bf 1305} (2013)
   135},
\href{http://arxiv.org/abs/0905.1473}{{ arXiv:0905.1473 [hep-th]}}.
%%CITATION = ARXIV:0905.1473;%%.

\bibitem{Mason:2009qx}
L.~Mason and D.~Skinner, ``{Dual Superconformal Invariance, Momentum Twistors
  and Grassmannians},''
  \href{http://dx.doi.org/10.1088/1126-6708/2009/11/045}{{\em JHEP} {\bf 0911}
  (2009)  045},
\href{http://arxiv.org/abs/0909.0250}{{ arXiv:0909.0250 [hep-th]}}.
%%CITATION = ARXIV:0909.0250;%%.

\bibitem{Elvang:2013cua}
H.~Elvang and Y.-t. Huang, ``{Scattering Amplitudes},''
\href{http://arxiv.org/abs/1308.1697}{{ arXiv:1308.1697 [hep-th]}}.
%%CITATION = ARXIV:1308.1697;%%.

\bibitem{Beisert:2006ez}
N.~Beisert, B.~Eden, and M.~Staudacher, ``{Transcendentality and Crossing},''
  \href{http://dx.doi.org/10.1088/1742-5468/2007/01/P01021}{{\em J. Stat.
  Mech.} {\bf 0701} (2007)  P01021},
\href{http://arxiv.org/abs/hep-th/0610251}{{ arXiv:hep-th/0610251 [hep-th]}}.
%%CITATION = HEP-TH/0610251;%%.

\bibitem{ArkaniHamed:2012nw}
N.~Arkani-Hamed, J.~L. Bourjaily, F.~Cachazo, A.~B. Goncharov, A.~Postnikov,
  {\em et al.}, ``{Scattering Amplitudes and the Positive Grassmannian},''
\href{http://arxiv.org/abs/1212.5605}{{ arXiv:1212.5605 [hep-th]}}.
%%CITATION = ARXIV:1212.5605;%%.

\bibitem{Parker:2015cia}
D.~Parker, A.~Scherlis, M.~Spradlin, and A.~Volovich, ``{Hedgehog Bases for
  $A_n$ Cluster Polylogarithms and An Application to Six-Point Amplitudes},''
  \href{http://dx.doi.org/10.1007/JHEP11(2015)136}{{\em JHEP} {\bf 11} (2015)
  136},
\href{http://arxiv.org/abs/1507.01950}{{ arXiv:1507.01950 [hep-th]}}.
%%CITATION = ARXIV:1507.01950;%%.

\bibitem{CaronHuot:2011ky}
S.~Caron-Huot, ``{Superconformal symmetry and two-loop amplitudes in planar
  $\mathcal{N}=4$ super Yang-Mills},''
  \href{http://dx.doi.org/10.1007/JHEP12(2011)066}{{\em JHEP} {\bf 1112} (2011)
   066},
\href{http://arxiv.org/abs/1105.5606}{{ arXiv:1105.5606 [hep-th]}}.
%%CITATION = ARXIV:1105.5606;%%.

\bibitem{Caron-Huot:2018dsv}
S.~Caron-Huot, L.~J. Dixon, M.~von Hippel, A.~J. McLeod, and G.~Papathanasiou,
  ``{The Double Pentaladder Integral to All Orders},''
  \href{http://dx.doi.org/10.1007/JHEP07(2018)170}{{\em JHEP} {\bf 07} (2018)
  170},
\href{http://arxiv.org/abs/1806.01361}{{ arXiv:1806.01361 [hep-th]}}.
%%CITATION = ARXIV:1806.01361;%%.

\bibitem{YorgosAmps17}
G.~Papathanasiou. Talk at {\it amplitudes 2017},
  \url{https://indico.ph.ed.ac.uk/event/26/contributions/342/attachments/284/319/Papathanasiou_Amplitudes2017.pdf}.

\bibitem{DFGPrivate}
{J.~Drummond and {\"O}.~G{\"u}rdo{\u{g}}an, private communication}.

\bibitem{Harrington:2015bdt}
T.~Harrington and M.~Spradlin, ``{Cluster Functions and Scattering Amplitudes
  for Six and Seven Points},''
  \href{http://dx.doi.org/10.1007/JHEP07(2017)016}{{\em JHEP} {\bf 07} (2017)
  016},
\href{http://arxiv.org/abs/1512.07910}{{ arXiv:1512.07910 [hep-th]}}.
%%CITATION = ARXIV:1512.07910;%%.

\bibitem{Alday:2010vh}
L.~F. Alday, J.~Maldacena, A.~Sever, and P.~Vieira, ``{Y-system for Scattering
  Amplitudes},'' \href{http://dx.doi.org/10.1088/1751-8113/43/48/485401}{{\em
  J. Phys.} {\bf A43} (2010)  485401},
\href{http://arxiv.org/abs/1002.2459}{{ arXiv:1002.2459 [hep-th]}}.
%%CITATION = ARXIV:1002.2459;%%.

\bibitem{Yang:2010az}
G.~Yang, ``{Scattering amplitudes at strong coupling for 4K gluons},''
  \href{http://dx.doi.org/10.1007/JHEP12(2010)082}{{\em JHEP} {\bf 12} (2010)
  082},
\href{http://arxiv.org/abs/1004.3983}{{ arXiv:1004.3983 [hep-th]}}.
%%CITATION = ARXIV:1004.3983;%%.

\bibitem{Golden:2018gtk}
J.~Golden and A.~J. Mcleod, ``{Cluster Algebras and the Subalgebra
  Constructibility of the Seven-Particle Remainder Function},''
  \href{http://dx.doi.org/10.1007/JHEP01(2019)017}{{\em JHEP} {\bf 01} (2019)
  017},
\href{http://arxiv.org/abs/1810.12181}{{ arXiv:1810.12181 [hep-th]}}.
%%CITATION = ARXIV:1810.12181;%%.

\bibitem{SageMath}
W.~Stein and D.~Joyner, ``{SAGE}: System for algebra and geometry
  experimentation,'' {\em ACM SIGSAM Bulletin} {\bf 39} (2005) no. 2, 61--64.
  \url{http://www.sagemath.org}.

\bibitem{DelDuca:2011ne}
V.~Del~Duca, C.~Duhr, and V.~A. Smirnov, ``{The massless hexagon integral in D
  = 6 dimensions},''
  \href{http://dx.doi.org/10.1016/j.physletb.2011.07.079}{{\em Phys.Lett.} {\bf
  B703} (2011)  363--365},
\href{http://arxiv.org/abs/1104.2781}{{ arXiv:1104.2781 [hep-th]}}.
%%CITATION = ARXIV:1104.2781;%%.

\bibitem{Dixon:2011ng}
L.~J. Dixon, J.~M. Drummond, and J.~M. Henn, ``{The one-loop six-dimensional
  hexagon integral and its relation to MHV amplitudes in N=4 SYM},''
  \href{http://dx.doi.org/10.1007/JHEP06(2011)100}{{\em JHEP} {\bf 1106} (2011)
   100},
\href{http://arxiv.org/abs/1104.2787}{{ arXiv:1104.2787 [hep-th]}}.
%%CITATION = ARXIV:1104.2787;%%.

\bibitem{Anastasiou:2013srw}
C.~Anastasiou, C.~Duhr, F.~Dulat, and B.~Mistlberger, ``{Soft triple-real
  radiation for Higgs production at N3LO},''
  \href{http://dx.doi.org/10.1007/JHEP07(2013)003}{{\em JHEP} {\bf 07} (2013)
  003},
\href{http://arxiv.org/abs/1302.4379}{{ arXiv:1302.4379 [hep-ph]}}.
%%CITATION = ARXIV:1302.4379;%%.

\bibitem{Panzer:2014caa}
E.~Panzer, ``{Algorithms for the symbolic integration of hyperlogarithms with
  applications to Feynman integrals},''
  \href{http://dx.doi.org/10.1016/j.cpc.2014.10.019}{{\em Comput. Phys.
  Commun.} {\bf 188} (2015)  148--166},
\href{http://arxiv.org/abs/1403.3385}{{ arXiv:1403.3385 [hep-th]}}.
%%CITATION = ARXIV:1403.3385;%%.

\bibitem{Apery:1979}
R.~Ap\'{e}ry, ``{Irrationalit\'e de $\zeta(2)$ et $\zeta(3)$},''
  {\em Ast\'erisque.} {\bf
  61} (1979)  11--13.

\bibitem{FSZ}
S.~Fischler, J.~Sprang, and W.~Zudilin, ``{Many odd zeta values are
  irrational},'' \href{http://dx.doi.org/10.1112/S0010437X1900722X}{{\em
  Compositio Math} {\bf 155} (2019)  938--952},
  \href{http://arxiv.org/abs/1803.08905}{{ arXiv:1803.08905 [math.NT]}}.

\bibitem{2011arXiv1101.4497D}
M.~{Deneufch{\^a}tel}, G.~H.~E. {Duchamp}, V.~H.~N. {Minh}, and A.~I.
  {Solomon}, ``{Independence of hyperlogarithms over function fields via
  algebraic combinatorics},''{\em arXiv e-prints} (Jan, 2011)  ,
  \href{http://arxiv.org/abs/1101.4497}{{ arXiv:1101.4497 [math.CO]}}.

\bibitem{Cachazo:2008hp}
F.~Cachazo, M.~Spradlin, and A.~Volovich, ``{Leading Singularities of the
  Two-Loop Six-Particle MHV Amplitude},''
  \href{http://dx.doi.org/10.1103/PhysRevD.78.105022}{{\em Phys. Rev.} {\bf
  D78} (2008)  105022},
\href{http://arxiv.org/abs/0805.4832}{{ arXiv:0805.4832 [hep-th]}}.
%%CITATION = ARXIV:0805.4832;%%.

\bibitem{DelDuca:2009au}
V.~Del~Duca, C.~Duhr, and V.~A. Smirnov, ``{An Analytic Result for the Two-Loop
  Hexagon Wilson Loop in $\mathcal{N} = 4$ SYM},''
  \href{http://dx.doi.org/10.1007/JHEP03(2010)099}{{\em JHEP} {\bf 1003} (2010)
   099},
\href{http://arxiv.org/abs/0911.5332}{{ arXiv:0911.5332 [hep-ph]}}.
%%CITATION = ARXIV:0911.5332;%%.

\bibitem{DelDuca:2010zg}
V.~Del~Duca, C.~Duhr, and V.~A. Smirnov, ``{The Two-Loop Hexagon Wilson Loop in
  $\mathcal{N} = 4$ SYM},''
  \href{http://dx.doi.org/10.1007/JHEP05(2010)084}{{\em JHEP} {\bf 1005} (2010)
   084},
\href{http://arxiv.org/abs/1003.1702}{{ arXiv:1003.1702 [hep-th]}}.
%%CITATION = ARXIV:1003.1702;%%.

\bibitem{Prlina:2018ukf}
I.~Prlina, M.~Spradlin, and S.~Stanojevic, ``{All-loop singularities of
  scattering amplitudes in massless planar theories},''
  \href{http://dx.doi.org/10.1103/PhysRevLett.121.081601}{{\em Phys. Rev.
  Lett.} {\bf 121} (2018) no. 8, 081601},
\href{http://arxiv.org/abs/1805.11617}{{ arXiv:1805.11617 [hep-th]}}.
%%CITATION = ARXIV:1805.11617;%%.

\bibitem{Zagier:1994}
D.~Zagier, ``{Values of Zeta Functions and Their Applications},''
  \href{http://dx.doi.org/10.1007/978-3-0348-9112-7_23}{{\em Progress in
  Mathematics} {\bf 120} (1994)  497--512}.

\bibitem{Duhr:2019tlz}
C.~Duhr and F.~Dulat, ``{PolyLogTools - Polylogs for the masses},''
\href{http://arxiv.org/abs/1904.07279}{{ arXiv:1904.07279 [hep-th]}}.
%%CITATION = ARXIV:1904.07279;%%.

\bibitem{Ablinger:2011te}
J.~Ablinger, J.~{Bl\"umlein}, and C.~Schneider, ``{Harmonic Sums and
  Polylogarithms Generated by Cyclotomic Polynomials},''
  \href{http://dx.doi.org/10.1063/1.3629472}{{\em J.Math.Phys.} {\bf 52} (2011)
   102301},
\href{http://arxiv.org/abs/1105.6063}{{ arXiv:1105.6063 [math-ph]}}.
%%CITATION = ARXIV:1105.6063;%%.

\bibitem{Basso:2013vsa}
B.~Basso, A.~Sever, and P.~Vieira, ``{Spacetime and Flux Tube S-Matrices at
  Finite Coupling for $\mathcal{N}=4$ Supersymmetric Yang-Mills Theory},''
  \href{http://dx.doi.org/10.1103/PhysRevLett.111.091602}{{\em Phys.Rev.Lett.}
  {\bf 111} (2013) no. 9, 091602},
\href{http://arxiv.org/abs/1303.1396}{{ arXiv:1303.1396 [hep-th]}}.
%%CITATION = ARXIV:1303.1396;%%.

\bibitem{Basso:2013aha}
B.~Basso, A.~Sever, and P.~Vieira, ``{Space-time S-matrix and Flux tube
  S-matrix II. Extracting and Matching Data},''
  \href{http://dx.doi.org/10.1007/JHEP01(2014)008}{{\em JHEP} {\bf 1401} (2014)
   008},
\href{http://arxiv.org/abs/1306.2058}{{ arXiv:1306.2058 [hep-th]}}.
%%CITATION = ARXIV:1306.2058;%%.

\bibitem{Basso:2014hfa}
B.~Basso, J.~Caetano, L.~Cordova, A.~Sever, and P.~Vieira, ``{OPE for all
  Helicity Amplitudes},''
\href{http://arxiv.org/abs/1412.1132}{{ arXiv:1412.1132 [hep-th]}}.
%%CITATION = ARXIV:1412.1132;%%.

\bibitem{Arkani-Hamed:2014dca}
N.~Arkani-Hamed, A.~Hodges, and J.~Trnka, ``{Positive Amplitudes In The
  Amplituhedron},'' \href{http://dx.doi.org/10.1007/JHEP08(2015)030}{{\em JHEP}
  {\bf 08} (2015)  030},
\href{http://arxiv.org/abs/1412.8478}{{ arXiv:1412.8478 [hep-th]}}.
%%CITATION = ARXIV:1412.8478;%%.

\bibitem{Dixon:2016apl}
L.~J. Dixon, M.~von Hippel, A.~J. McLeod, and J.~Trnka, ``{Multi-loop
  positivity of the planar $ \mathcal{N} $ = 4 SYM six-point amplitude},''
  \href{http://dx.doi.org/10.1007/JHEP02(2017)112}{{\em JHEP} {\bf 02} (2017)
  112},
\href{http://arxiv.org/abs/1611.08325}{{ arXiv:1611.08325 [hep-th]}}.
%%CITATION = ARXIV:1611.08325;%%.

\bibitem{CaronHuot:2012ab}
S.~Caron-Huot and K.~J. Larsen, ``{Uniqueness of two-loop master contours},''
  \href{http://dx.doi.org/10.1007/JHEP10(2012)026}{{\em JHEP} {\bf 10} (2012)
  026},
\href{http://arxiv.org/abs/1205.0801}{{ arXiv:1205.0801 [hep-ph]}}.
%%CITATION = ARXIV:1205.0801;%%.

\bibitem{Nandan:2013ip}
D.~Nandan, M.~F. Paulos, M.~Spradlin, and A.~Volovich, ``{Star Integrals,
  Convolutions and Simplices},''
  \href{http://dx.doi.org/10.1007/JHEP05(2013)105}{{\em JHEP} {\bf 05} (2013)
  105},
\href{http://arxiv.org/abs/1301.2500}{{ arXiv:1301.2500 [hep-th]}}.
%%CITATION = ARXIV:1301.2500;%%.

\bibitem{Bourjaily:2015jna}
J.~L. Bourjaily and J.~Trnka, ``{Local Integrand Representations of All
  Two-Loop Amplitudes in Planar SYM},''
  \href{http://dx.doi.org/10.1007/JHEP08(2015)119}{{\em JHEP} {\bf 08} (2015)
  119},
\href{http://arxiv.org/abs/1505.05886}{{ arXiv:1505.05886 [hep-th]}}.
%%CITATION = ARXIV:1505.05886;%%.

\bibitem{Bourjaily:2017wjl}
J.~L. Bourjaily, E.~Herrmann, and J.~Trnka, ``{Prescriptive Unitarity},''
  \href{http://dx.doi.org/10.1007/JHEP06(2017)059}{{\em JHEP} {\bf 06} (2017)
  059},
\href{http://arxiv.org/abs/1704.05460}{{ arXiv:1704.05460 [hep-th]}}.
%%CITATION = ARXIV:1704.05460;%%.

\bibitem{Bourjaily:2017bsb}
J.~L. Bourjaily, A.~J. McLeod, M.~Spradlin, M.~von Hippel, and M.~Wilhelm,
  ``{Elliptic Double-Box Integrals: Massless Scattering Amplitudes beyond
  Polylogarithms},''
  \href{http://dx.doi.org/10.1103/PhysRevLett.120.121603}{{\em Phys. Rev.
  Lett.} {\bf 120} (2018) no. 12, 121603},
\href{http://arxiv.org/abs/1712.02785}{{ arXiv:1712.02785 [hep-th]}}.
%%CITATION = ARXIV:1712.02785;%%.

\bibitem{Bourjaily:2018ycu}
J.~L. Bourjaily, Y.-H. He, A.~J. Mcleod, M.~Von~Hippel, and M.~Wilhelm,
  ``{Traintracks through Calabi-Yau Manifolds: Scattering Amplitudes beyond
  Elliptic Polylogarithms},''
  \href{http://dx.doi.org/10.1103/PhysRevLett.121.071603}{{\em Phys. Rev.
  Lett.} {\bf 121} (2018) no. 7, 071603},
\href{http://arxiv.org/abs/1805.09326}{{ arXiv:1805.09326 [hep-th]}}.
%%CITATION = ARXIV:1805.09326;%%.

\bibitem{Bourjaily:2018yfy}
J.~L. Bourjaily, A.~J. McLeod, M.~von Hippel, and M.~Wilhelm, ``{Bounded
  Collection of Feynman Integral Calabi-Yau Geometries},''
  \href{http://dx.doi.org/10.1103/PhysRevLett.122.031601}{{\em Phys. Rev.
  Lett.} {\bf 122} (2019) no. 3, 031601},
\href{http://arxiv.org/abs/1810.07689}{{ arXiv:1810.07689 [hep-th]}}.
%%CITATION = ARXIV:1810.07689;%%.

\bibitem{Broedel:2017kkb}
J.~Broedel, C.~Duhr, F.~Dulat, and L.~Tancredi, ``{Elliptic polylogarithms and
  iterated integrals on elliptic curves. Part I: general formalism},''
  \href{http://dx.doi.org/10.1007/JHEP05(2018)093}{{\em JHEP} {\bf 05} (2018)
  093},
\href{http://arxiv.org/abs/1712.07089}{{ arXiv:1712.07089 [hep-th]}}.
%%CITATION = ARXIV:1712.07089;%%.

\bibitem{Broedel:2018iwv}
J.~Broedel, C.~Duhr, F.~Dulat, B.~Penante, and L.~Tancredi, ``{Elliptic symbol
  calculus: from elliptic polylogarithms to iterated integrals of Eisenstein
  series},'' \href{http://dx.doi.org/10.1007/JHEP08(2018)014}{{\em JHEP} {\bf
  08} (2018)  014},
\href{http://arxiv.org/abs/1803.10256}{{ arXiv:1803.10256 [hep-th]}}.
%%CITATION = ARXIV:1803.10256;%%.

\end{thebibliography}\endgroup
\providecommand{\href}[2]{#2}\begingroup\raggedright\endgroup

\end{fmffile}
\end{document}